\documentclass[useAMS,usenatbib]{mnras}

\usepackage{natbib}
\usepackage{tabu,booktabs}
\usepackage{amsmath}
\usepackage{xcolor}
\usepackage[english]{babel}
\usepackage{graphicx}
\usepackage{xspace}

\usepackage{longtable}
\usepackage{threeparttablex}
\usepackage{array}

\newcommand{\C}{\texttt{Chempy}}
\newcommand{\G}{\texttt{Gasoline2}}

\def \etal {et~al.~}

\newcommand{\hMpc}{{\ifmmode{h^{-1}{\rm Mpc}}\else{$h^{-1}$Mpc}\fi}}
\newcommand{\Mpc}{{\ifmmode{{\rm Mpc}}\else{Mpc}\fi}}
\newcommand{\hkpc}{{\ifmmode{h^{-1}{\rm kpc}}\else{$h^{-1}$kpc}\fi}}
\newcommand{\kpc}{{\ifmmode{ {\rm kpc} }\else{{\rm kpc}}\fi}}
\newcommand{\kms}{{\ifmmode{ {\rm km\,s^{-1}} }\else{ ${\rm km\,s^{-1}}$ }\fi}}
\newcommand{\hMsun}{{\ifmmode{h^{-1}{\rm {M_{\odot}}}}\else{$h^{-1}{\rm{M_{\odot}}}$}\fi}}
\newcommand{\Msun}{{\ifmmode{{\rm M}_{\odot}}\else{${\rm M}_{\odot}$}\fi}}
\newcommand{\Mhalo}{{\ifmmode{M_{\rm halo}}\else{$M_{\rm halo}$}\fi}}
\newcommand{\Rvir}{{\ifmmode{R_{\rm vir}}\else{$R_{\rm vir}$}\fi}}
\newcommand{\Mvir}{{\ifmmode{M_{\rm vir}}\else{$M_{\rm vir}$}\fi}}
\newcommand{\Mstar}{{\ifmmode{M_{\rm star}}\else{$M_{\rm star}$}\fi}}
\newcommand{\Vrot}{{\ifmmode{V_{\rm rot}}\else{$V_{\rm rot}$}\fi}}
\newcommand{\ltsima}{$\; \buildrel < \over \sim \;$}
\newcommand{\gtsima}{$\; \buildrel > \over \sim \;$}
\newcommand{\lsim}{\lower.5ex\hbox{\ltsima}}
\newcommand{\gsim}{\lower.5ex\hbox{\gtsima}}

\def\lesssim{\mathrel{\hbox{\rlap{\hbox{\lower4pt\hbox{$\sim$}}}\hbox{$<$}}}}
\def\gtrsim{\mathrel{\hbox{\rlap{\hbox{\lower4pt\hbox{$\sim$}}}\hbox{$>$}}}}

\newcommand{\beq}{\begin{equation}}
\newcommand{\eeq}{\end{equation}}
\def\beqa{\begin{eqnarray}}
\def\eeqa{\end{eqnarray}}
\def\LCDM{\ensuremath{\Lambda}CDM}

\def \kms {\ifmmode  \,\rm km\,s^{-1} \else $\,\rm km\,s^{-1}  $ \fi }
\def \kpc {\ifmmode  {\rm kpc}  \else ${\rm  kpc}$ \fi  }  
\def \hkpc {\ifmmode  {h^{-1}\rm kpc}  \else ${h^{-1}\rm kpc}$ \fi  }  
\def \hMpc {\ifmmode  {h^{-1}\rm Mpc}  \else ${h^{-1}\rm Mpc}$ \fi  }  
\def \Mpch {\ifmmode  {h^{-1}\rm Mpc}  \else ${h^{-1}\rm Mpc}$ \fi  }  
\def \Msun {\ifmmode {\rm M}_{\odot} \else ${\rm M}_{\odot}$ \fi} 
\def \hMsun {\ifmmode h^{-1}\,\rm M_{\odot} \else $h^{-1}\,\rm M_{\odot}$ \fi}

\def \LCDM {\ifmmode \Lambda{\rm CDM} \else $\Lambda{\rm CDM}$ \fi}
\def \sig8 {\ifmmode \sigma_8 \else $\sigma_8$ \fi} 
\def \OmegaM {\ifmmode \Omega_{\rm m} \else $\Omega_{\rm m}$ \fi} 
\def \Omegab {\ifmmode \Omega_{\rm b} \else $\Omega_{\rm b}$ \fi} 
\def \OmegaL {\ifmmode \Omega_{\rm \Lambda} \else $\Omega_{\rm \Lambda}$\fi} 
\def \Deltavir {\ifmmode \Delta_{\rm vir} \else $\Delta_{\rm vir}$ \fi}
\def \rhocrit {\ifmmode \rho_{\rm crit} \else $\rho_{\rm crit}$ \fi}
\def \rhou {\ifmmode \rho_{\rm u} \else $\rho_{\rm u}$ \fi}
\def \zc {\ifmmode z_{\rm c} \else $z_{\rm c}$ \fi}

\def\aj{AJ}%
\def\araa{ARA\&A}%
\def\apj{ApJ}%
\def\apjl{ApJ}%
\def\apjs{ApJS}%
\def\ao{Appl.~Opt.}%
\def\aap{A\&A}%
\def\aapr{A\&A~Rev.}%
\def\mnras{MNRAS}%
\def\pasa{PASA}%
\def\pasp{PASP}%
\def\nphysa{Nucl.~Phys.~A}%
\def\head{ .ps \vbox to 0pt{\vss \hbox to 0pt{\hskip 440pt\rm
      LA-UR-10-07069\hss} \vskip 25pt}} 

\def \spose#1{\hbox  to 0pt{#1\hss}}  
\def \lta{\mathrel{\spose{\lower 3pt\hbox{$\sim$}}\raise 2.0pt\hbox{$<$}}}
\def \gta{\mathrel{\spose{\lower 3pt\hbox{$\sim$}}\raise 2.0pt\hbox{$>$}}}

\title[Simulated diversity of chemical abundance patterns]{The challenge of simultaneously matching the observed diversity of chemical abundance patterns in cosmological hydrodynamical simulations}

\author[T. Buck \etal] {Tobias Buck$^{1}$\thanks{E-mail: tbuck@aip.de}, Jan Rybizki$^{2}$, Sven Buder$^{3,4}$,  Aura Obreja$^{5}$, Andrea V. Macci\`o$^{6,2}$,
\newauthor{Christoph Pfrommer$^{1}$, Matthias Steinmetz$^{1}$, Melissa Ness$^{7}$}\\
$^1$Leibniz-Institut f\"ur Astrophysik Potsdam (AIP), An der Sternwarte 16, D-14482 Potsdam, Germany\\
$^2$Max-Planck-Institut f\"ur Astronomie, K\"onigstuhl 17, 69117 Heidelberg, Germany\\
$^3$Research School of Astronomy \& Astrophysics, Australian National University, ACT 2611, Australia\\
$^4$Center of Excellence for Astrophysics in Three Dimensions (ASTRO-3D), Australia\\
$^5$Universit\"ats-Sternwarte M\"unchen, Scheinerstraße 1, D-81679 M\"unchen, Germany\\
$^6$New York University Abu Dhabi, PO Box 129188, Saadiyat Island, Abu Dhabi, United Arab Emirates\\
$^7$Department of Astronomy, Columbia University, 550 West 120th Street, New York, NY, USA
}

\begin{document}

\date{Accepted XXXX . Received XXXX; in original form XXXX}

\pagerange{\pageref{firstpage}--\pageref{lastpage}} \pubyear{2021}

\maketitle

\label{firstpage}

\begin{abstract}
With the advent of large spectroscopic surveys the amount of high quality chemo-dynamical data in the Milky Way (MW) increased tremendously.
Accurately and correctly capturing and explaining the detailed features in the high-quality observational data is notoriously difficult for state-of-the-art numerical models.
In order to keep up with the quantity and quality of observational datasets, improved prescriptions for galactic chemical evolution need to be incorporated into the simulations. 
Here we present a new, flexible, time resolved chemical enrichment model for cosmological simulations. Our model allows to easily change a number of stellar physics parameters such as the shape of the initial mass function (IMF), stellar lifetimes, chemical yields or SN Ia delay times. We implement our model into the \texttt{Gasoline2} code and perform a series of cosmological simulations varying a number of key parameters, foremost evaluating different stellar yield sets for massive stars from the literature. 
We find that total metallicity, total iron abundance and gas phase oxygen abundance are robust predictions from different yield sets and in agreement with observational relations. On the other hand, individual element abundances, especially $\alpha$-elements show significant differences across different yield sets and none of our models can simultaneously match constraints on the dwarf and MW mass scale. This offers a unique way of observationally constraining model parameters. For MW mass galaxies we find for most yield tables tested in this work a bimodality in the $[\alpha$/Fe] vs. [Fe/H] plane of rather low intrinsic scatter potentially in tension with the observed abundance scatter. 
\end{abstract}

%%%%%%%%%%%%%%%%%%%%%%%%%%%%%%%%%%%%%%%%%%%%%%%%%%%
\noindent
\begin{keywords}

Galaxy: structure --- galaxies: evolution --- galaxies: abundances --- galaxies:
  formation --- Galaxy: abundances --- methods: numerical
 \end{keywords}

%%%%%%%%%%%%%%%%%%%%%%%%%%%%%%%%%%%%%%%%%%%%%%%%%%%

%%%%%%%%%%%%%%%%%%%%%%%%%%%%%%%%%%%%%%%%%%%%%%%%%%%
\section{Introduction} \label{sec:introduction}
%%%%%%%%%%%%%%%%%%%%%%%%%%%%%%%%%%%%%%%%%%%%%%%%%%%

Galaxies are the lighthouses in the vast ocean of the Universe \citep{Mo2010}. Their structure and appearance is the result of the interplay of several generations of stars which are shaped by different epochs of gas accretion and expulsion. Therefore galactic structures encode valuable information about the fundamental physical processes governing galaxy formation. Especially the abundance of chemical elements in stars, the inter-stellar medium (ISM) and the circum-galactic medium (CGM) serves as an ideal tracer of the integrated galactic history. 

The observed diversity of chemical elements with atomic numbers larger or equal to C is the result of stellar nucleosynthesis and the distribution of those elements in the Universe a consequence of gas expulsion from star forming regions and galaxies via processes dubbed feedback \citep[e.g.][]{Maiolino2019}. Those chemical elements are ideal observational tracers to determine the evolutionary state of stars or the physical conditions such as density, temperature or ionisation state of the gas. They further contribute importantly to the dynamics of the whole system as main coolants which enable the dissipation of energy by the emission of radiation. Thus the production and evolution of heavy elements is 
one of the fundamental ingredients of modern models for the formation and evolution of both stars and galaxies \citep[e.g.][]{Vogelsberger2020}. 

The stellar lifetimes and the production of new chemical elements (in the reminder of this article referred to as metals) strongly depends on the mass of a star and its own chemical composition. When a generation of stars evolves it will pollute the ISM with chemical elements fused in their cores, predominantly during their last evolutionary phases. As a consequence, the successive generations of stars born from the enriched ISM will attain a higher mass fraction of heavy elements with respect to their progenitors. Thereby chemical evolution does not proceed in a linear manner due to the strong mass dependence of stellar evolution timescales. For example, high mass stars evolve much more quickly compared to low and intermediate mass stars ($\tau\sim4$ Myr for a $40\Msun$ star and $\tau\sim10$ Gyr for a $1\Msun$ star).  

Therefore, higher mass stars first enrich the ISM with heavy chemical elements while intermediate and low mass stars lock the chemical elements produced in past generations for much longer times \citep[up to $1-10$ Gyr, e.g.][]{Marigo2001,Thielemann2003} with the immediate consequence that the chemical characteristics of second generation stars will be mainly influenced by the dying massive stars. Consequently, including the (time resolved) formation, distribution, exchange and recycling of chemical elements into any simulation of the formation and evolution of galaxies is critical for our understanding.

Since the first implementation of chemical enrichment by the metal release of core-collapse supernovae (CC-SN) into a smoothed particle hydrodynamics (SPH) code \citep{Steinmetz1994} much progress in the modelling of galactic chemical evolution has been made \citep[e.g.][]{Raiteri1996,Recchi2001,Mosconi2001,Kawata2001,Lia2002,Tornatore2007,Oppenheimer2008,Hirschmann2016}.
By now there exist several models that distinguish between elements originating from SN~Ia and CC-SN and in some cases also from asymptotic-giant branch (AGB) stars and the winds from their progenitors \citep[e.g.][]{Scannapieco2005,Wiersma2009,Naiman2018}.

Additionally, much progress in the collection of high quality observational data has been made and a correlation between the observed abundance patterns of chemical elements in stars and the ISM with the mass of a galaxy, the position within a galaxy and the birth epoch of the stars \citep[e.g.][]{Baade1944,Wallerstein1962,Hayden2015} could be established. Furthermore, there exist a great wealth of data for the chemical abundances of the CGM at low- and high-redshift. 
But most notably with the advent of large spectroscopic surveys the amount of high quality chemo-dynamical data in the Milky Way (MW) increased tremendously \citep{Boeche2011,Anders2014,Buder2019,Xiang2019,Steinmetz2020} exhibiting distinct chemical abundance patterns of the stellar disk \citep[e.g.][]{Boeche2013,Hayden2015}. Because the surface abundance of stars is assumed to reflect the chemical composition of the star's birth gas cloud the measured chemical abundances present a fossil record of the Galaxy's formation history. The study of the chemo-kinematic structures of the MW in combination with inferred stellar ages \citep{Bensby2014,Martig2016,Feuillet2016,Mackereth2017,Ness2019} enabled the field of Galactic Archeology \citep{Freeman2002} and has lead to unique insights into the formation and evolution of the MW's stellar disk \citep[e.g.][]{Minchev2013,Sestito2021}.

Accurately capturing and explaining the detailed features present in the observational data is notoriously difficult but recent numerical models begin to reproduce observed features \citep[e.g.][]{Grand2018,Hilmi2020,Khoperskov2020,Agertz2020,Renaud2020}. By construction chemical evolution models encompass a large number of stellar physics parameters such as the shape of the initial mass function (IMF), stellar lifetimes or chemical yields or SN~Ia delay times. Many of those parameters are poorly constrained a priori \citep{Cote2016} and chosen ad-hoc. Recently, \citet{Gutcke2019} for instance studied the influence of a metallicity dependent high mass slope of the IMF on galactic abundance patterns. Perhaps the most important external or theoretical input for chemical evolution models are the nucleosynthetic yields for the various enrichment channels \citep{Romano2010,Cote2016b}. In the past, theoretical yields have produced mismatches with the observations and the physical shortcomings of stellar nucleosynthetic yield models are still under debate \citep[e.g.][]{Nomoto2013,Pignatari2016,Mueller2016}.

Thus, in order to cope with the large range of parameters to explore, various galactic chemical evolution (GCE) models have been developed \citep[e.g.][]{Tinsley1979,Chiappini1997,Rybizki2017,Ritter2018b,Kobayashi2020}. These models range from analytical ones with simplifying assumptions \citep[e.g.][]{Weinberg2017,Spitoni2017,Spitoni2020} to hydrodynamical simulations including detailed chemical enrichment as well as galactic dynamics \citep[e.g.][]{Kawata2003,Few2012,Buck2020,Brook2020}. All these analytical models made possible to test a large set of parameters over a wide range of values, and marginalise over parameters for simple astrophysical inferences. Yet, due to their simplifying assumptions those models lack the possibility of including all the complex dynamics present in full hydrodynamical simulations which might potentially explain some of their shortcomings as seen \citet[e.g. in][]{Blancato2019}.  

In this work we present a new hybrid approach where essential chemical evolution parameters are first constrained by fitting a galactic chemical evolution model to observational data \citep{Philcox2018} and then employed in the chemical enrichment prescription in cosmological hydrodynamical simulations of the formation of galaxies from dwarf to MW mass.

This paper is organised as follows: in \S2 we present the details of the chemical evolution model followed by the description of a novel implementation of chemical enrichment into a cosmological simulation code in \S3. Readers which are already familiar with the concept of simple stellar population models and their usage in cosmological simulations might readily skip those two sections. We continue in \S4 we analyse the results of the numerical simulations with special emphasise on the abundance patterns of the stellar disks in light of current galactic archeological studies. We end this study with a discussion and summary of our results in \S5.

%%%%%%%%%%%%%%%%%%%%%%%%%%%%%%%%%%%%%%%%%%%%%%%%%%%
\section{Method: A flexible chemical evolution model} \label{sec:method}
%%%%%%%%%%%%%%%%%%%%%%%%%%%%%%%%%%%%%%%%%%%%%%%%%%%

State-of-the-art computational resources still do not enable cosmological simulations to resolve individual stars. Instead, the standard approach is to discretise the stellar density field into tracer particles which represent a population of stars. The population of stars contained in those stellar particles is described by a single age, $\tau_{\rm star}$, a metallicity, $Z_{\rm star}$ and an IMF specifying the number of stars in a given mass bin. Such stellar populations are usually referred to as simple stellar population (SSP).
In order to determine how the SSP interacts with the surrounding ISM one has to determine the (time resolved) amount of energy, metals and mass released by the stellar particle, usually referred to as stellar feedback. 

There are many stellar evolution models in the literature employing slightly different stellar lifetimes functions, IMFs or metal mass return fractions. Instead of deciding for one specific model we make use of the flexible chemical evolution code \C\footnote{A  Python implementation of the current version of \C\ can be found on GitHub \url{https://github.com/jan-rybizki/Chempy}.} \citep{Rybizki2017} in order to synthesise the final stellar evolution model used in the simulations. This means we are using \C\, to fix the SSP parameters to synthesise the yield tables to be used for a specific run. In future work \C\, can be further used to fit the SSP parameters to observed data and thus constrain the SSP models employed in hydrodynamical simulations\footnote{Note, we use the part of \C\, that calculates the IMF integrated metal yield of an SSP. This is equivalent to \texttt{SYGMA} \citep{Ritter2018b} in the NUPYCEE collaboration.}.

In summary, \C\ is a general purpose tool to link the parameters of a stellar (chemical) evolution model $\theta$ (e.g. the IMF high-mass slope and/or a specific yield set) to the likelihood of an observations $\mathcal{O}$ (such as stellar abundances) via its model predictions $d$. A detailed prescription of the modelling procedure is given in the code paper \citep{Rybizki2017} but for completeness we describe the basic working principle of \C\ below. Since our aim is to use \C\ to create an SSP model which we then can utilise as an input to cosmological simulations we focus here on the SSP part of \C\ while only briefly explaining its ISM parameters which are of little interest for us.

\subsection{The Simple Stellar Population model}
\label{sec:ssp}

At the core \C~calculates the time evolution of the chemical yields for an SSP. In order to specify the stellar physics of the chemical evolution model at least seven parameters need to be set (see table \ref{tab:params}) and additional parameters for the stellar nucleosynthetic yields have to be chosen (see table \ref{tab:yields}). Currently those include the relative mass return fractions from three nucleosynthetic channels: Supernova of type Ia (SN\,Ia), core-collapse supernova (CC-SNe) and asymptotic giant branch stars (AGB).

An SSP is fully characterised by its age, mass and initial element composition, $\mathrm{SSP}\left(\tau_\mathrm{star},m,[\mathrm{X/H}]\right)$. By assuming some IMF, a stellar lifetime function and specific nucleosynthetic yields, the stellar (energetic and mass) feedback is simultaneously fixed. In the case of numerical simulations the total mass of an SSP is set by the stellar particle mass and its initial elemental abundance is inherited from the composition of the ISM from which the star particle is formed. For this work we are focussing on fixed feedback fractions and do not consider stochastically sampling the IMF, though it can be accounted for in \C.

%%%%%%%%%%%%%%%%% TABLE 1 %%%%%%%%%%%%%%%%%%%%%%%%%%%
\begin{table}
\begin{center}
\caption{Free stellar evolution parameters, $\theta$, used in \C\ together with the fiducial values adopted in this work.}
\label{tab:params}
%\begin{tiny}
\begin{tabular}{ r l c }
  \hline  \hline
	$\theta$ & description & $\overline{\theta}_\mathrm{fiducial}$\\
  \hline
  	IMF type & functional form of IMF & \\
  	$\alpha_\mathrm{IMF}$ & Chabrier high-mass slope & $-2.3$  \\
	 & IMF mass range & $0.1-100$ \Msun \\
	 & CC-SN mass range & $8-40$ \Msun \\
	 & SNIa delay time exponent& $-1.12$ \\
  	$\log_{10}\left(\mathrm{N}_\mathrm{Ia}\right)$ & normalisation of SN Ia rate& $-2.9$ \\
   	$\tau_\mathrm{Ia}$ & SN\,Ia delay time in Gyr & $40$\,Myr\\
	$Z_{\rm SSP}$ & metallicity of the SSP & $10^{-5}Z_\odot - 2Z_\odot$\\
  \hline
\end{tabular}
%\end{tiny}
\end{center}
\end{table}
%%%%%%%%%%%%%%%%%%%%%%%%%%%%%%%%%%%%%%%%%%%%%%%%%%%%%

\subsubsection{Stellar Initial Mass Function}
The amount, $N$, of stars of given stellar mass $m$ included in the SSP are set by the functional form of the IMF which in the fiducial case is assumed to follow the \citet[][]{Chabrier2003} IMF
\begin{align}
\label{eq:imf}
&\frac{{\rm{d}}N}{{\rm{d}}m}= 0.158 \exp\left[-\frac{\left(\log m/\Msun-\log 0.079\right)^2}{2\times0.69^2}\right], \!\!\!\! &m<1\Msun \nonumber\\
&\frac{{\rm{d}}N}{{\rm{d}}m}=A ~\left(m/\Msun\right)^{-\left(1+\alpha_\mathrm{IMF}\right)},\!\!\!\! &m>1\Msun
\end{align}
where A is determined by a smooth transition of the two regimes and the high-mass slope ($\alpha_\mathrm{IMF}$) is one of \C's basic free parameters. For this work we assume $\alpha_\mathrm{IMF}$ to be universal with respect to stellar metallicity and cosmic time. The parameter $\alpha_\mathrm{IMF}$ sets the ratio of low- to high-mass stars and controls the number of high-mass stars. Thus for our purposes the exact choice of $\alpha_\mathrm{IMF}$ strongly influences the number of CC-SN and therefore the available feedback energy and elemental composition of the feedback. The total range of stellar masses over which the IMF extends can also be set and we have chosen a range from $m_{\rm min}=0.1$ to $m_{\rm max}=100\Msun$ (see also table \ref{tab:params}).

\begin{figure*}
\begin{center}
\begin{minipage}{.495\textwidth}
\centering
\includegraphics[width=\textwidth]{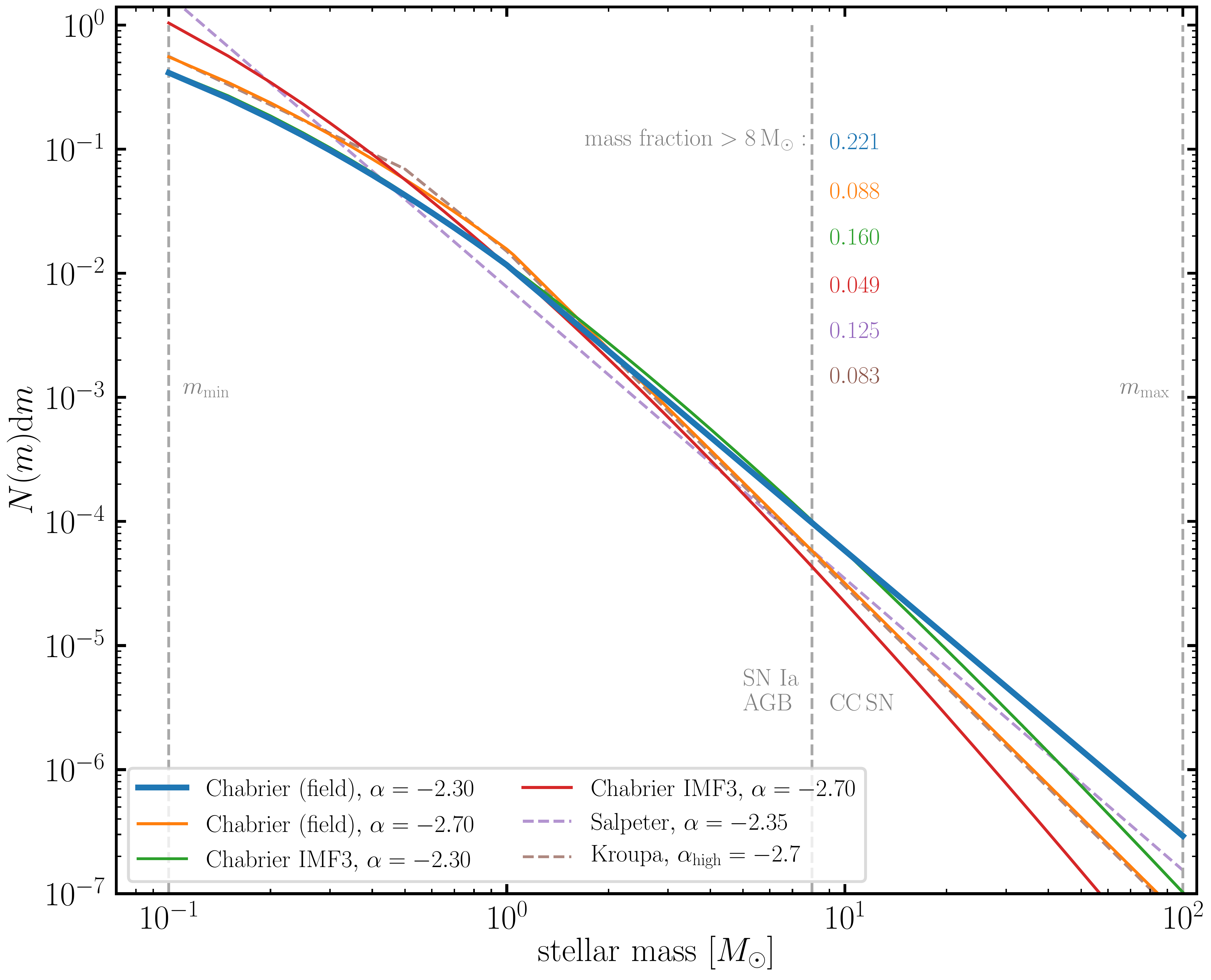}
\end{minipage}
\begin{minipage}{.495\textwidth}
\centering
\includegraphics[width=\textwidth]{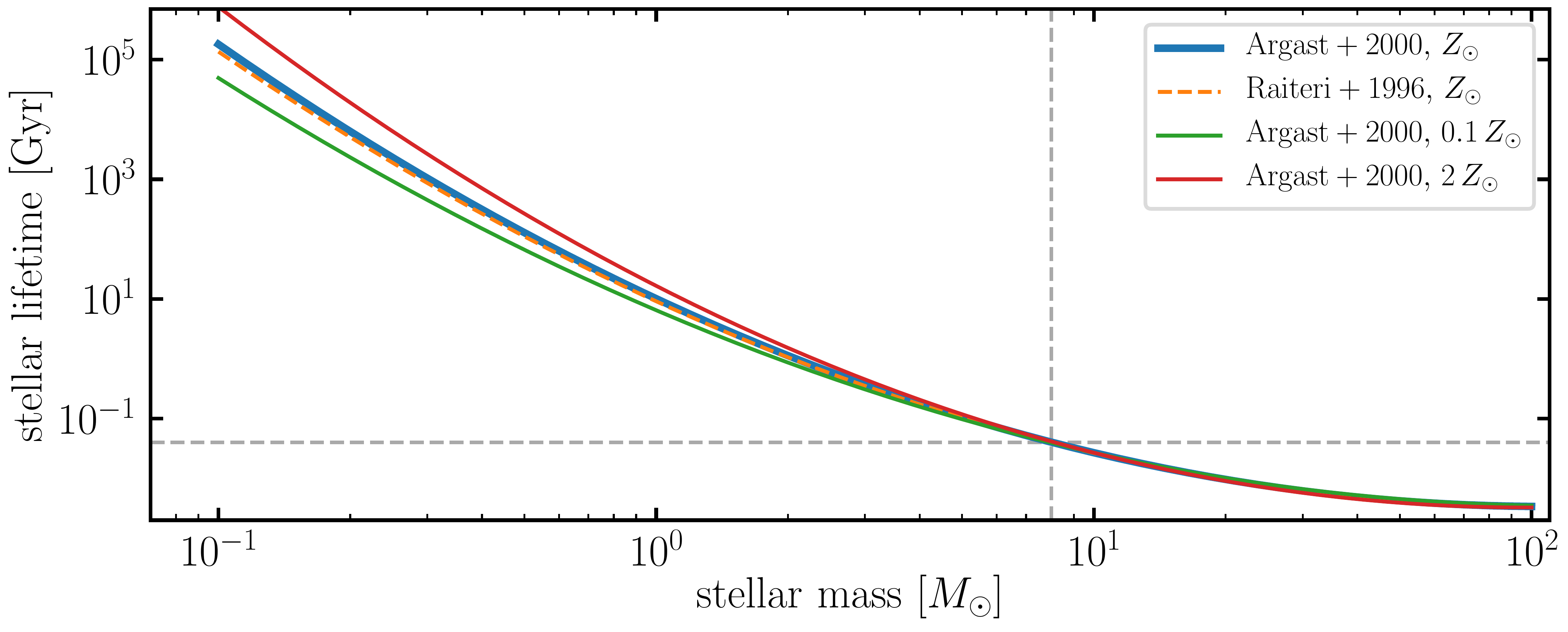}
\includegraphics[width=\textwidth]{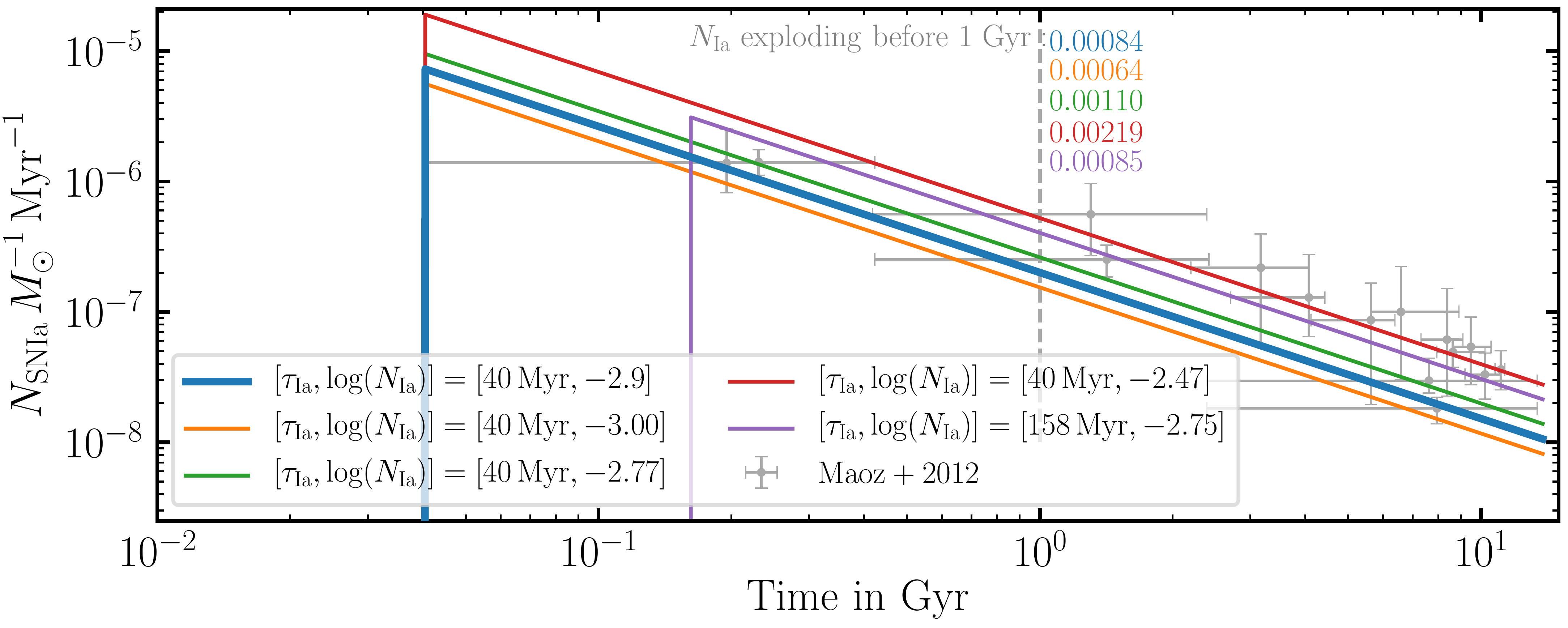}
\end{minipage}
\end{center}
\vspace{-.35cm}
\caption{The simple stellar population model parameters: The stellar initial mass function (IMF, left panel), the stellar lifetime function (upper right panel) and the SNIa delay time distribution function (lower right panel). In the left panel we compare different IMFs as indicated in the legend. For each IMF the coloured numbers indicate the mass fraction of star above $8 \Msun$ up to the assumed maximum mass of $100 \Msun$. The upper right panel compares stellar lifetimes as a function of stellar mass for models from \citet{Argast2000} for three different metallicities with a model from \citet{Raiteri1996} at solar metallicity. Gray dashed lines indicate the lifetime of an $8 \Msun$ star which is $\sim40$ Myr. The lower right panel shows the empirical SNIa delay time distribution \citep{Maoz2010} with parameters taken from \citet{Maoz2012} and \citet{MaozMannucci2012}. Gray data points show the observational data taken from \citet{Maoz2012}.}
\label{fig:IMF}
\end{figure*}

\C~offers the possibility to flexibly choose the functional form of the IMF. Next to the \citet{Chabrier2003} IMF there are the \citet{Chabrier2001}, the \citet{Kroupa1993}, \citet{Salpeter1955} and a generic broken power law IMF \citep[e.g.][]{Miller1979} implemented. In the left panel of Fig. \ref{fig:IMF} we show a comparison of our fiducial IMF with a high mass slope of $\alpha_\mathrm{IMF}=-2.3$ (blue line), its 2$\sigma$ deviation $\alpha_\mathrm{IMF}=-2.7$ (orange line), with the other IMFs implemented in \C.

%%%%%%%%%%%%%%%%%% FIGURE 1 %%%%%%%%%%%%%%%%%%%%%%%%%%%
%%%%%%%%%%%%%%%%%%%%%%%%%%%%%%%%%%%%%%%%%%%%%%%%%%%%

\subsubsection{Stellar Lifetimes}
Newly synthesised elements are released into the ISM at the end of a star's lifetime either via AGB star winds or via SN explosions. In order to determine when those events take place we need to specify a stellar lifetime function which in general are a strong function of initial stellar mass and only weakly depend on stellar metallicity (see e.g. upper right panel of Fig. \ref{fig:IMF}). \C\ implements the stellar lifetime functions of \citet{Argast2000} and \citet{Raiteri1996} where the former is our fiducial choice. The upper right panel of Fig. \ref{fig:IMF} shows stellar lifetimes as a function of stellar mass at solar metallicity for both parameterizations (blue solid and orange dashed line) in comparison to intrinsic lifetime variations predicted for different initial metallicities (red and green lines). This clearly shows that the metallicity dependence of stellar lifetimes is stronger than the differences between different authors. 

Most massive stars will end their lives as a CC-SNe and feed back mass and energy into to the ISM. The exact stellar mass which divides exploding from non-exploding stars is still under debate \citep[e.g.][]{Doherty2017} but usually assumed to be around $m_{\rm CC-SN}=8\Msun$ (vertical gray dashed line in Fig. \ref{fig:IMF}). Similarly there is evidence that the highest mass stars might end their lives with a direct collapse into a black hole not releasing any energy or elements into the surrounding ISM. In this work, we assume that all stars more massive than $40\Msun$ will follow this path and set the maximum mass of stars that explode as CC-SN to $40\Msun$. 

\begin{figure*}
\begin{center}
\begin{minipage}{.495\textwidth}
\centering
\includegraphics[width=\textwidth]{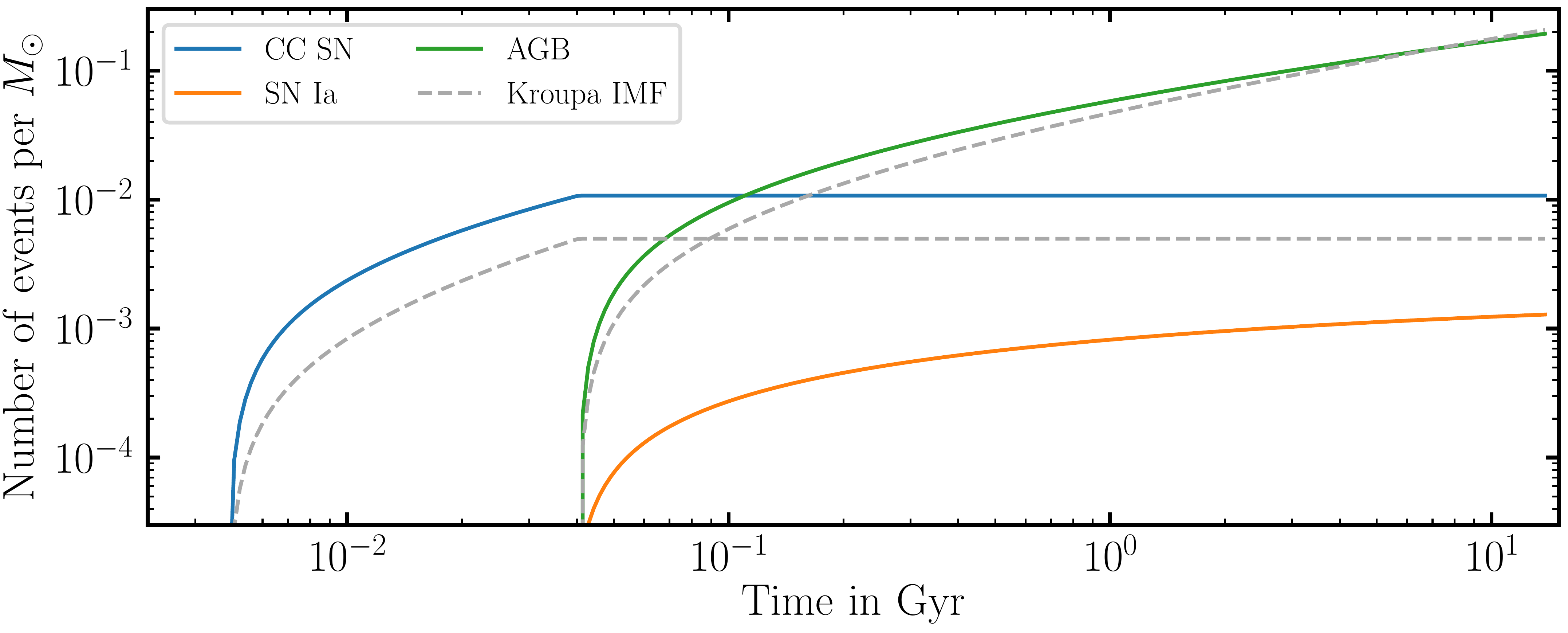}
\includegraphics[width=\textwidth]{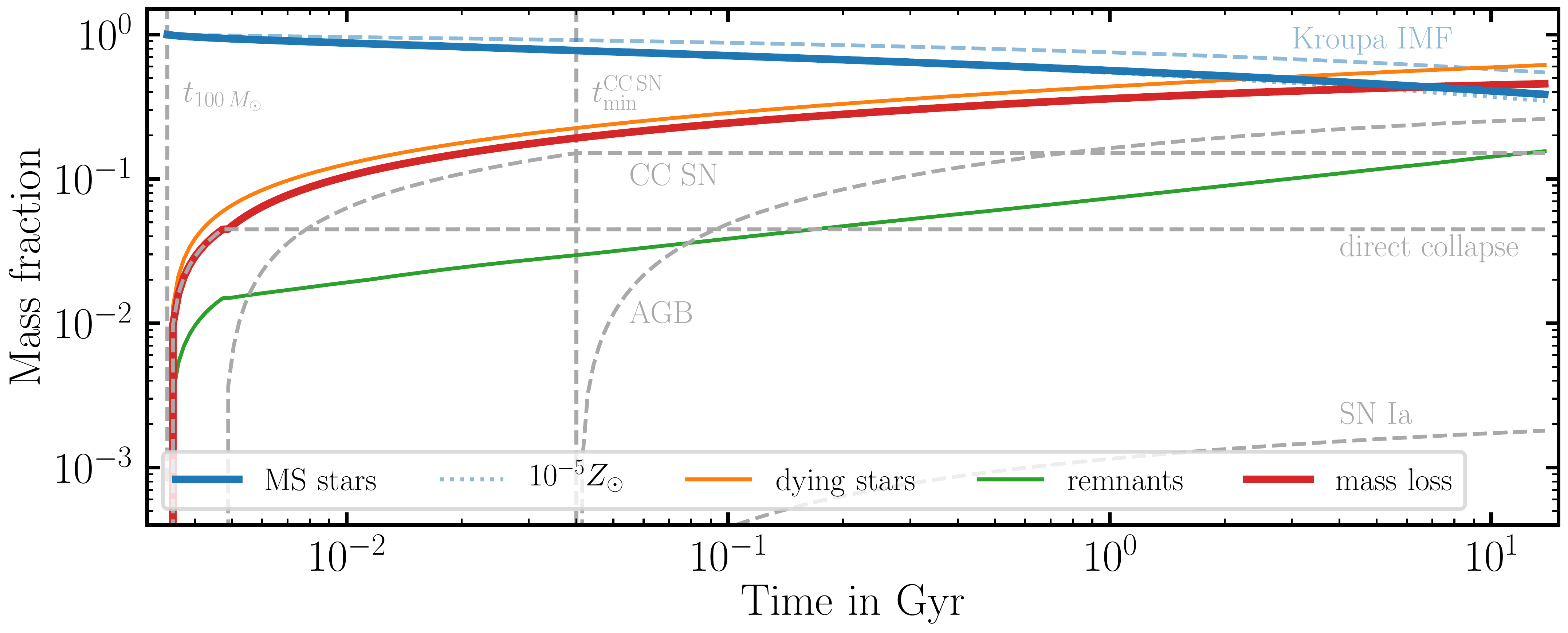}
\end{minipage}
\begin{minipage}{.495\textwidth}
\centering
\includegraphics[width=\textwidth]{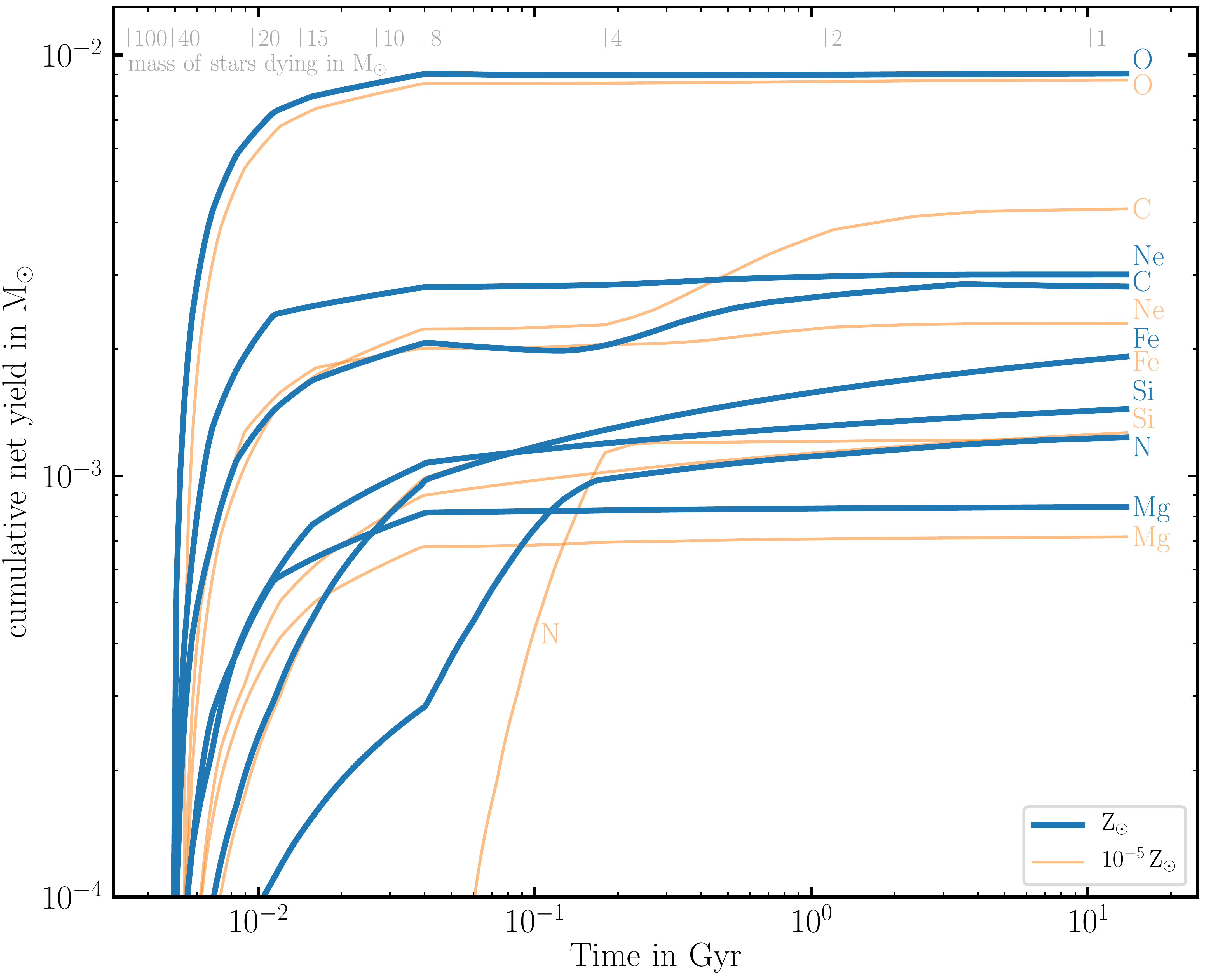}
\end{minipage}
\end{center}
\vspace{-.35cm}
\caption{The simple stellar model: The upper left panel shows the number of CC-SN, SN~Ia and AGB stars as a function of time at solar metallicity assuming a fiducial Chabrier IMF (solid coloured lines) while gray dashed lines assume a Kroupa IMF. The lower left panel shows the corresponding mass fraction as a function of time of different components of the SSP. The solid blue line shows the mass fraction in main sequence stars as a function of SSP age assuming a Chabrier IMF while the dashed blue line assumes a Kroupa IMF. The red solid line shows the total mass loss of the population (assuming a Chabrier IMF) which is lower than the mass fraction of dying stars (orange line) by the amount of mass in stellar remnants (green line, e.g. stellar mass black holes or white dwarfs). Gray dashed lines show the mass loss for each individual channel (CC-SN, AGB stars and SN~Ia). The two vertical dashed lines show the lifetime of a $100 \Msun$ star and the lifetime of the minimum mass star exploding as CC-SN ($8 \Msun$ here). The right hand panel shows the corresponding cumulative ejected SSP mass split into different elements for two different metallicities assuming stellar yields taken from \citet{Chieffi2004,Karakas2016,Seitenzahl2013}.}
\label{fig:massloss}
\end{figure*}

Combining the IMF with the stellar lifetime function we are able to calculate the time evolution of CC-SN explosions and the number of stars in the AGB phase for a given SSP. In the upper left panel of Fig. \ref{fig:massloss} we compare our fiducial model (Chabrier IMF) evolution to an alternative model using the Kroupa IMF. The most important difference is the number of high mass stars given by the different high mass slopes of the IMFs which manifests itself in a different number of CC-SN. The Chabrier IMF results in roughly twice the number of CC-SN compared to the Kroupa IMF.

%%%%%%%%%%%%%%%%%% FIGURE 2 %%%%%%%%%%%%%%%%%%%%%%%%%%%
%%%%%%%%%%%%%%%%%%%%%%%%%%%%%%%%%%%%%%%%%%%%%%%%%%%%

\subsubsection{SN~Ia delay times}
The progenitor stars as well as the explosion mechanism resulting in SN\,Ia explosions is less well understood. Therefore we treat the fraction of stars and the mass range in which stars explode as SN\,Ia empirically. We parametrise the delay time distribution (DTD) of the SN\,Ia explosions with a power-law following \citet{Maoz2010} (see also table \ref{tab:params}):
\begin{equation}
N_{\rm SNIa}\propto t^{-1.12}.
\end{equation}
The normalisation, the number of SN\,Ia per solar mass over a time of 15\,Gyr ($\mathrm{N}_\mathrm{Ia}$), and the time delay until the first SN\,Ia events occur ($\tau_\mathrm{Ia}$) are free parameters obtained from observations \citep[e.g.][]{Maoz2012}. The lower right panel of Fig. \ref{fig:IMF} shows various realisations of the delay time distribution compared to the data on SN\,Ia explosions by \citet{Maoz2012}. Our default model assumes the SN\,Ia normalisation ($\mathrm{N}_\mathrm{Ia}$) from \citet[tab.\,1]{MaozMannucci2012} and the delay time ($\tau_\mathrm{Ia}$) is taken to be the minimum lifetime of stars that might explode as SN\,Ia, $\tau_{\rm Ia}=40$ Myr for $8\Msun$ stars. The resulting number of SN\,Ia explosions as a function of time is shown in Fig. \ref{fig:massloss}. Through our empirical choice of parameterization this is independent of the exact IMF chosen.

We like to note here, that \C\ is able to calculate the distribution of stars along the IMF and the SN\,Ia explosions as a stochastic process. However, as soon as the SSP mass is sufficiently large ($\gsim$$10^4$ M$_\odot$) this converges rapidly to the analytic solution. Therefore and for the sake of computational simplicity we employed the faster analytic version in this work. 

\subsubsection{Stellar Massloss}
\label{sec:massloss}

We consider three different channels contributing to the mass and elemental enrichment: Supernova of type Ia (SN\,Ia), core-collapse supernova (CC-SNe) and asymptotic giant branch stars (AGB). For the latter two channels we consider metallicity and mass dependent enrichment, i.e. the elemental enrichment and the remnant mass depend on stellar metallicity as well as the mass of dying stars\footnote{The energy release of CC-SNe on the other hand is fixed to the fiducial value of $10^{51}$ erg per explosion and we do not consider energetic feedback from AGB winds for now.} and as such explicitly on stellar lifetime. We further assume that all relevant materials are only ejected at the end of the star's lifetime. SN\,Ia on the other hand follow a different explosion path and we assume no metallicity dependence. The equations of how to calculate the massloss of a given SSP between times $t_0$ and $t_1$ are e.g. given in section 4, equation 2 and 3 of \cite{Wiersma2009}. 
The resulting time release of mass into the ISM and the accompanied massloss of the SSP particle is shown in the lower left panel of Fig. \ref{fig:massloss}. The blue solid line shows the evolving mass of main sequence stars in the SSP for our fiducial model (Chabrier IMF), the dashed blue line shows an alternative model assuming a Kroupa IMF. Comparing the two lines we see that the choice of the IMF has a significant impact on the expected massloss, less so the initial metallicity of the SSP which is shown with a blue dotted line and almost follows the fiducial solid blue line.

The thin orange line shows the amount of mass of dying stars as a function of time. Not all the mass of dying stars is returned to the ISM as the exploding stars leave behind a stellar remnant (green line, black hole or white dwarf depending on the star's mass). The red solid line thus shows the amount of mass returned to the ISM which results from subtracting the remnant mass from the mass of dying stars. Thin gray dashed lines split the massloss into the different channels traced by our model. Note, after a Hubble time most mass returned to the ISM comes from AGB star winds followed by CC-SN. For the most massive stars undergoing direct black hole collapse we assume a mass fraction of $0.75$ is returned to the ISM. SN\,Ia explosions feed back only very little absolute mass. However, as we will see later their contribution to the iron mass is still significant. 

%%%%%%%%%%%%%%%%%% FIGURE 3 %%%%%%%%%%%%%%%%%%%%%%%%%%%
\begin{figure*}
\begin{center}
\includegraphics[width=\textwidth]{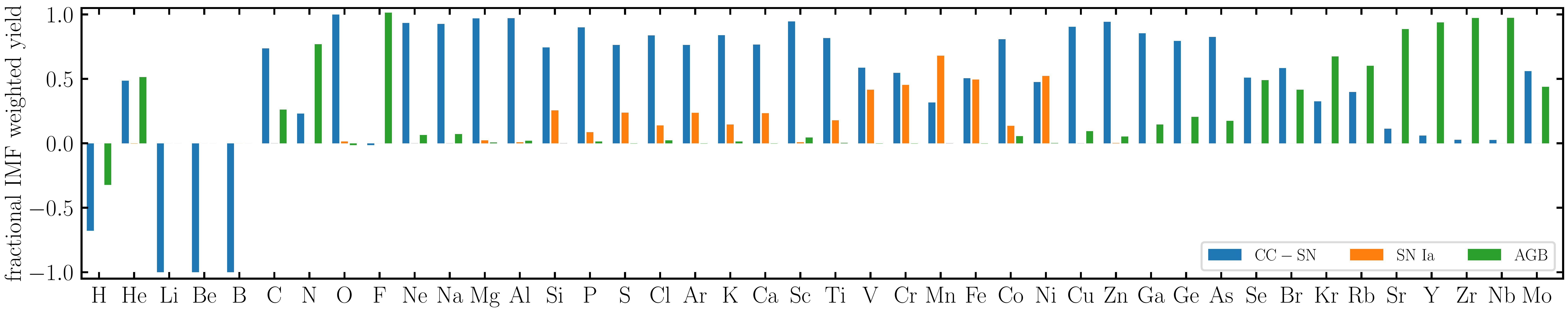}
\includegraphics[width=\textwidth]{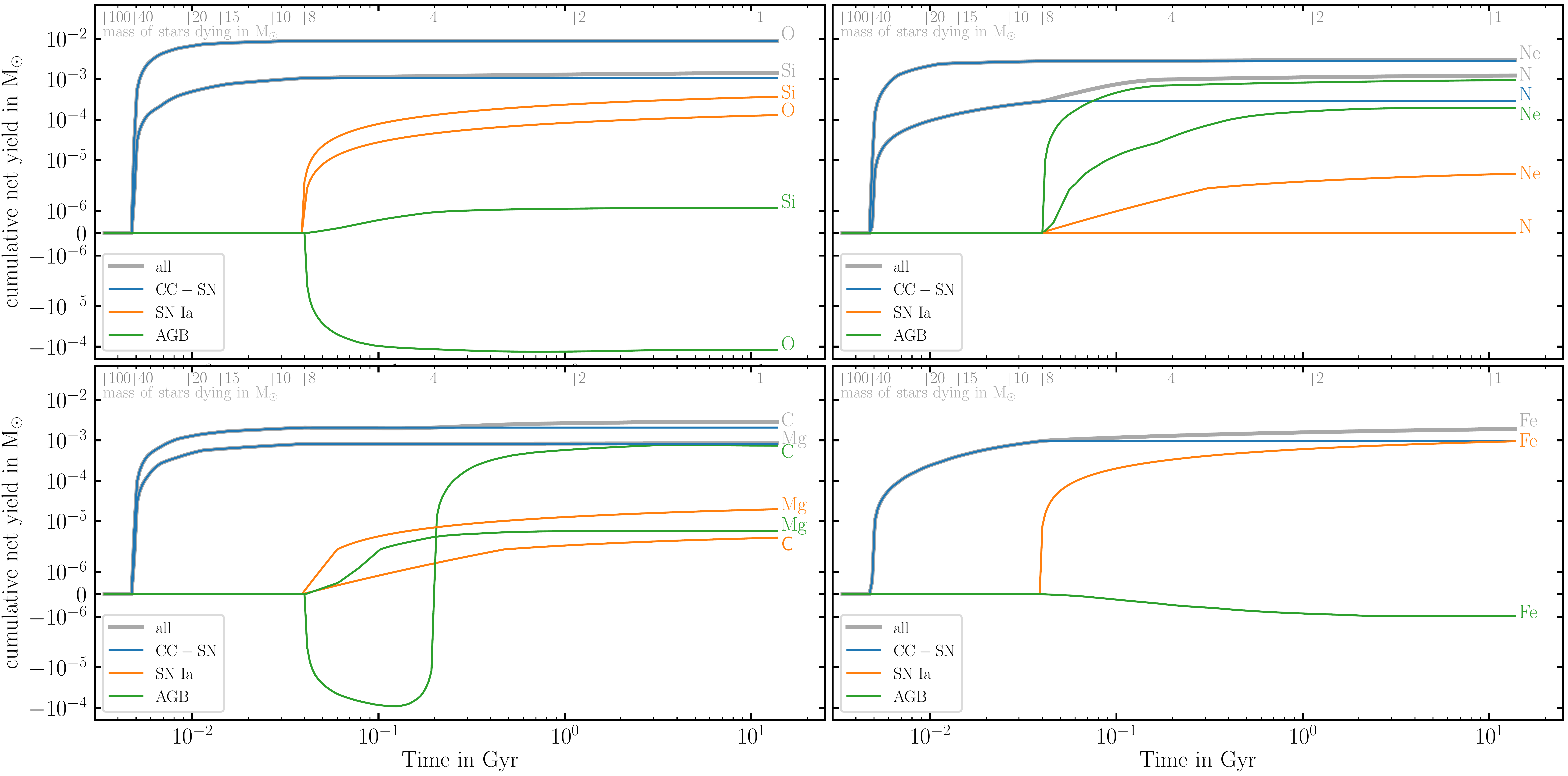}
\end{center}
\vspace{-.35cm}
\caption{Example tracks of traced elements. The upper panel shows the fractional element return of the whole SSP for 42 example elements (out of a total of 81 tracked elements) split into the three different channels of elements produced by CC-SN (blue), AGB (green) stars and SNIa (orange) after a Hubble time. The histograms show the relative contribution of different processes to the total metal return of the SSP; e.g. CC-SN and SN~Ia contribute equally to the production of iron. Negative values as e.g. for hydrogen indicate destruction/fusion of elements. The four bottom panels show the element return of the SSP as a function of time for different elements as indicated in each panel. We split the total element return (gray line) into the three different channels showing the contribution by CC-SN in blue, by AGB stars in green and by SNIa in orange. Note, the negative iron yield for AGB stars reflects the iron destruction through neutron capture-process.}
\label{fig:yieldhist}
\end{figure*}
%%%%%%%%%%%%%%%%%%%%%%%%%%%%%%%%%%%%%%%%%%%%%%%%%%%%

\subsubsection{Nucleosynthetic yields}

The elemental enrichment in \C\ is calculated according to publicly available yield tables from the literature. A list of implemented yield tables is presented in table \ref{tab:yields} but it is straight forward to extend this list further using \C's python functions. Our fiducial yield set for example combines tables by \citet{Karakas2016} for AGB stars with tables from \citet{Chieffi2004} for CC-SN and yields from \citet{Seitenzahl2013} for SN\,Ia\footnote{We use their N100 model calculated in 3D superseding the 1D W7 model of \cite{Iwamoto1999} which had old electron capture rates overproducing Ni and underproducing Mn abundances. This is remedied with the \citet{Seitenzahl2013} models which best reproduce observables by \citet{Sim2013}}. 

Stellar nucleosynthesis changes the relative fraction of elements inside a star by different destruction and production rates for various elements. Light elements such as hydrogen or helium are preferentially destroyed in favour of more massive elements. Similarly massive elements such as iron do not take part in further fusion processes and can be regarded as the end product of stellar nucleosynthesis. Therefore we split the elemental enrichment of stars into two parts: the first part describes the amount of newly synthesised elements and the second part describes the initial abundance of elements which passes unaltered through the star. This second part describes the elemental composition of the birth ISM which is locked up in the star's atmosphere and released back into the ISM at the end of the star's life \citep[see also discussion in section 4 of][]{Wiersma2009}. We refer to the yield sets following this distinction of newly produced elements from the birth elemental abundance of the stars as net-yields while the tables including the birth composition are referred to as gross-yields. Formulating the elemental mass return in this way ensures that the yield tables are independent of the exact initial chemical composition of the underlying CC-SN models and contain only the newly synthesised elements. On the other hand, in the gross-yield formulation tables might contain elements which were neither synthesised in the star nor available in the ISM it was formed from simply because they were part of the initial abundances of the CC-SN explosion model. All yield tables implemented in \C~list the element return in the net-yield formulation and exist for various initial stellar metallicities (see table \ref{tab:yields}).

%%%%%%%%%%%%%%%%% TABLE 2 %%%%%%%%%%%%%%%%%%%%%%%%%%%
\begin{table}
\begin{center}
\caption{Yield tables implemented in Chempy.}
\label{tab:yields}
\begin{minipage}{.5\textwidth}
%\scriptsize
\center
\begin{tabular}{l c c }
		\hline\hline
		Yield Table & Masses & Metallicities \\
		\hline
		\multicolumn{3}{c}{CC SN}\\
		\hline
		\citet{Portinari1998} & [6,120] & [0.0004,0.05] \\
		\citet{Francois2004} & [11,40] & [0.02] \\
		\citet{Chieffi2004} & [13,35] & [0,0.02] \\
		\citet{Nomoto2013} & [13,40] & [0.001,0.05] \\
		\citet{Frischknecht2016} & [15,40] & [0.00001,0.0134] \\
		West \& Heger (in prep.) & [13,30] & [0,0.3] \\
		\citet{Ritter2018} & [12,25] & [0.0001,0.02] \\
		\citet{Limongi2018}\footnote{Using the rotation parametrization of \citet{Prantzos2018}} & [13,120] & [0.0000134,0.0134] \\
		\hline
		\multicolumn{3}{c}{SN$_{\rm Ia}$}\\
		\hline
		\citet{Iwamoto1999} & [1.38] & [0,0.02] \\
		\citet{Thielemann2003} & [1.374] & [0.02] \\
		\citet{Seitenzahl2013} & [1.40] & [0.02] \\
		\hline
		\multicolumn{3}{c}{AGB}\\
		\hline
		\citet{Karakas2010} & [1,6.5] & [0.0001,0.02] \\
		\citet{Ventura2013} & [1,6.5] & [0.0001,0.02] \\
		\citet{Pignatari2016} & [1.65,5] & [0.01,0.02] \\
		\citet{Karakas2016} & [1,8] & [0.001,0.03] \\
		TNG\footnote{The Ilustris-TNG \citep{Pillepich2018,Naiman2018} yield set for AGB stars is a mixture of yields taken from \citet{Karakas2010,Doherty2014} and \citet{Fishlock2014}} & [1,7.5] & [0.0001,0.02] \\
		\hline
		\multicolumn{3}{c}{Hypernova}\\
		\hline
		\citet{Nomoto2013} & [20,40] & [0.001,0.05] \\
        \hline
\end{tabular}
\end{minipage}
\end{center}
\end{table}
%%%%%%%%%%%%%%%%%%%%%%%%%%%%%%%%%%%%%%%%%%%%%%%%%%%%%

The right hand panel of Fig. \ref{fig:massloss} shows the cumulative time evolution of the net-yields of 7 different elements returned by an SSP of total mass $1\Msun$. Thick blue line show an SSP of solar metallicity while thin orange lines show results for an SSP of $10^{-5}Z_\odot$. These panels show that O, Ne and Mg are all released early on by massive stars with short lifetimes. C, N or Fe on the other hand have significant late time contributions from lower mass stars and SN~Ia explosions. Comparing blue and orange lines we can further appreciate significant evolution of the net-yields with stellar metallicity. Note, C is the only element which shows larger yields at low metallicity possibly explaining the observations of carbon-enhanced metal-poor stars in the MW \citep[e.g.][]{Starkenburg2014}. 

Figure \ref{fig:yieldhist} shows in more detail the origin of the differences in the time release of elements for our fiducial yield set. In this figure we split the cumulative yield return into the contributions from different nucleosynthetic channels (CC-SN in blue, SN\,Ia in orange and AGB winds in green). The top panel shows a histogram of the fractional net-yield contribution of each nucleosynthetic channel integrated over the IMF and over a Hubble time of a subset of 42 out of a total of 81 elements traced by \C\footnote{See Tutorial 8 of the \C~package at \url{https://github.com/jan-rybizki/Chempy/tree/master/tutorials}}. 

The values for H, Li, Be, and B are negative indicating their preferential consumption inside stars. For the other elements, these figures show that most of them originate from a combination of different channels with varying relative contributions. For example, for iron and Ni we see that CC-SN and SN\,Ia contribute equal amounts of Fe to the enrichment (similar conclusions can be made for Cr, V). O, Na, Mg, Al and Cu are almost only produced by CC-SN while F stems entirely from AGB winds. For He on the other hand we see that CC-SN and AGB winds contribute roughly equal amounts of mass to the enrichment. C is mostly produced in CC-SN but has roughly one quarter contribution from AGB winds while N shows the exact opposite. Mn again is mostly released by SN\,Ia with a one third contribution by CC-SN. Finally, Si, S, Ar or Ca originate from CC-SN explosions with roughly one quarter contribution by SN\,Ia.

While for some elements different channels might contribute equally to the elemental return there can be a significant time delay in the onset of the release of elements of up to 100 Myr (roughly one dynamical time inside the stellar disk of an L$^{\star}$ galaxy) between the release of elements from different sources. This is highlighted in the lower four panels showing the cumulative net-yields of different elements as a function of time (gray lines) split into the contributions from different nucleosynthetic channels (CC-SN in blue, SN\,Ia in orange and AGB winds in green). While CC-SN start releasing their elements as soon as the most massive stars explode ($>40\Msun$, $t_{\rm min}\sim4$ Myr after birth of the SSP), SN\,Ia and AGB star winds release their elements much later ($\sim40$ Myr after birth). Interestingly, the release of O by SN\,Ia explosions is roughly offset by the destruction of oxygen in AGB stars. Similarly there exists a period of time when the AGB net-yield of C is negative due to high mass AGB stars\footnote{Note, this is only true for the net-yields. The total mass return to the ISM will be the sum of newly synthesised elements referred to as the net-yields and the initial elemental abundance. Here, some of the initial C will be fused into higher mass elements effectively lowering the C abundance of the initial composition.} and only becoming positive when lower mass stars enter the AGB phase.   

\begin{figure*}
\begin{center}
\includegraphics[width=\textwidth]{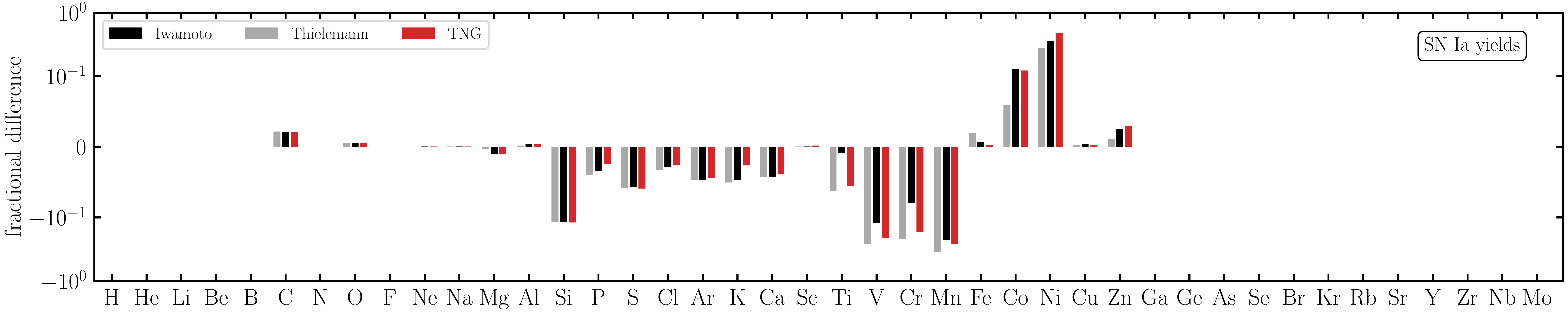}
\includegraphics[width=\textwidth]{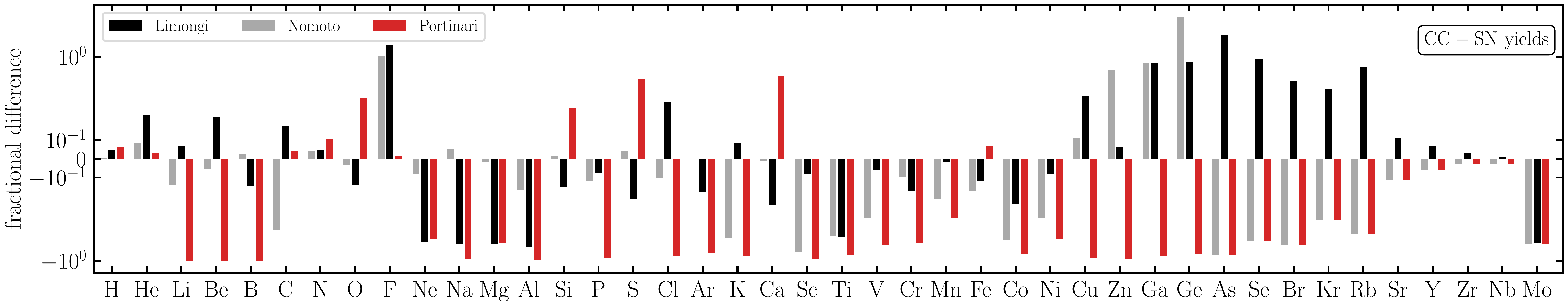}
\includegraphics[width=\textwidth]{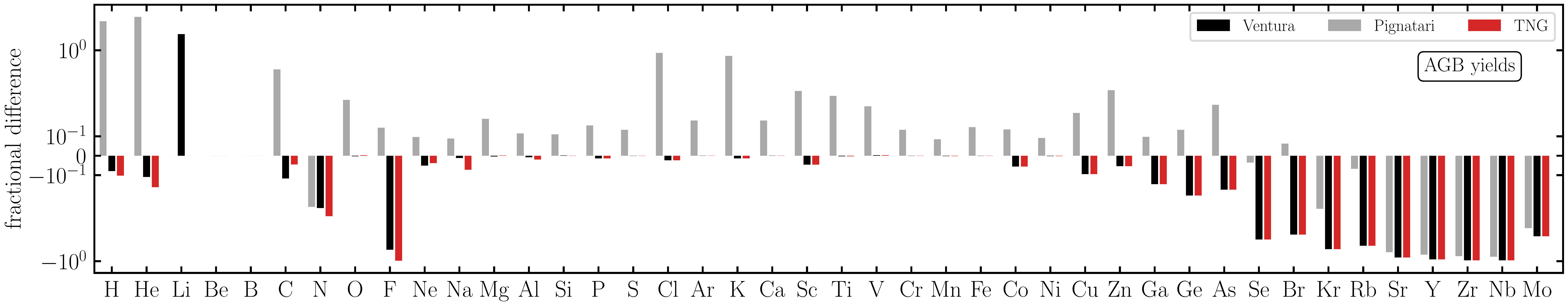}
\end{center}
\vspace{-.35cm}
\caption{IMF weighted relative differences between our fiducial yield set using CC SN yields from \citet{Chieffi2004}, AGB star yields from \citet{Karakas2016} and SNIa yields from \citet{Seitenzahl2013} and other yield sets implemented in \C~ as indicated in the legend. The top panel shows SNIa yields by \citet{Iwamoto1999,Thielemann2003} and the Illustris-TNG yields \citep[see text for more explanation,][]{Pillepich2018,Naiman2018}. The two middle panels compare CC SN yields from \citet{Limongi2018,Nomoto2013,Portinari1998}. The bottom panel compares AGB yields by \citet{Ventura2013,Pignatari2016} and the Illustris-TNG yields with our fiducial yield set. An additional comparison of CC SN yields from \citet{Frischknecht2016}, West \& Heger (in prep.) and the Nugrid collaboration \citep{Ritter2018} is shown in Fig. \ref{fig:yield_comparison1} in the appendix.}
\label{fig:yield_comparison}
\end{figure*}

While the time evolution is qualitatively similar for all net-yields taken from different sources in the literature their quantitative agreement is not. There is no real consensus on how much mass of a given element is synthesised in a specific star (as a function of metallicity and mass) nor the relative contribution of the different channels to a single element. In fact, there can be a difference of up to a factor of a few between different yield sets. This fact is highlighted in Fig. \ref{fig:yield_comparison} which compares the fractional difference between IMF averaged net-yields from different yield tables in the literature in comparison to our fiducial yield set. The upper panel compares SN\,Ia yields, the middle panel CC-SN yields (see also Fig. \ref{fig:yield_comparison1} for additional yield tables) and the lower panel compares AGB wind yields. Note, the y-axis is in logarithmic scale which shows that for some elements the difference can be up to a factor of a few.

At first sight this might give the impression that in general chemical yields might be unreliable. However, investigating this figure more closely we see that for the most abundant elements (H, He, O, C, Ne, Fe, N, Si, Mg, S) the differences between different yield tables is smaller than $\lesssim20\%$ across all channels \citep[see also][for a more in depth discussion of uncertainties related to specific stellar yields]{Romano2010}. This intrinsic uncertainty should be kept in mind when comparing model predictions to observations. We further note that those are only the uncertainties attributed to the stellar models and nucleosynthetic processes. Additional uncertainties originate when using a different IMF (see e.g. discussion of Fig. \ref{fig:IMF}).  

%%%%%%%%%%%%%%%%%% FIGURE 4 %%%%%%%%%%%%%%%%%%%%%%%%%%%
%%%%%%%%%%%%%%%%%%%%%%%%%%%%%%%%%%%%%%%%%%%%%%%%%%%%

\subsubsection{Coupling the SSP model to numerical simulations}
\label{sec:feedback}

In figures \ref{fig:massloss} and \ref{fig:yieldhist} we have shown that there is substantial time evolution of the chemical enrichment for a single SSP over its lifetime. Therefore it is necessary to explicitly follow the time release of elements in numerical simulations instead of applying an instantaneous recycling approximation. The various elements follow distinct tracks which make it relatively complicated to accurately approximate their evolution by a small set of analytic functions. Therefore a lookup table approach seems the most straightforward implementation in order to make use of the chemical enrichment in numerical simulations.

The usage of net-yield tables in \C~makes the application of the chemical enrichment prescriptions formulated in \C~in numerical simulations relatively easy. Those simulations usually trace the initial elemental abundance of gas and stellar particles such that the combined elemental return for an element $i$, $\Delta m^{\rm i}_{\rm tot}$ in a timestep $\Delta t$ from a stellar particle of mass $m_{\rm star}$ follows via 
\begin{equation}
\Delta m^{i}_{\rm tot} = \Delta m^{i}_{\rm new} + \Delta m^{i}_{\rm init}
\end{equation}
where $\Delta m^{i}_{\rm new}$ is the newly synthesised mass of element $i$ taken from the net-yield tables at the appropriate stellar metallicity that is released during this timestep and $\Delta m^{i}_{\rm init}$ is the amount of initial mass of that element inherited from the birth gas cloud that is released in this timestep. Note that $\Delta m^{i}_{\rm new}$ might be negative for elements such as hydrogen which gets fused into higher mass elements. The sum of $\Delta m^{\rm i}_{\rm tot}$ over all elements $i$ will sum up exactly to the total mass lost by the SSP. In general, mass conservation demands that the sum over all elements in the net-yield table should return $0$ as all newly synthesised elements result from the destruction of some lighter mass elements.

%%%%%%%%%%%%%%%%%%%%%%%%%%%%%%%%%%%%%%%%%%%%%%%%%%%
\section{Application: Cosmological zoom-in simulations} \label{sec:simulation}
%%%%%%%%%%%%%%%%%%%%%%%%%%%%%%%%%%%%%%%%%%%%%%%%%%%

%%%%%%%%%%%%%%%%% TABLE 3 %%%%%%%%%%%%%%%%%%%%%%%%%%%
\begin{table*}
\begin{center}
\caption{Combinations of yield sets for CC-SN, SN\,Ia and AGB star yields and SSP parameters probed in this work. Our fiducial model is referred to as \textit{new fid} while other combinations varying CC-SN yields are referred to as \textit{alt, alt2, alt3}. A variation of AGB star yields is referred to \textit{alt4}. Other physics parameter variations of the fiducial yield set are referred to as \textit{steepIMF} for a stepper high mass slope of the IMF, \textit{longDelay} for a longer SN~Ia delay time and \textit{highNorm} for a higher SN~Ia normalisation.}
\label{tab:runs}
\begin{minipage}{\textwidth}
%\scriptsize
\center
\begin{tabular}{ l c c c c c c}
  \hline  \hline
	Name & CC-SN yield & SN\,Ia yield & AGB yield & IMF & SN\,Ia & SN\,Ia \\
	 & & & & slope & delay time & normalisation \\
  \hline
  	new fid 	&  \citet{Chieffi2004}  & \citet{Seitenzahl2013} & \citet{Karakas2016} &  -2.3 & 40 Myr &$-2.9$ \\
	alt		& \citet{Limongi2018} & \citet{Seitenzahl2013} & \citet{Karakas2016}  & -2.3 & 40 Myr & $-2.9$\\
	alt2		& West \& Heger (in prep.) & \citet{Seitenzahl2013} & \citet{Karakas2016} & -2.3 & 40 Myr & $-2.9$\\
	alt3		& \citet{Nomoto2013} & \citet{Seitenzahl2013} & \citet{Karakas2016} & -2.3 & 40 Myr & $-2.9$\\
  \hline
	alt4		& \citet{Nomoto2013} & \citet{Seitenzahl2013} & TNG & -2.3 & 40 Myr & $-2.9$\\   %in all the folders this was simply called alt, changed for paper
	steepIMF &  \citet{Chieffi2004}  & \citet{Seitenzahl2013} & \citet{Karakas2016} & -2.7 & 40 Myr & $-2.9$\\ 
	longDelay &  \citet{Chieffi2004}  & \citet{Seitenzahl2013} & \citet{Karakas2016} & -2.3 & 158 Myr & $-2.9$ \\ 
	highNorm &  \citet{Chieffi2004}  & \citet{Seitenzahl2013} & \citet{Karakas2016} & -2.3 & 40 Myr & $-2.75$\\
  \hline
\end{tabular}
\end{minipage}
\end{center}
\end{table*}
%%%%%%%%%%%%%%%%%%%%%%%%%%%%%%%%%%%%%%%%%%%%%%%%%%%%%

We have taken the above approach and implemented the time release of elements as synthesised by \C~into the modern smoothed particle hydrodynamics (SPH) code \G~ \citep{Wadsley2017}. The python code exemplifying how to use \C~to synthesise the yield tables can be found at \url{https://github.com/TobiBu/chemical_enrichment.git}.
For this study we perform 23 cosmological zoom-in simulations of galaxies ranging from $10^6-10^{11} \Msun$ in stellar mass. Energetic feedback prescriptions and initial conditions are adopted from the NIHAO project \citep{Wang2015} and for completeness we briefly discuss them below. Basic parameters of each central galaxy are listed in table \ref{tab:props}. 

The NIHAO galaxies adopt cosmological parameters from the \cite{Planck}, namely: \OmegaM=0.3175, \OmegaL=0.6825, \Omegab=0.049, H${_0}$ = 67.1\kms\Mpc$^{-1}$, \sig8 = 0.8344. Zoom-in initial conditions are created using a modified version of the \texttt{GRAFIC2} package \citep{Bertschinger2001,Penzo2014} by selecting galaxies for re-simulation from cosmological dark matter only boxes selecting only on the dark halo mass in combination with an isolation criterion for numerical reasons. Formally, we only consider haloes that have no other halo with mass greater than one-fifth of the virial mass within $3$ virial radii.

The adopted force softening for dark matter, gas and star particles are listed in table \ref{tab:props}. Simulated galaxies from the NIHAO project generally match the observed properties of central galaxies \citep[see e.g.][]{Dutton2017,Obreja2016,Obreja2019a,Buck2017,Buck2018} and MW/M31 satellites \citep{Buck2019c}. They further show good numerical convergence in terms of resolution \citep{Buck2019b,Buck2020} and feedback prescriptions \citep[e.g.][]{Obreja2019}. In particular, the adopted feedback prescriptions result in the formation of a chemical bimodality of the stellar disk of L$^\star$ galaxies \citep{Buck2020b}.

\subsection{Hydrodynamics}

The hydrodynamics solver \G\ \citep{Wadsley2017} employs a modern formulation of the smooth particle hydrodynamics method which improves multi-phase mixing and removes spurious numerical surface tension. The simulations employ the Wendland C2 smoothing kernel \citep{Dehnen2012} with a number of 50 neighbour particles for the calculation of the smoothed hydrodynamic properties. The treatment of artificial viscosity uses the signal velocity as described in \cite{Price2008} and a timestep limiter as  described in \cite{Saitoh2009} has been implemented such that cool particles behave correctly when hit by a hot blastwave. All simulations employ a pressure floor following \citet{Agertz2009} to keep the Jeans mass of the gas resolved by at least $4$ SPH kernel masses to suppress artificial fragmentation. This is equivalent to the criteria proposed in \citet{Richings2016} and fulfils the \citet{Truelove1997} criterion at all times. Finally, we adopted a metal diffusion algorithm between SPH particles as described in \cite{Wadsley2008} and extensively discussed in section 2.2  of \citet{Shen2010}.

Gas cooling in the temperature range from 10 to $10^9$K is implemented via hydrogen, helium, and various metal-lines using pre-calculated \texttt{cloudy} \citep[version 07.02;][]{Ferland1998} tables \citep{Shen2010,Obreja2019}. Heating from cosmic reionization is implemented via a meta-galactic UV background \citep{Haardt2012}. 

%%%%%%%%%%%%%%%%% TABLE 4 %%%%%%%%%%%%%%%%%%%%%%%%%%%
\begin{table*}
\begin{center}
\caption{Simulation properties of the main galaxies: \normalfont{We state the total stellar mass, $M_{\rm star}$ and the total amount of gas, $M_{\rm gas}$, within the virial radius. We further report the projected stellar half light radius, $R_{\rm half}$, the total metallicity, $Z/Z_{\odot}$, iron abundance [Fe/H] and gas phase oxygen abundance measured within $2R_{\rm half}$. The last column presents the stellar, gas and dark matter mass resolution.}}
\label{tab:props}
\scriptsize
\begin{tabular}{l c c c c c c c c c c c}
		\hline\hline
		simulation & yield set & $M_{\rm star}$ & $M_{\rm gas}$ & $R_{\rm half}$ & $\log(Z/Z_{\odot})$ & [Fe/H] & $12 + \log(\rm{O/H})$ & $m_{\rm star}/m_{\rm gas}/m_{\rm dark}$ \\
		  & & [$10^{10}\Msun$] & [$10^{10}\Msun$] & [kpc] &  &  &  & [$10^5\Msun$] \\
		\hline
		g2.79e12 & new fid & 18.23 & 21.70 & 2.81 & -1.62 & 0.17 & 9.24 & 1.06/3.18/17.35  \\
		g8.26e11 & new fid & 4.00 & 8.00 & 2.93 & -1.74 & -0.06 & 9.01 & 1.06/3.18/17.35  \\
		g8.26e11 & alt & 4.10 & 7.51 & 2.67 & -1.80 & -0.14 & 8.99 & 1.06/3.18/17.35 \\
		g8.26e11 & alt2 & 3.52 & 7.69 & 2.61 & -1.95 & -0.17 & 8.74 & 1.06/3.18/17.35  \\
		g8.26e11 & alt3 & 4.22 & 7.80 & 2.69 & -1.74 & -0.13 & 9.04 & 1.06/3.18/17.35  \\
		g8.26e11 & alt4 & 3.91 & 7.65 & 2.68 & -1.81 & -0.17 & 9.04 & 1.06/3.18/17.35 \\
		g8.26e11 & steepIMF & 4.29 & 7.93 & 2.47 & -2.01 & -0.22 & 8.61 & 1.06/3.18/17.35 \\
		g8.26e11 & longDelay & 4.19 & 7.62 & 2.68 & -1.73 & -0.11 & 9.05 & 1.06/3.18/17.35\\
		g8.26e11 & highNorm & 4.29 & 7.71 & 2.64 & -1.71 & 0.03 & 9.06 & 1.06/3.18/17.35\\
		g7.55e11 & new & 3.75 & 6.81 & 2.71 & -1.77 & -0.09 & 8.97 & 1.06/3.18/17.35  \\
		g2.19e11 & new fid & 0.08 & 0.92 & 2.78 & -2.72 & -1.05 & 8.02 & 0.13/0.39/2.17  \\
		g1.57e11 & new fid & 0.10 & 1.00 & 4.67 & -2.78 & -1.12 & 8.06 & 0.13/0.39/2.17  \\
		g4.99e10 & new fid & 0.01 & 0.19 & 2.93 & -3.18 & -1.51 & 7.53 & 0.04/0.11/0.64  \\
		g2.83e10 & new fid & 0.003 & 0.10 & 1.77 & -3.39 & -1.75 & 7.35 & 0.04/0.11/0.64  \\
		g2.83e10 & alt & 0.004 & 0.12 & 1.76 & -3.34 & -1.79 & 7.47 & 0.04/0.11/0.64  \\
		g2.83e10 & alt2 & 0.003 & 0.14 & 1.63 & -3.57 & -1.81 & 7.02 & 0.04/0.11/0.64  \\ % in sim folder called alt3
		g2.83e10 & alt3 & 0.003 & 0.13 & 2.18 & -3.27 & -1.74 & 7.52 & 0.04/0.11/0.64 \\ % in sim folder called alt4
		g2.83e10 & alt4 & 0.003 & 0.12 & 1.99 & -3.49 & -1.85 & 7.45 & 0.04/0.11/0.64  \\ % in sim folder called alt
		g2.83e10 & longDelay & 0.003 & 0.10 & 2.04 & -3.35 & -1.77 & 7.34 & 0.04/0.11/0.64 \\
		g2.83e10 & highNorm & 0.003 & 0.11 & 1.75 & -3.33 & -1.64 & 7.35 & 0.04/0.11/0.64 \\
		g2.83e10 & steepIMF & 0.006 & 0.09 & 1.67 & -3.35 & -1.59 & 7.24 & 0.04/0.11/0.64  \\
		g2.83e10 & low fb & 0.005 & 0.16 & 2.02 & -3.19 & -1.56 & 7.51 & 0.04/0.11/0.64 \\
		g7.05e09 & new fid & 0.0002 & 0.01 & 0.45 & -3.51 & -1.85 & 7.29 & 0.01/0.03/0.19  \\
        \hline
\end{tabular}
\end{center}
\end{table*}
%%%%%%%%%%%%%%%%%%%%%%%%%%%%%%%%%%%%%%%%%%%%%%%%%%%%%

\subsection{Star Formation and Feedback}

Star formation is implemented following \citet{Stinson2006} in which dense ($n_{\rm  th}  > 50 m_{\rm  gas}/\epsilon_{\rm gas}^3 = 10.3$cm$^{-3}$, where $m_{\rm gas}$ denotes the gas particle mass, $\epsilon_{\rm gas}$ the gas gravitational softening and the value of 50 refers to the number of neighbouring particles) and cold (T $< 15,000$K) gas is eligible to form stars. A systematic study of the impact of the star formation threshold density on the galaxy properties has recently been conducted by \citet{Dutton2019,Dutton2020} where only simulations with a high ($n>10$ cm$^{-3}$) threshold density reproduce the observed spatial clustering of young star clusters \citep{Buck2019}. 

The gas fulfilling the above requirements will be stochastically converted into star particles according to:
$\frac{\Delta  m_{\rm star}}{\Delta t}=c_{\rm star}\frac{m_{\rm  gas}}{t_{\rm dyn}}$, where  $\Delta m_{\rm star}$  is the mass of the star particle formed, $m_{\rm  gas}$ the gas particle mass, $\Delta  t$ the timestep between star formation events (here: $8 \times 10^5$  yr) and $t_{\rm dyn}$ is the gas particles dynamical time. The variable $c_{\rm star}$ refers to the star formation efficiency, i.e. the fraction of gas that will be converted into stars during the time $t_{\rm dyn}$. We set this parameter to $c_{\rm star}=0.1$ which reproduces the \citet{Kennicutt1998} Schmidt relation \citep{Stinson2006}. The initial stellar particle mass at birth is fixed to one third of the initial gas mass and successive mass loss is implemented as described in section \ref{sec:massloss}

We model the energy input from stellar winds and photoionisation from luminous young stars following the 'early stellar feedback' prescriptions in \citet{Stinson2013} and inject the resulting feedback energy as pure thermal feedback. The energy input from CC-SN and SN\,Ia blastwaves is implemented following \citet{Stinson2006} using self-consistent supernovae rates calculated by the stellar evolution models described in section \ref{sec:method}. For both feedback channels the efficiency parameters were chosen such that one MW mass galaxy respects the abundance matching relations \citep{Behroozi2013,Kravtsov2018,Moster2018} at all redshifts. 

Elemental feedback from CC-SN, SN\,Ia and AGB stars is implemented as described in section \ref{sec:feedback}. We tabulate the time resolved, mass dependent element release of a single stellar population as a function of initial metallicity in a grid of 50 metallicity bins logarithmically spaced between $10^{-5}-0.05$ in metallicity with each metal bin resolved by 100 time bins logarithmically spaced in time from $0-13.8$ Gyr. All our simulations track the evolution of the 10 most abundant elements by default (H, He, O, C, Ne, Fe, N, Si, Mg, S) while any other element present in the yield tables can additionally be tracked. Our fiducial combination of yield tables uses SN\,Ia yields from \citet{Seitenzahl2013}, CC-SN yields from \citet{Chieffi2004} and AGB star yields from \citet{Karakas2016}. Additionally we simulated a dwarf galaxy and a MW-mass galaxy with varying CC-SN yields (see table \ref{tab:runs} for more details) as those yields should have the most effect on the final chemical composition (see e.g. \ref{fig:yieldhist}). For this work we define the abundances of element X as [X/H] $=\log\left(m_{\rm X}/m_{\rm H}\right) - \log\left(X_{\rm{X,\odot}}/X_{\rm{H,{\odot}}}\right)$ where $m_{\rm X}$ and $m_{\rm H}$ are the gas or star particle's mass of element X and hydrogen mass while $X_{\rm{X,\odot}}$ and $X_{\rm{H,{\odot}}}$ denote the solar mass fractions of element X and hydrogen taken from \citet{Asplund2009}. Finally,
the abundance of $\alpha$ elements is defined as the (mass weighted) sum of O, Mg, Si, and S as such the [$\alpha$/Fe] abundance as [(O+Mg+Si+S)/Fe].  

%%%%%%%%%%%%%%%%%%%%%%%%%%%%%%%%%%%%%%%%%%%%%%%%%%%
\section{Results: Galactic chemical composition}
\label{sec:results}
%%%%%%%%%%%%%%%%%%%%%%%%%%%%%%%%%%%%%%%%%%%%%%%%%%%

We evaluate the new chemical enrichment model implemented in \G\, by investigating the stellar mass-metallicity relation (left panel of Fig. \ref{fig:mass_metal}), the stellar [Fe/H] vs. stellar mass relation (right panel of Fig. \ref{fig:mass_metal}) and the gas phase oxygen abundance (Fig. \ref{fig:oxygen}). We compare our simulations to observations from \citet{Gallazzi2005}, \citet{Panter2008}, \citet{Kirby2013} and \citet{Tremonti2004}, \citet{Zahid2013} and \citet{Berg2012}, respectively. We further investigate changes with respect to the fiducial NIHAO suite.

\subsection{Stellar mass metallicity relation}
\label{sec:mzrel}

%%%%%%%%%%%%%%%%%% FIGURE 5 %%%%%%%%%%%%%%%%%%%%%%%%%%%
\begin{figure*}
\begin{center}
\includegraphics[width=.49\textwidth]{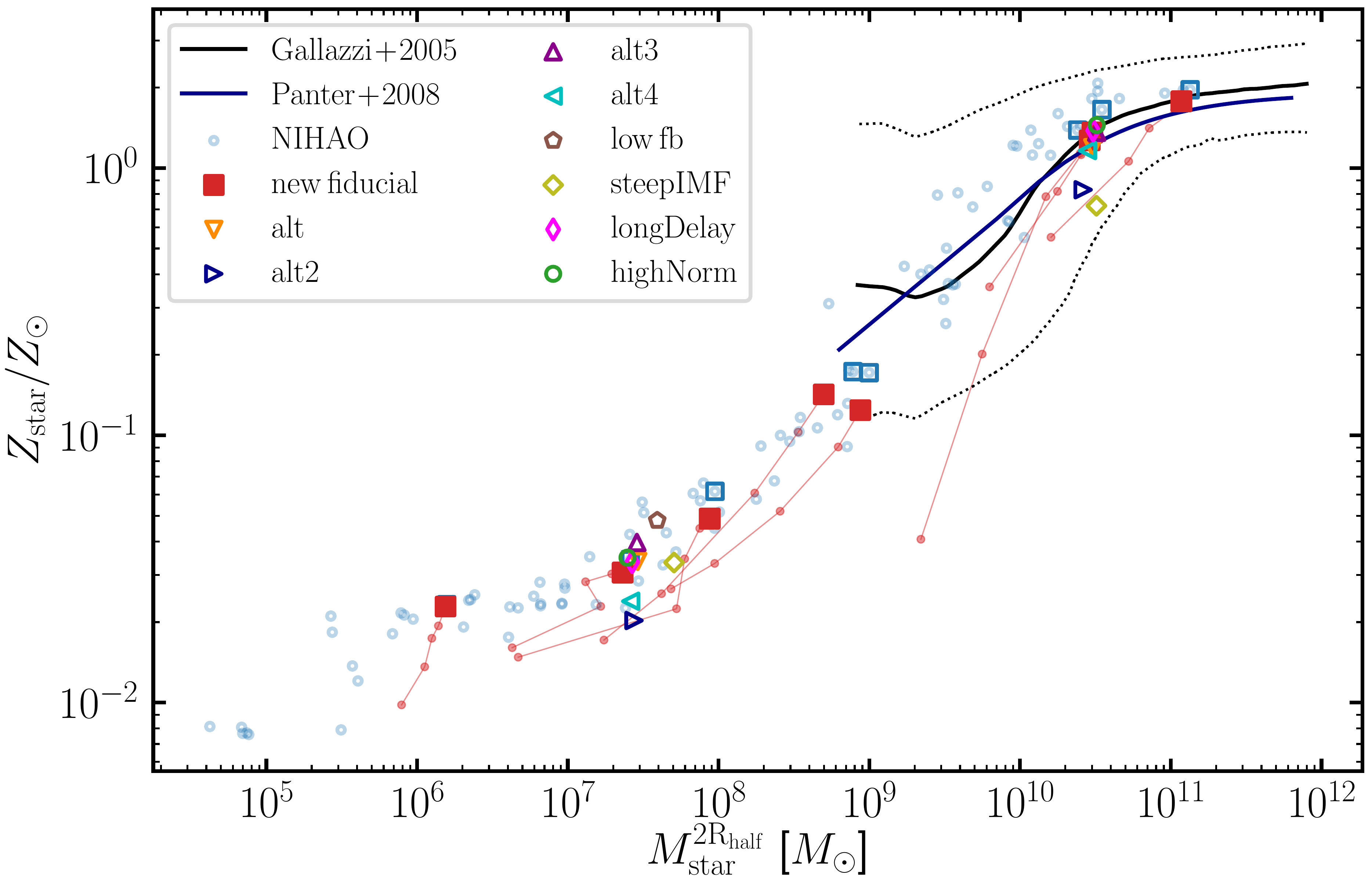}
\includegraphics[width=.49\textwidth]{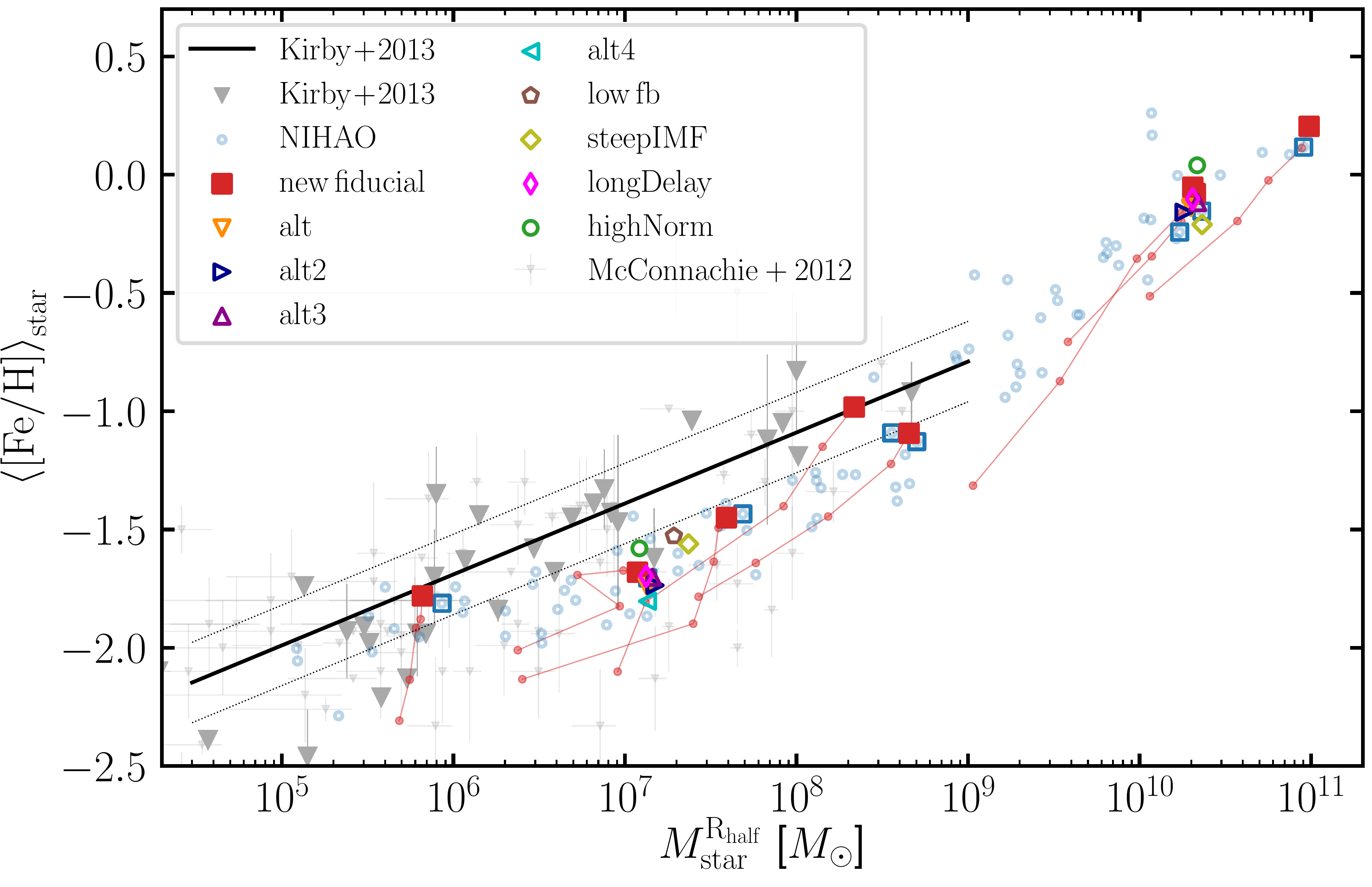}
\end{center}
\vspace{-.35cm}
\caption{Stellar mass metallicity relation (left panel) and stellar mass vs. stellar iron abundance (right panel) of the new chemical enrichment implemented in \G~(red points) in comparison to observational data from \citet{Gallazzi2005}, \citet{Panter2008} and \citet{Kirby2013}, respectively. In the right panel faint gray  points show additional observational data from \citet{McConnachie2012}. We highlight the temporal evolution of our model galaxies with thin red dots connected with lines. Faint blue data points show results from the NIHAO sample highlighting the galaxies selected for re-simulation with blue squares. Open coloured symbols show additional runs with varying yield tables or stellar physics parameters as detailed in table \ref{tab:runs}.}
\label{fig:mass_metal}
\end{figure*}
%%%%%%%%%%%%%%%%%%%%%%%%%%%%%%%%%%%%%%%%%%%%%%%%%%%%

The left panel of Fig. \ref{fig:mass_metal} shows the stellar mass metallicity relation for the new chemical enrichment model (coloured symbols) in comparison to observational data as indicated in the legend as well as results from the fiducial NIHAO implementation (faint blue dots). Stellar masses and metallicity are measured for all stars within 2 times the projected stellar half light radius, $2\,R_{\rm{half}}$. The right hand side panel shows the stellar iron abundance vs. stellar mass both measured within the projected half light radius of the galaxies in comparison with the relation for Local group dwarf galaxies measured by \citet{Kirby2013}.

Let us first compare the results of the new chemical enrichment (red squares) with their corresponding NIHAO galaxies (blue open squares). We see that for almost all galaxies stellar masses, total metallicities and stellar iron abundances agree well between new and old runs. Deviations between individual galaxies are within the expected scatter introduced by stochastic star formation recipes \citep[see also][]{Keller2019,Genel2019}. Note, the agreement between stellar masses measured at two different radii further implies that the new chemical enrichment does not lead to drastic changes in stellar mass profiles. Comparing this further with the runs where we tested different yield sets from the literature (open coloured symbols), we see that the latter have a modest influence on the total metallicity or iron abundance but less so on the total stellar mass. We have checked that this differences in metallicities are solely due to different yields and not to a different star formation history (see Fig. \ref{fig:sfh} in the appendix).

Let us now turn to compare the simulation results to the observed relations. For the total metallicity observed relations exist for the stellar mass range $10^9-10^{12}\Msun$ while for the iron abundance Local Group dwarfs constrain the relation in the mass range $\sim10^4-10^9\Msun$. While the total metallicity of our simulations agrees well with the observed mass-metallicity relation, the iron abundance around $\sim10^7$ \Msun falls slightly below the observed relation although within the scatter of individual observed dwarf galaxies. This is especially true for both the old NIHAO implementation and the new chemical enrichment model regardless of the yield table used. Note, we highlight the temporal evolution of our model galaxies with thin red dots connected with lines. This shows that the galaxies mostly evolve along the redshift zero relation.

%%%%%%%%%%%%%%%%%% FIGURE 6 %%%%%%%%%%%%%%%%%%%%%%%%%%%
\begin{figure}
\begin{center}
\includegraphics[width=\columnwidth]{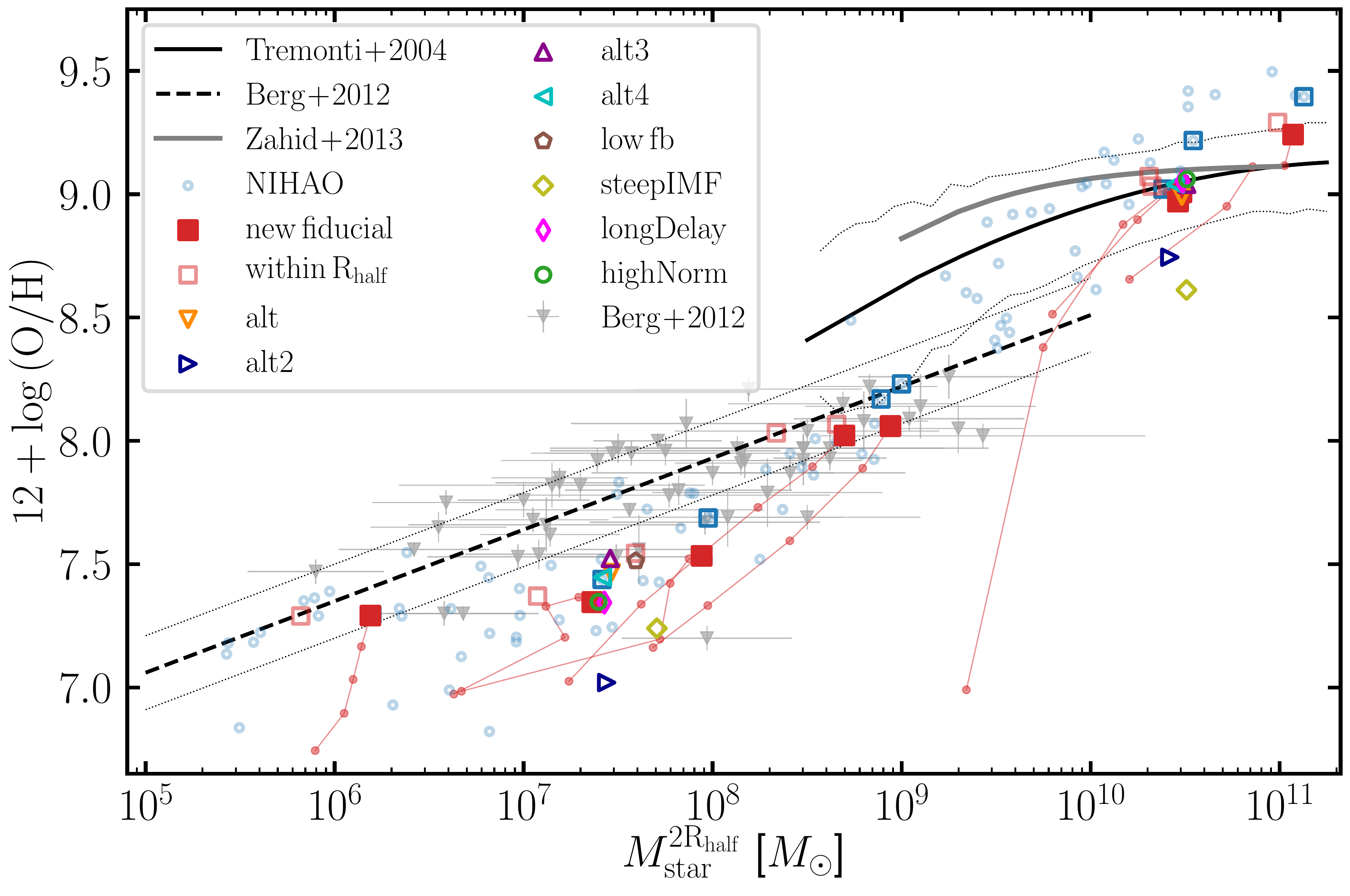}
\end{center}
\vspace{-.35cm}
\caption{Stellar mass gas phase oxygen abundance of the new chemical enrichment implemented in \G~(red points) in comparison with observational data from \citet{Tremonti2004}, \citet{Zahid2013} and \citet{Berg2012}, respectively. Again, we highlight the temporal evolution of our model galaxies with thin red dots connected with lines. Faint blue data points show results from the NIHAO simulations where we highlight the galaxies selected for re-simulation with blue squares. Open coloured symbols show additional runs with varying yield tables or stellar physics parameters as detailed in table \ref{tab:runs}.}
\label{fig:oxygen}
\end{figure}
%%%%%%%%%%%%%%%%%%%%%%%%%%%%%%%%%%%%%%%%%%%%%%%%%%%%

\subsection{Stellar mass vs. gas oxygen abundance}

Fig. \ref{fig:oxygen} compares the gas phase oxygen abundance of our simulations to the observed relation of \citet{Tremonti2004}, \citet{Zahid2013} and \citet{Berg2012}. At the high mass end, the new chemical enrichment implementation results in lower oxygen abundances compared to the old NIHAO simulations. Our new results are in better agreement with the observations. At the low mass end we find no differences in the oxygen abundance between the different implementations and both agree with the observed relation. At intermediate stellar masses of $10^6-10^8\Msun$ we find similar results as for the stellar iron abundance. Both the new chemical enrichment model and the old NIHAO implementation under-predict the gas phase oxygen abundance. We have tested if this is affected by the spatial selection applied (2$R_{\rm half}$  vs. 1$R_{\rm half}$ shown in faint red open squares) but find only a mild dependence of the oxygen abundance on the spatial selection. However, the stellar mass measured for a smaller apertures shifts towards lower masses which slightly alleviates the tension between simulations and observations (especially for the models with larger O abundance like e.g. \textit{alt, alt3}). 

Finally, we find that only a change in the strength of stellar feedback (here a reduction of the coupling efficiency of the early stellar feedback from 13\% to 6\%, magenta open pentagon) or a change in the high mass slope of the IMF (olive open diamond) towards a steeper slope and as such a reduced amount of CC-SN tend to increase the iron abundance and bring the simulations closer to the observed relations at intermediate stellar masses. Although we find that a steeper IMF high mass slope in combination with our fiducial yield table under-predicts the gas oxygen abundance stronger than the fiducial model. 

Our findings that the mass metallicity relation seems to constrain feedback mechanism in dwarf galaxies agrees with recent findings from \citet{Agertz2020a}. Those authors find that explosive feedback expels too many metals from low mass dwarf galaxies such that those feedback mechanisms result in a lower metallicity at fixed stellar mass while other dwarf galaxy scaling relations are unaffected by this. Our tests here seem to confirm their findings although a reduced feedback strength also results in roughly a factor of two larger stellar masses for the dwarf galaxy tested here. A larger suite of  model galaxies at this mass scale is certainly needed to investigate this issue further and potentially constrain feedback mechanisms at this galaxy mass range. Also, care must be taken in comparing simulation results of isolated galaxies with relations obtained for Local Group galaxies which evidently live in a galactic environment influenced by the two major galaxies MW and Andromeda. This might bias Local Group galaxies which might have experienced accretion of pre-enriched gas or suffered some kind of stripping \citep[e.g.][]{Buck2019c} by ram pressure and tidal forces. 

%%%%%%%%%%%%%%%%%% FIGURE 7 %%%%%%%%%%%%%%%%%%%%%%%%%%%
\begin{figure}
\begin{center}
\includegraphics[width=0.48\columnwidth]{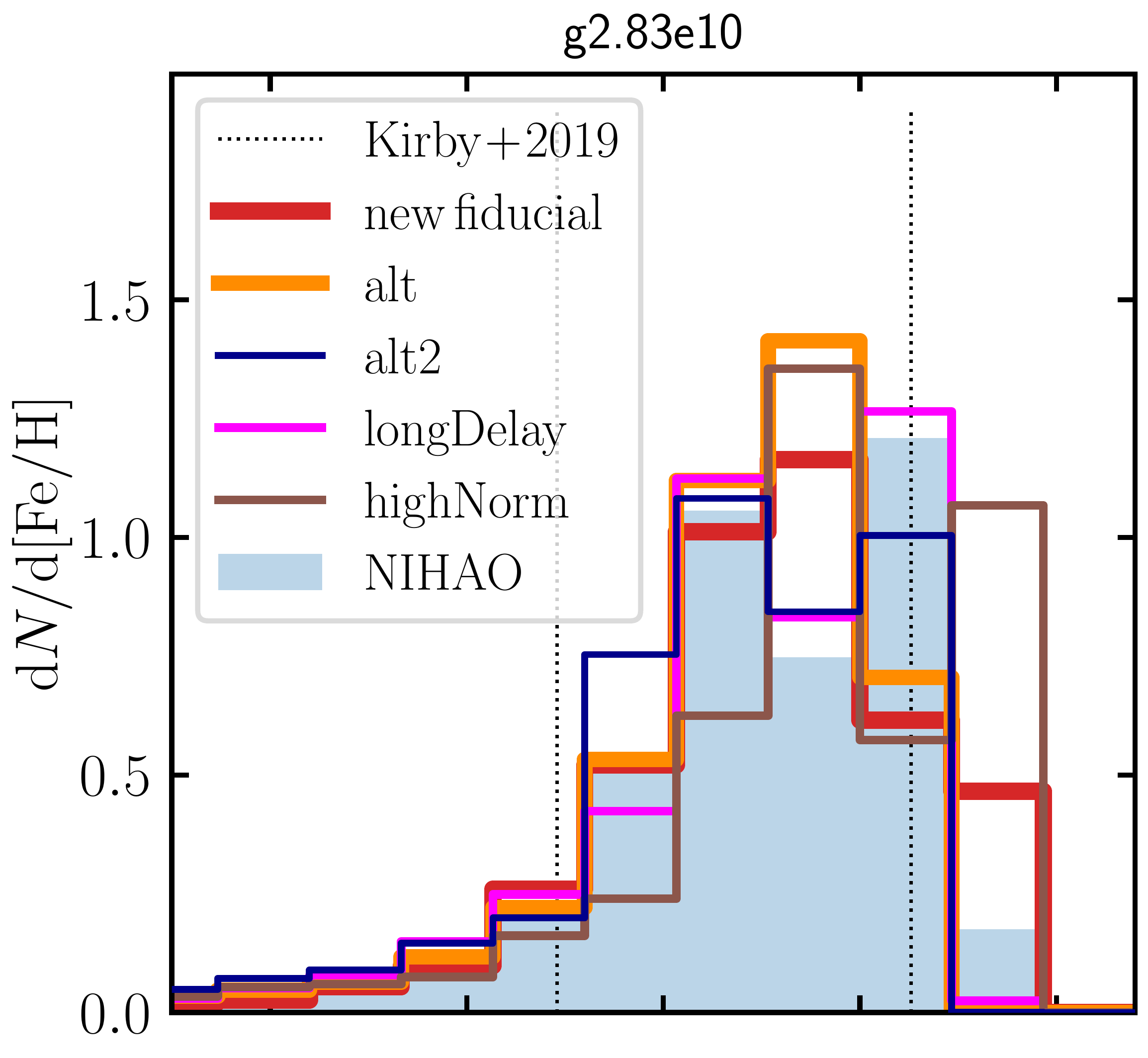}
\includegraphics[width=0.495\columnwidth]{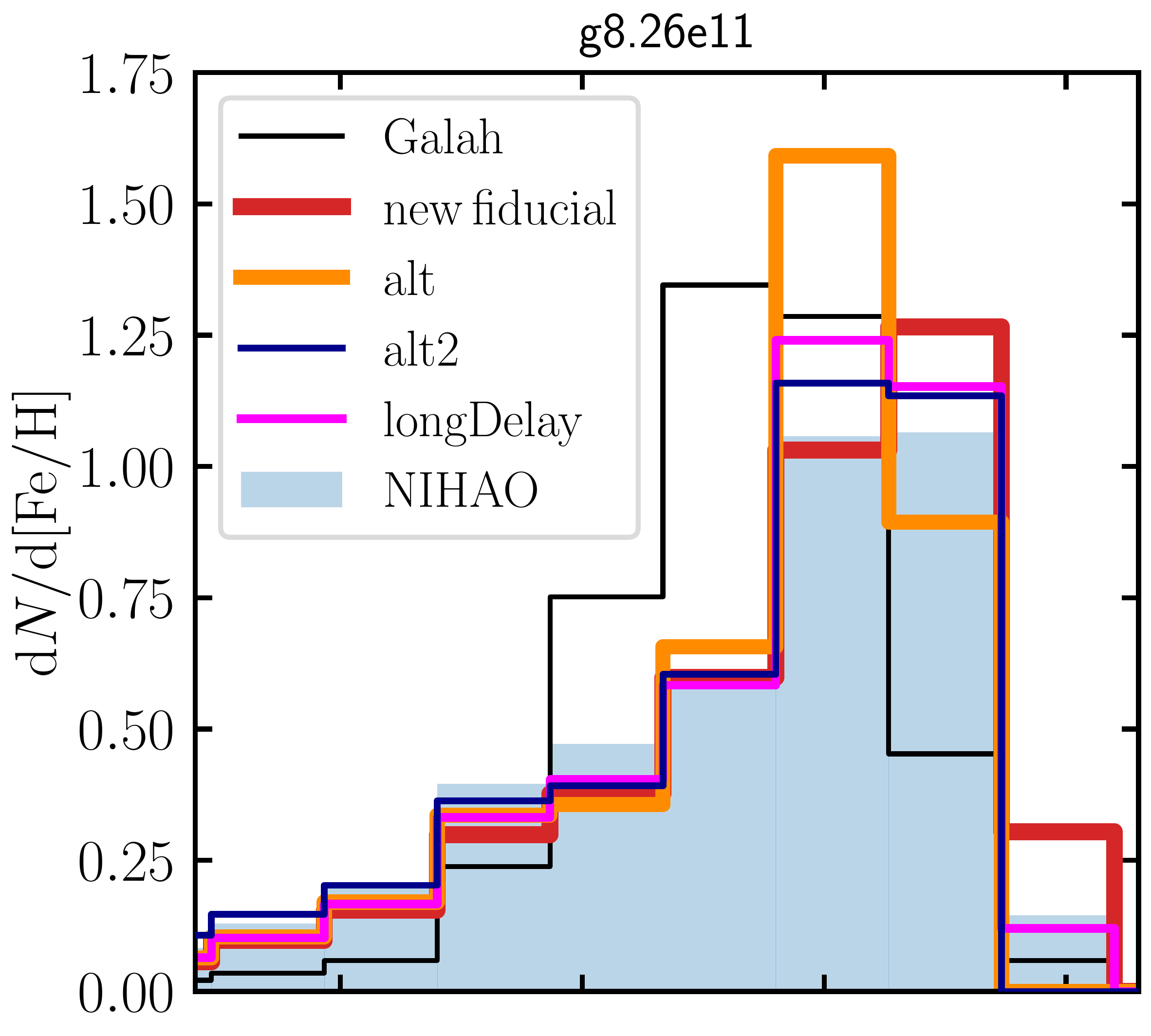}
\includegraphics[width=0.48\columnwidth]{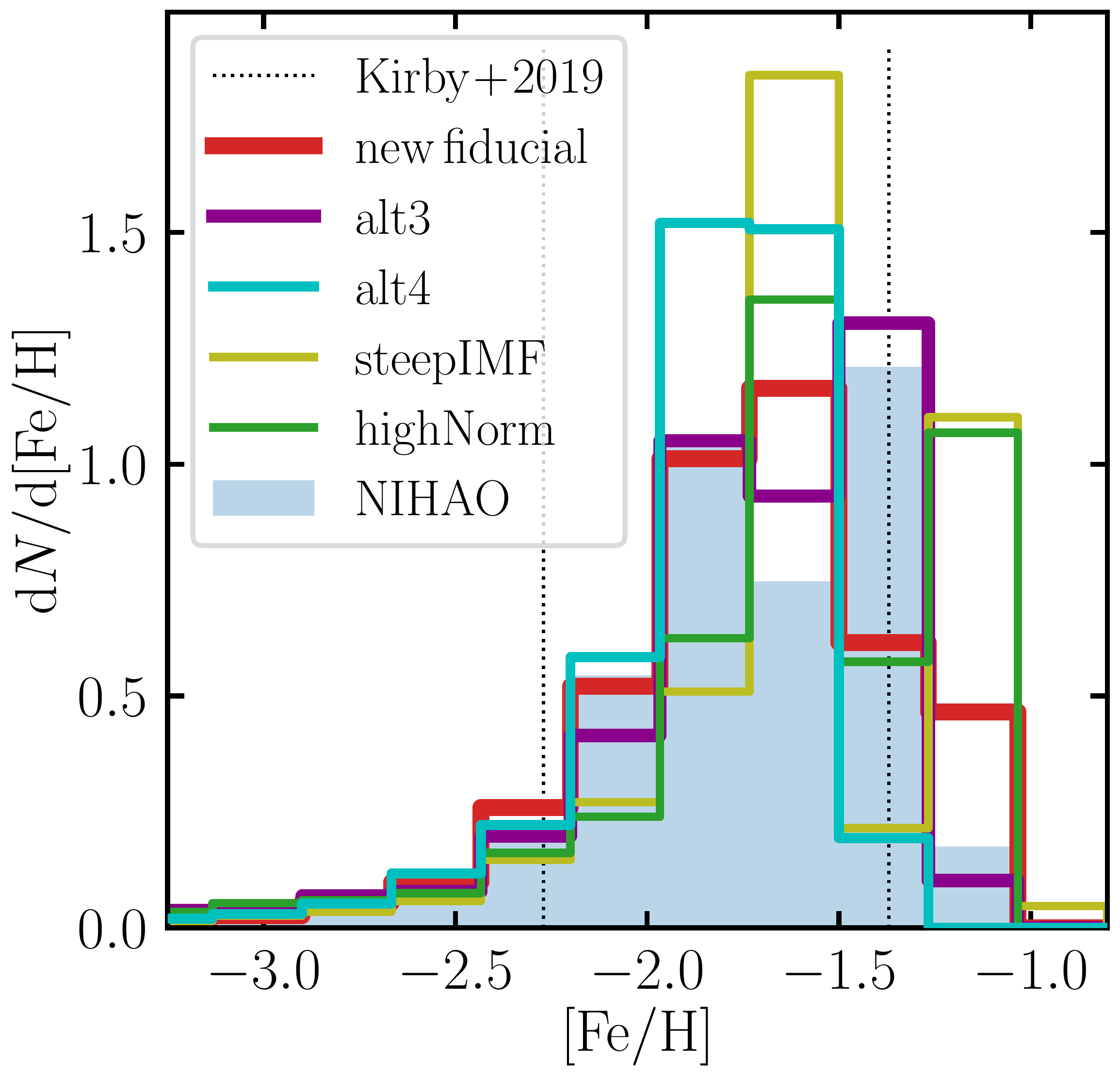}
\includegraphics[width=0.495\columnwidth]{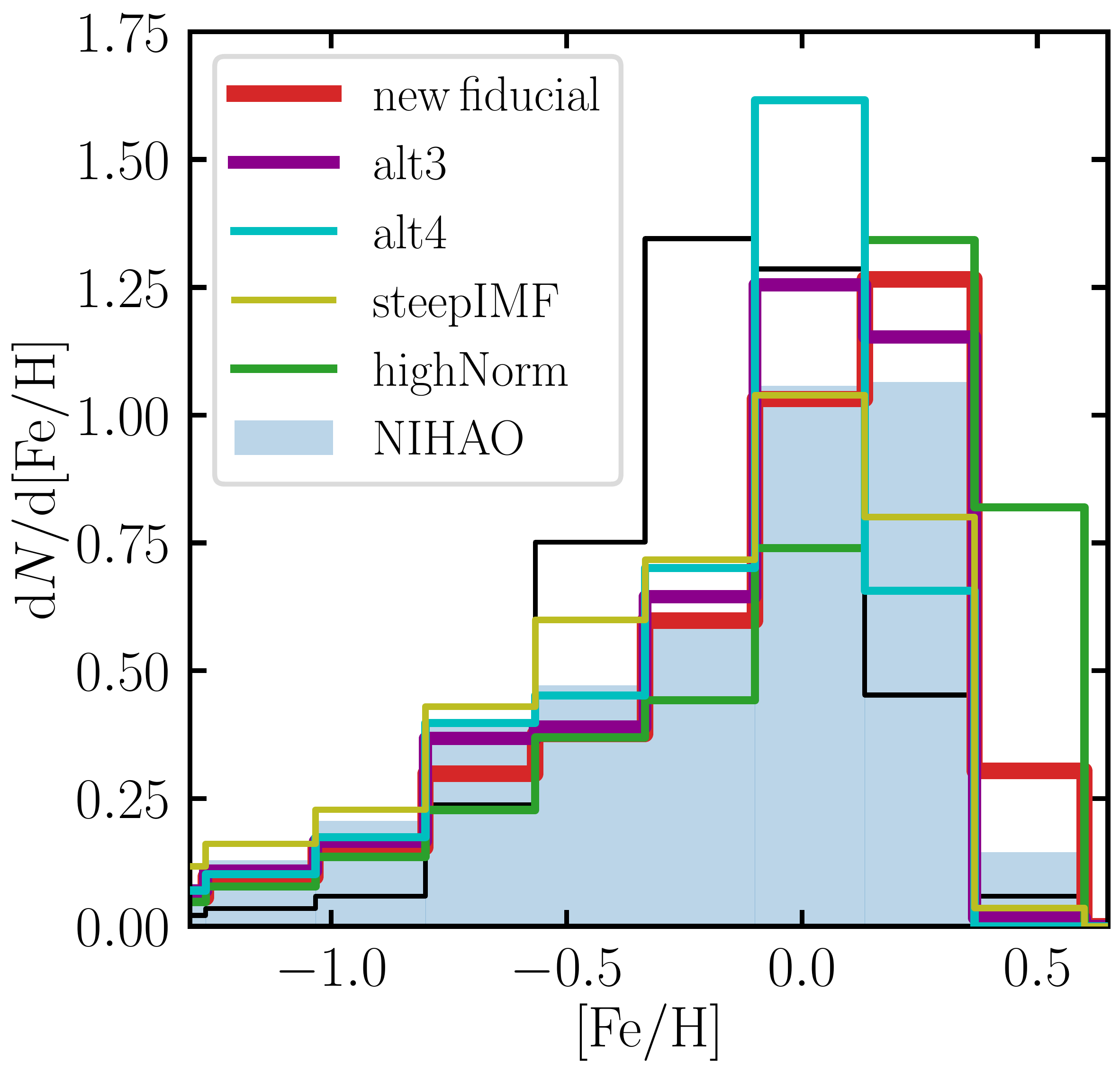}
\end{center}
\vspace{-.35cm}
\caption{Distribution functions of stellar iron abundance for the dwarf galaxy g2.83e10 (left panels) and the MW mass galaxy g8.26e11 (right panels) for various different yield sets tested in this work. For comparison, blue filled histograms show the results for the fiducial NIHAO implementation. In the left panels, the thin dotted vertical lines indicate the range of mean iron abundance of 5 dwarf galaxy satellites of M31 as measured by \citet[][table 4]{Kirby2019}. In the right panels the thin black histogram indicates the observed iron abundance distribution of solar neighbourhood stars from GALAH-DR3 \citep{Buder2020}.}
\label{fig:metal_hist}
\end{figure}
%%%%%%%%%%%%%%%%%%%%%%%%%%%%%%%%%%%%%%%%%%%%%%%%%%%%

 \subsection{Metallicity and  $\alpha$-element distributions}

Section \ref{sec:mzrel} revealed that the average stellar metallicity of the simulations is in agreement with the observed mass-metallicity relation. However, it is further instructive to not only investigate zero order moments of the abundance distribution but look at the distribution functions. In the next two subsections we investigate in detail the metallicity distribution functions and the $\alpha$ element distribution functions for different yield sets probed in this work. Contrary to the previous section we do not apply any spatial selection and use all star particles within the virial radius of the galaxies.  

\subsubsection{Metallicity distribution}

Figure \ref{fig:metal_hist} shows the distribution of stellar iron abundance for the dwarf galaxy \texttt{g2.83e10} (left column) and the MW mass galaxy \texttt{g8.26e11} (right column) for all 8 different models. Figure \ref{fig:z_hist} in the appendix shows the distribution function of total metallicity for the two galaxies. In all panels the filled blue histogram shows the fiducial NIHAO implementation while in the left hand panels in Fig. \ref{fig:metal_hist} the vertical thin dotted lines show the range of mean iron abundance of 5 dwarf galaxy satellites of M31 as measured by \citet[][table 4]{Kirby2019} and the black thin histograms in the right hand panels show results from the DR3 of the GALAH survey based on a magnitude limited selection and thus only selecting stars in the solar neighbourhood.  

The distribution functions of all models share a common feature: They are skewed towards low metallicities/iron abundance. All models calculated for this work show a rather extended tail of low metallicity stars. This common feature is a manifestation of the star formation history which follows the same shape in all 8 models (see fig. \ref{fig:sfh}). The maxima of our distribution functions of [Fe/H] for the dwarf galaxy (left panels of Fig. \ref{fig:metal_hist}) agree well with the measured limits by \citet{Kirby2019} for all our models. However, the maxima of \textit{longDelay}, \textit{alt3} and NIHAO are close to the observed upper limit. We furthermore see significant secondary peaks of \textit{highNorm} and \textit{steepIMF} above the limit. 

%%%%%%%%%%%%%%%%%% FIGURE 8 %%%%%%%%%%%%%%%%%%%%%%%%%%%
\begin{figure*}
\begin{center}
\includegraphics[width=.4775\textwidth]{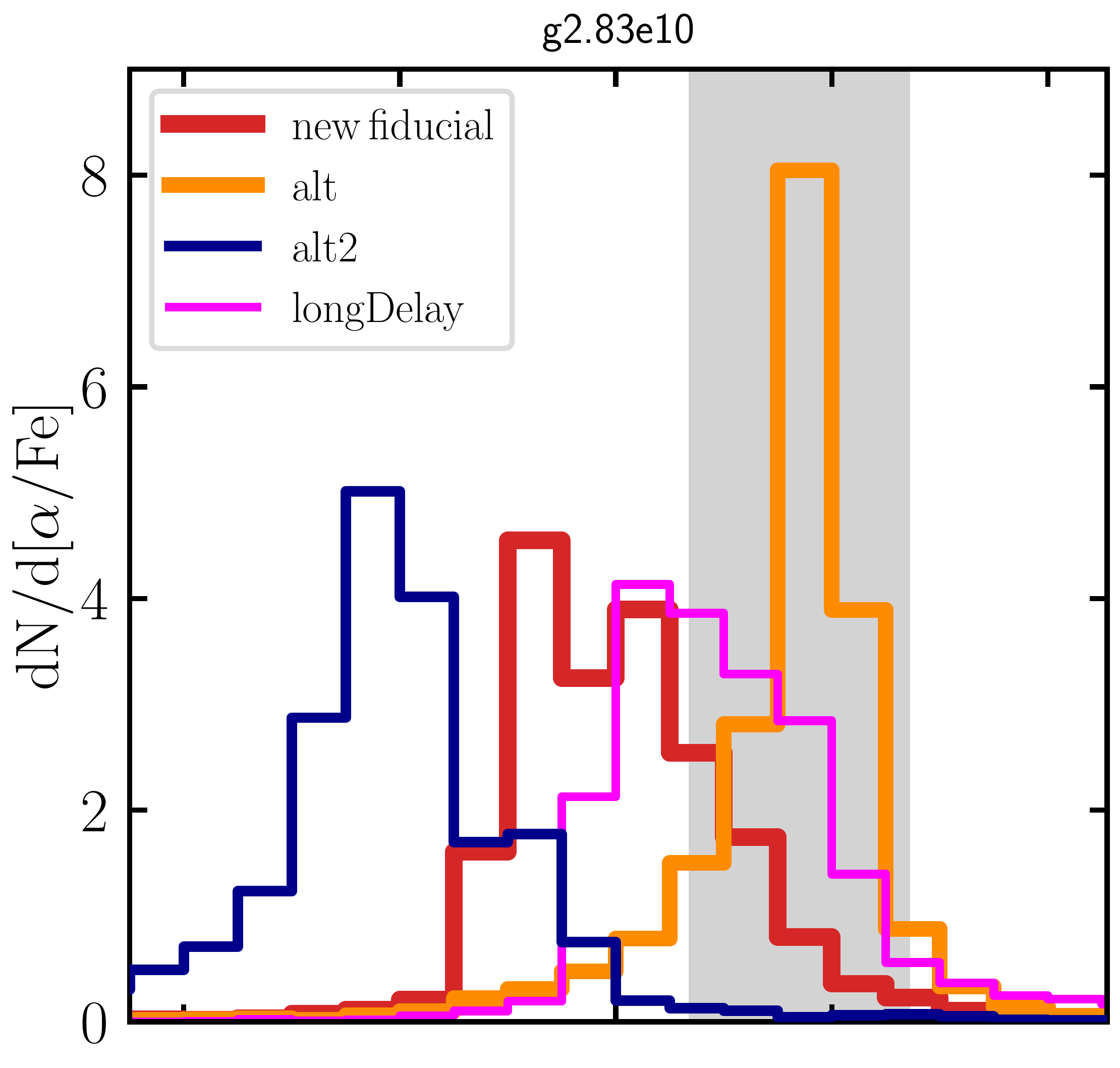}
\includegraphics[width=.49\textwidth]{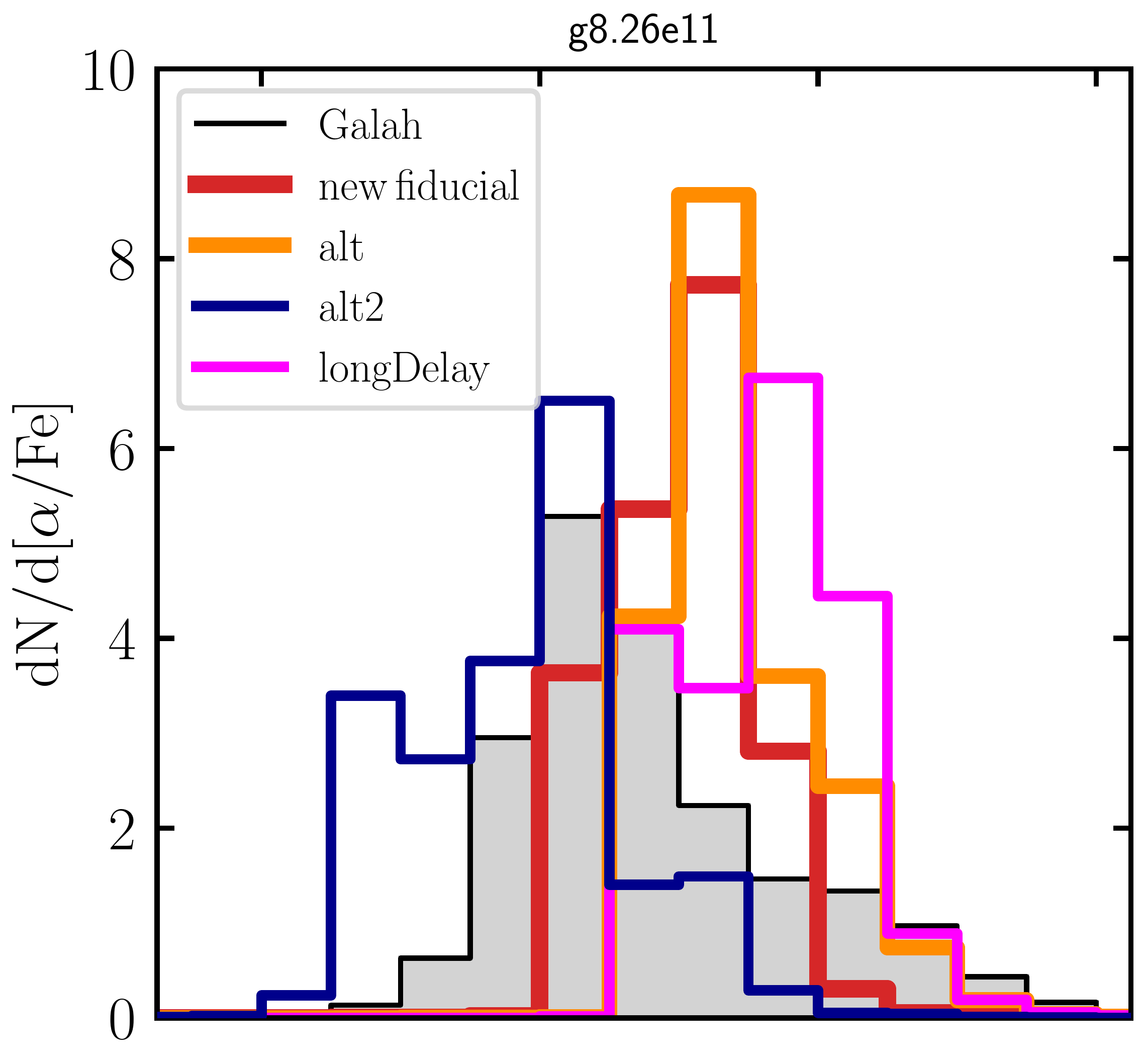}
\includegraphics[width=.4775\textwidth]{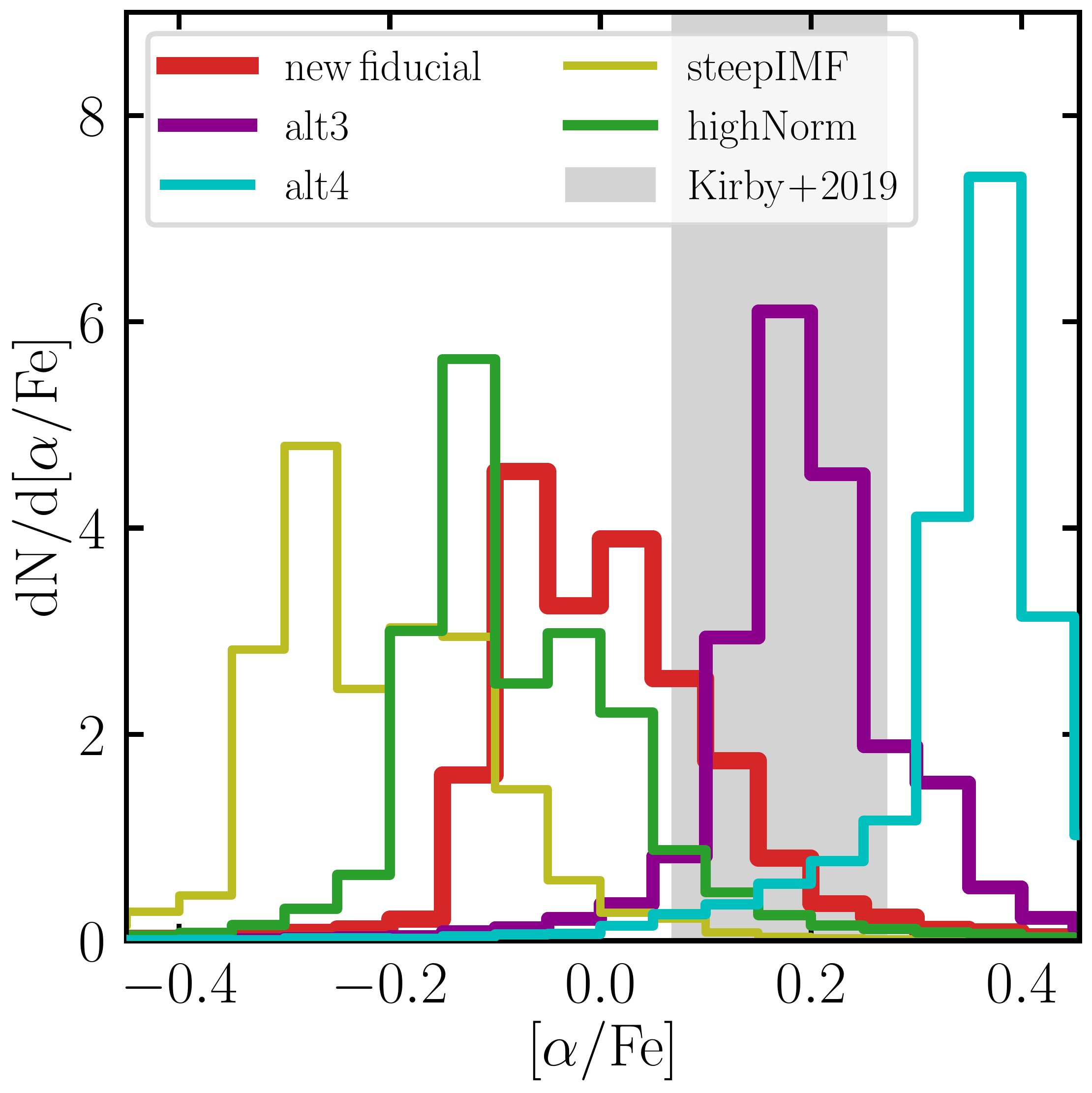}
\includegraphics[width=.49\textwidth]{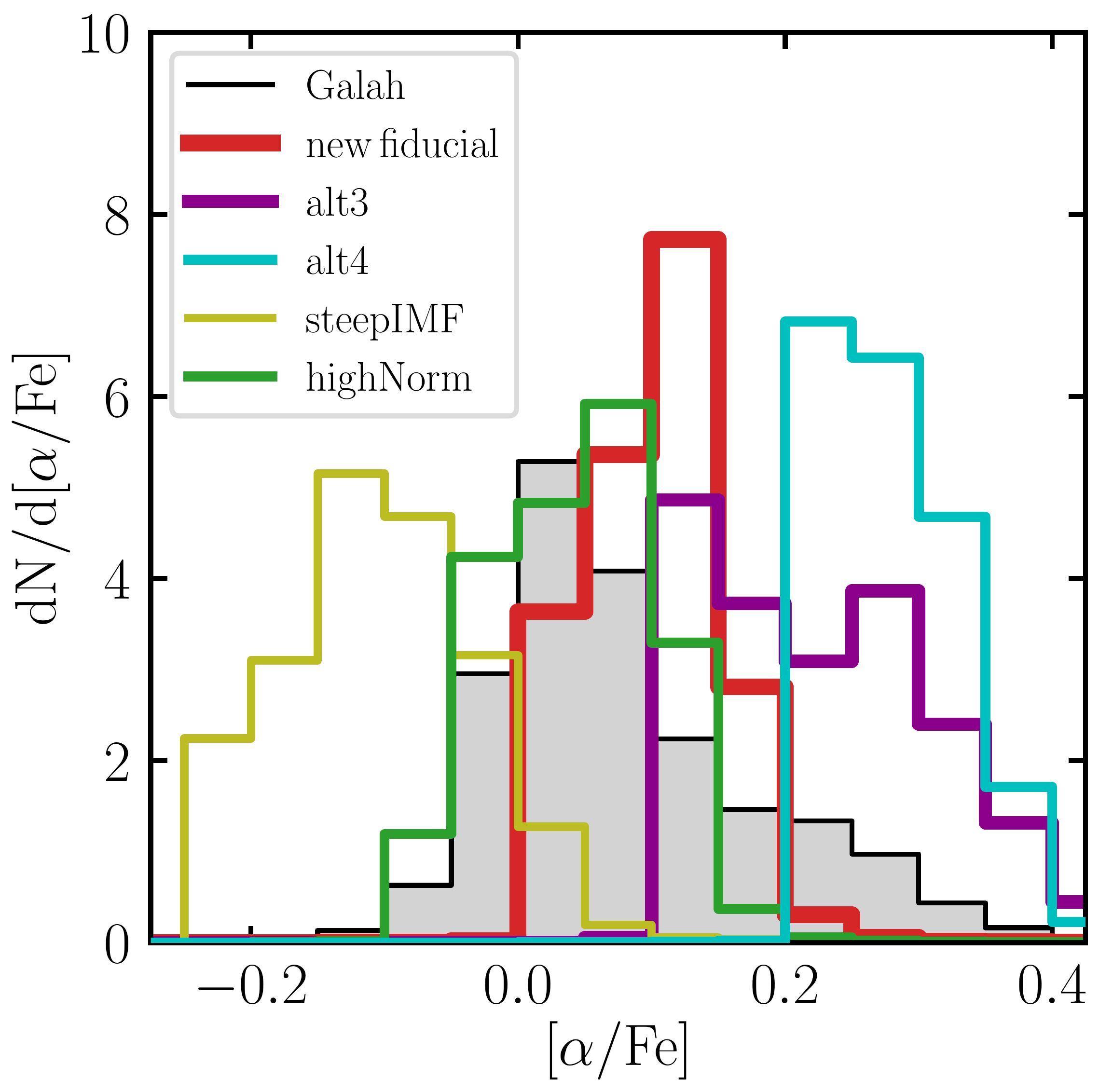}
\end{center}
\vspace{-.35cm}
\caption{Distribution functions of $\alpha$-element abundance ([$\alpha$/Fe]$\equiv$[(O+Mg+Si+Ca+S+Ti)/Fe]) for the dwarf galaxy \texttt{g2.83e10} (left panels) and the MW mass galaxy \texttt{g8.26e11} (right panels) for various different yield sets tested in this work. In the left hand panels, the gray vertical bar shows the mean $\alpha$-abundance of 5 dwarf galaxy satellites of M31 as measured by \citet[][table 4]{Kirby2019}. In the right panels, the gray/black histograms show results for the solar neighbourhood from the GALAH DR3 \citep{Buder2020}.}
\label{fig:alpha_hist}
\end{figure*}
%%%%%%%%%%%%%%%%%%%%%%%%%%%%%%%%%%%%%%%%%%%%%%%%%%%%

The shape of our distribution functions for the MW mass galaxy (right panels of Fig.  \ref{fig:metal_hist}) agree qualitatively with the one measured by GALAH DR3 for the local MW, that is, a maximum around solar [Fe/H] with a skewed shape, extending towards lower [Fe/H]. Quantitatively, we do see, however, that the measured distribution of GALAH DR3 peaks slightly below zero, i.e. [Fe/H] $\sim-0.1$ dex, similar to most studies of the Solar vicinity \citep[e.g.][]{Casagrande2011,Katz2021}.
Furthermore, the observed distribution is less skewed in GALAH DR3 compared to our implementations. Selecting stars only from a solar neighbourhood in the simulations does not change this conclusion as this effects both the lowest and highest metallicity bins. Whether this discrepancy is due to a different SFH in the simulations and the MW or due to different SN\,Ia yields needs to be investigated further.  

In this work we did not vary the yield set for SN~Ia only their delay time and normalisation and keep the yieldset fixed to \citet{Seitenzahl2013}. For a more in depth exploration of the impact of different type Ia SN yields on MW chemical evolution see the recent work by \citet{Palla2021}. This work shows that Fe masses and [$\alpha$/Fe] abundance patterns are less affected by the detailed yield sets while for Fe-peak elements (which are not the focus of this paper) there seems to be a strong dependence on the white dwarf explosion mechanism. Here we find that a difference in delay time between our fiducial model ($\tau=40$ Myr) and the \textit{longDelay} runs ($\tau=160$ Myr) does not lead to significant differences in the iron abundance. However, a slight change in the SN~Ia normalisation ($-2.9$ vs. $-2.75$ for the \textit{highNorm} run) leads to a slight increase in the iron abundance for both the dwarf and the MW mass galaxy.

As previously noted, all models exhibit similar mean metallicities/iron abundances. Note, we do not recalibrate any simulation parameters in this work. This	 is important to note again as in the next subsection we turn to investigate the distribution of $\alpha$-elements relativ to iron, the [$\alpha$/Fe] ratio. Therefore we like to stress here that especially the iron abundance distributions of our models do not differ significantly and all differences discussed in the next section can be attributed to differences in yield tables (remember the similarity in star formation history).

\subsubsection{Diversity in $\alpha$-element distribution}

The left panels of Fig. \ref{fig:alpha_hist} show the distribution of stellar $\alpha$-element abundance for all 8 simulations of \texttt{g2.83e10} in comparison to observational data from \citet[][table 4]{Kirby2019} for 5 Andromeda satellite galaxies shown with the gray shaded area. The right panels of Fig. \ref{fig:alpha_hist} show the stellar $\alpha$-element abundance for the 8 models of the MW mass galaxy \texttt{g8.26e11} in comparison to the GALAH DR3 data. Here we define the $\alpha$-element abundance as [$\alpha$/Fe]$\equiv$[(O+Mg+Si+Ca+S+Ti)/Fe], whereas \citet{Kirby2019} uses the union of Mg, Si, Ca and Ti and GALAH DR3 uses an uncertainty-weighted average of Mg, Si, Ca, and Ti. In contrast to the iron abundance distribution functions shown in Fig. \ref{fig:metal_hist} the $\alpha$-element distribution shows significant diversity between different yield tables or feedback strengths. For the dwarf galaxy the peaks of the different $\alpha$-element distributions span a range of 0.5 dex between [$\alpha$/Fe]$\sim-0.25$ (\textit{alt2}) up to [$\alpha$/Fe]$\gsim0.3$ (\textit{alt4}). Comparing to the observations by \citet[][]{Kirby2019} of dwarf galaxies of similar stellar mass puts strong constraints on valid yield sets. Only the alternative yield set \textit{alt} and \textit{alt3} result in [$\alpha$/Fe] ratios in agreement with observed $\alpha$-enhancements. All other models under-predict the observed values except for the yield set \textit{alt4} which strongly over produces $\alpha$-elements. We have checked that differences in stellar mass between our model galaxies and the observed dwarfs do not influence our conclusions. The left panels of Fig. \ref{fig:mass_comp_hist} in the appendix show that the average $\alpha$-element abundance is not sensitive to the stellar mass of the dwarfs and only shows a mild dependence on stellar mass. This implies that cosmic variance that manifests itself with different star formation histories for the dwarf galaxies does not have a strong impact on the $\alpha$-element abundance distribution at constant stellar mass. This opens the possibility to constrain chemical enrichment models by comparing those results to accurate and correct measurements of stellar masses and abundances. 

Finally, we find that most of the differences of the different models are caused by differences in the O and Mg abundances as shown in Fig. \ref{fig:omgsis_hist} in the appendix.

%%%%%%%%%%%%%%%%%% FIGURE 9 %%%%%%%%%%%%%%%%%%%%%%%%%%%
\begin{figure*}
\begin{center}
\includegraphics[width=\textwidth]{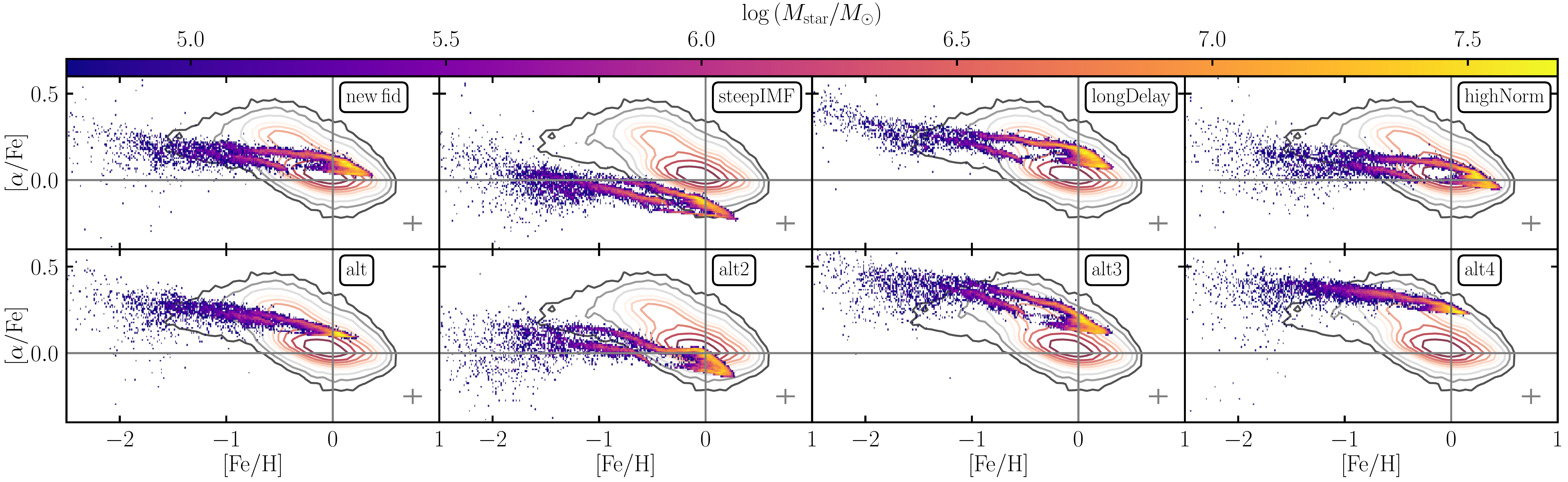}
\end{center}
\vspace{-.35cm}
\caption{Distribution of $\alpha$-element abundances as a function of metallicity, coloured by stellar mass. For the simulation we show only stellar particles within a radial range of $6.5<R<9.5$ kpc and absolute vertical height of $\vert z\vert<2$ kpc from the disk mid-plane. For comparison, contours show result for the solar neighbourhood from GALAH DR3 \citep{Buder2020}. Gray error bars in the lower right of each panel indicate the median observational uncertainty of the GALAH data. Gray lines at [Fe/H]=0 and [$\alpha$/Fe] indicate solar abundance values. See text for more explanation.}
\label{fig:alpha_tracks}
\end{figure*}
%%%%%%%%%%%%%%%%%%%%%%%%%%%%%%%%%%%%%%%%%%%%%%%%%%%%

Note that the differences in $\alpha$-element abundance is not caused by a difference in iron abundance as we have discussed in the previous section. Especially the \textit{steepIMF} and \textit{alt2} model which under-predict the $\alpha$-element abundance show iron abundances in the same range as the other models. Similarly in order to explain the over-production of $\alpha$-elements in the \textit{alt4} model by a miss-match in the iron abundance we would expect this model to under-produce iron compared to the other models. This is clearly not the case as the lower right panel of Fig. \ref{fig:metal_hist} shows. 

For the MW mass galaxy (right panels of Fig. \ref{fig:alpha_hist}) the differences are not as pronounced as for the dwarf but variations among the models are still quite noticeable. The peaks of the different $\alpha$-element distributions span a range of $\sim0.3$ dex between [$\alpha$/Fe]$\sim-0.15$ (\textit{steepIMF}) up to [$\alpha$/Fe]$\gsim0.2$ (\textit{alt3}). Qualitative comparison to observational data from the GALAH survey (gray histograms) suggest [$\alpha$/Fe] values around solar hinting at the conclusion that the \textit{longDelay} and \textit{alt, alt3, alt4} yield sets over-predict the $\alpha$-enhancement while the \textit{steepIMF} and \textit{alt2} model tend to underproduce the $\alpha$-enhancement. Only the models \textit{fiducial} and \textit{highNorm} are marginally in agreement with the GALAH data (see also Fig. \ref{fig:alpha_tracks}). This finding is at odds with the results for the dwarf galaxy where the best matching models are the \textit{alt, alt3} and \textit{longDelay} yield sets. Note, the exact value of the $\alpha$-enhancement does not depend strongly on the stellar mass of the galaxy as the right hand side of Fig. \ref{fig:mass_comp_hist} in the appendix shows.

Note, the comparison to GALAH is only of qualitative nature for several reasons. First, the GALAH survey selects for primarily nearby main sequence stars near the Sun. Our comparison is not entirely consistent given as we select stars across a much larger radial extent in the simulation and not sampled as per the GALAH survey. However, we have checked that only selecting solar neighbourhood stars does not affect our results in the right panels of Fig. \ref{fig:alpha_hist} and subsequent conclusions, as this only affects the edges of the alpha-distributions at very low and high $\alpha$-abundances. Second, we are comparing a generic MW mass model to MW data and it is a priori unclear if the particular model galaxy is a good MW analogue. Third, MW represents a single object and it is still unclear if the observed $\alpha$-enhancement is representative for all MW mass galaxies. Current theoretical models do not agree on this point and offer explanations for both scenarios \citep[see e.g.][]{Grand2018,Mackereth2018,Clarke2019,Buck2020b,Renaud2020b}. However, a quantitative comparison needs to take into account the exact selection function of the particular survey and account for measurement uncertainty on stellar abundances and metallicities. Therefore we defer a more thorough comparison to a dedicated future study.

Comparing the distribution functions of $\alpha$-elements for the dwarf galaxy in the left hand side of Fig. \ref{fig:alpha_hist} with those for the MW mass galaxy we find for all yield models a consistent shift of $\sim0.1$ dex to larger values of [$\alpha$/Fe] for the more massive galaxy. This is also consistent for the other galaxies run with the fiducial yield table in our sample (see Fig. \ref{fig:mass_comp_hist} in the appendix). On the other hand, for the observations, to zero order we find that the median [$\alpha$/Fe] decreases with stellar mass when going from dwarf to MW mass scale as might be expected from a larger contribution of SN~Ia to the enrichment in MW mass galaxies and varying star formation efficiencies in the dwarf galaxies \citep[e.g.][]{Nidever2020}. Of course, this is a simplified comparison as also in the dwarf regime we expect the $\alpha$-abundance to be a function of metalicity \citep[e.g. Fig. 8][]{Kirby2019}.
However, understanding the exact cause of this finding and how much of it can be attributed to different star formation efficiencies in the simulations and the observations needs a larger sample of dwarf galaxies and a more in depth analysis which is outside the scope of the current paper which focusses on different yield tables. Future work is needed in order to gauge if this finding poses a challenge to our current models and if so, researching the reason for this opposite trend of $\alpha$-enhancement with stellar mass in the simulations and the observations will open new possibilities for our understanding of star formation and feedback in (dwarf) galaxies.

%%%%%%%%%%%%%%%%%% FIGURE 10 %%%%%%%%%%%%%%%%%%%%%%%%%%%
\begin{figure*}
\begin{center}
\includegraphics[width=\textwidth]{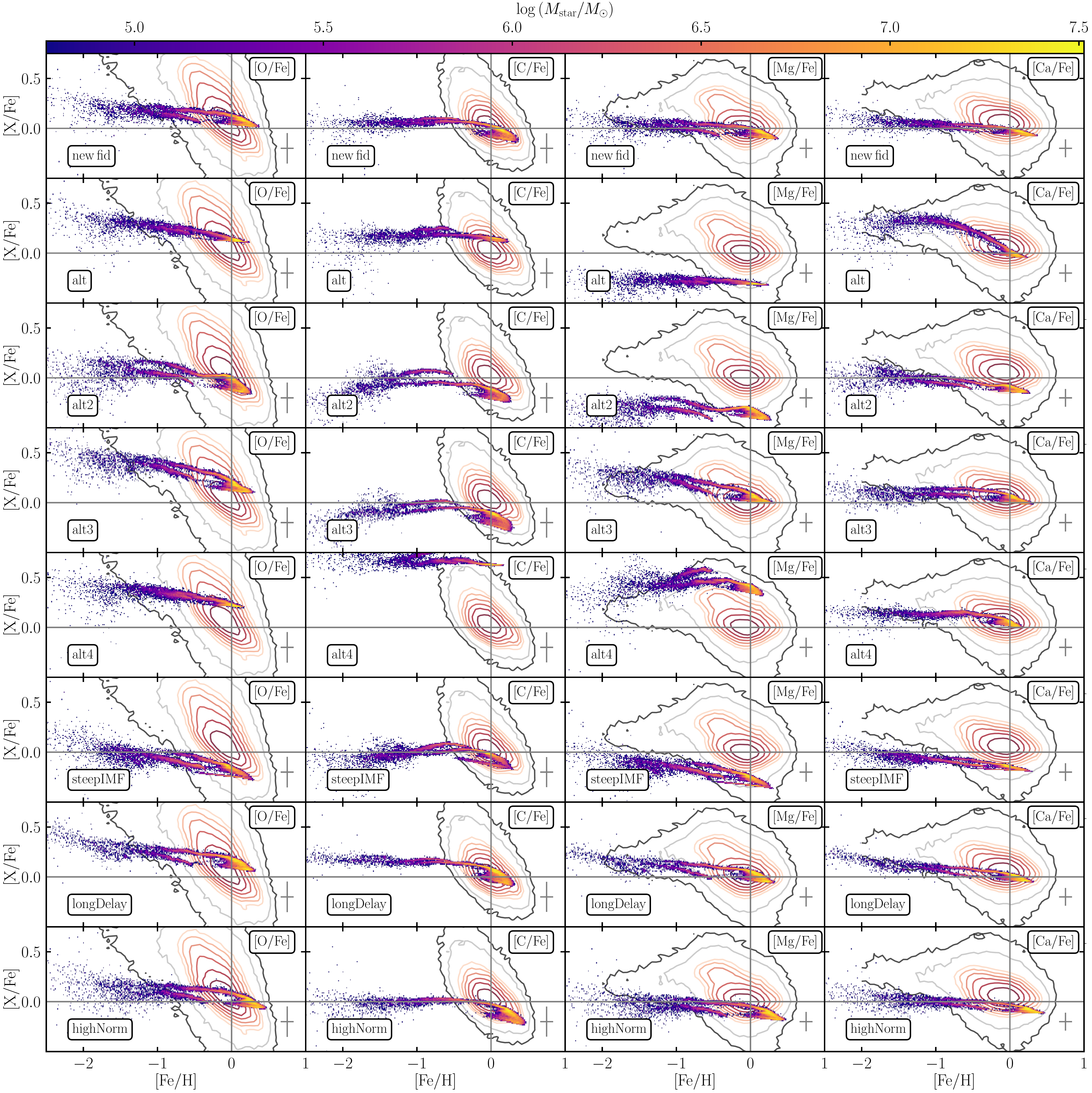}
\end{center}
\vspace{-.35cm}
\caption{Distribution of O, C, Mg and Ca (from left to right) abundances as a function of metallicity, coloured by stellar mass. We show only stellar particles within $6.5<R<9.5$ kpc and $\vert z\vert<2$ kpc. Different rows compare different yield sets as indicated in the lower left corner of each panel. For comparison, contours show the results of the GALAH survey \citep{Buder2020}. Gray error bars in the lower right of each panel indicate the median observational uncertainty of the GALAH data and gray lines indicate solar abundance values. Note, for the \textit{alt4} yield set the C abundance extends to outside the plotting range with [C/Fe]$\sim0.8$ (see also Fig. \ref{fig:omgsis_hist2}).}
\label{fig:comp}
\end{figure*}
%%%%%%%%%%%%%%%%%%%%%%%%%%%%%%%%%%%%%%%%%%%%%%%%%%%%

\subsection{Abundance tracks in [X/Fe] vs. [Fe/H]}
It is now well known that for the MW the evolution of the $\alpha$-abundance as a function of metallicity shows a very distinct trend \citep[e.g.][]{Tinsley1979,Wyse1995,Fuhrmann1998,Venn2004,Bensby2014,Hayden2015}. Low metallicity in-situ stars follow a plateau at high $\alpha$-abundances which obeys a knee at slightly sub-solar metallicities and bends towards lower $\alpha$-abundance at higher metallicity. Simultaneously there exist a lower $\alpha$-abundance sequence around solar $\alpha$-abundance across a wide range of metallicities. This peculiar tracks have been explained by various authors to arise from an exquisite interplay of star formation timescales, stellar yields and particular gas accretion history of the MW \citep[e.g.][]{Chiappini1997,Matteucci2012,Spitoni2020}. Therefore this plane of observables offers an ideal constraint for the interplay of yield sets and feedback model. In \citet{Buck2020b} we have shown that such a bimodal sequence is a rather generic feature of the NIHAO feedback scheme. In this section we use this finding to scrutinise different yield sets in their ability to reproduce (at fixed feedback parameters) the location and shape of the observed $\alpha$-abundance bimodality. Here two things are of importance: On the one hand, the enrichment to high $\alpha$-abundances of $\sim0.3$ dex at low metallicity and simultaneously solar values at solar metallicity should be achieved. And secondly, the shape/position of the knee/down bending should occur roughly at the observed metallicity as it constraints the turning point where the relative contribution of CC-SN to SN~Ia changes.

\begin{figure*}
\begin{center}
\includegraphics[width=\textwidth]{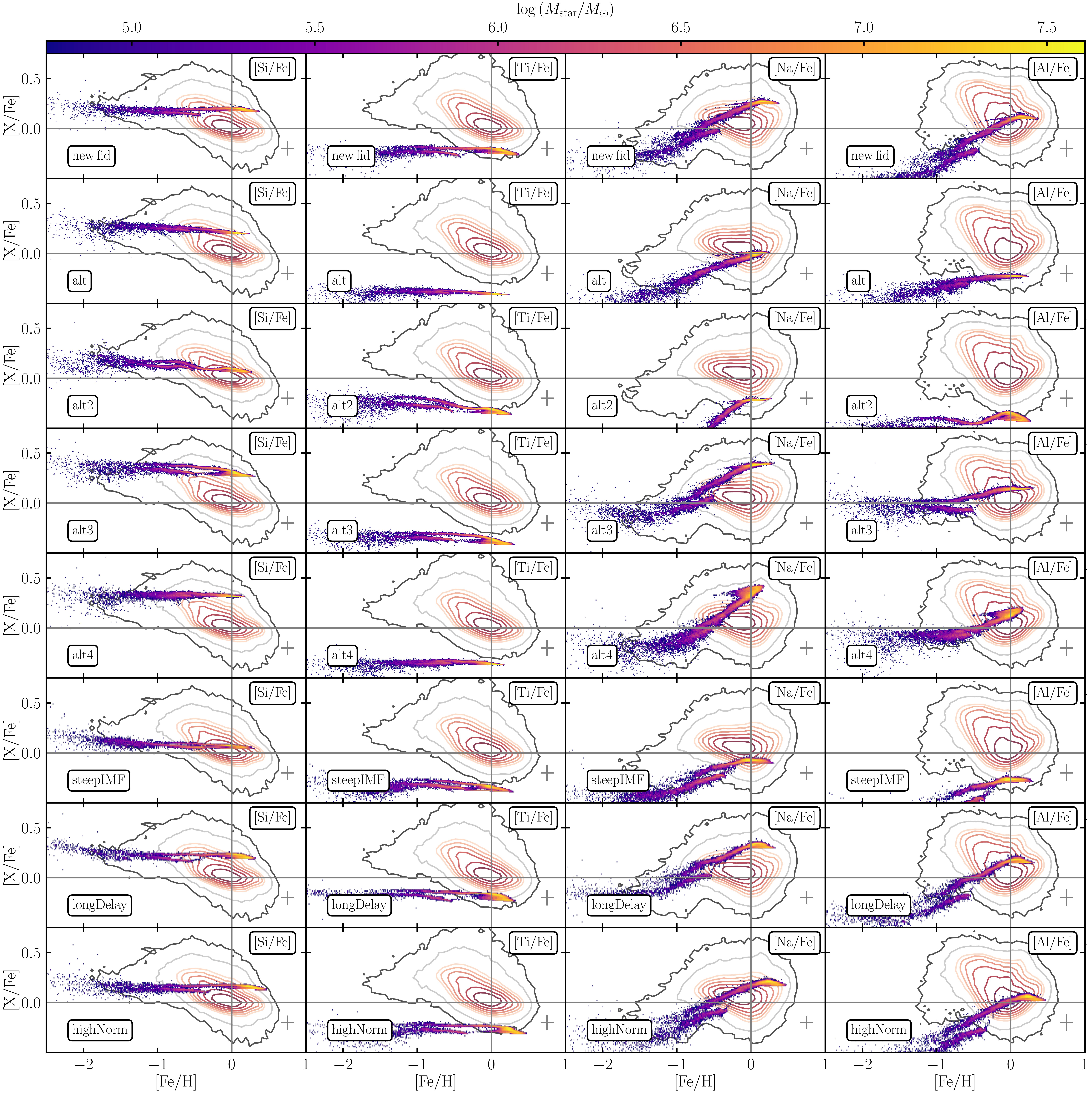}
\end{center}
\vspace{-.35cm}
\caption{Same as Fig. \ref{fig:comp} but this time for Si, Ti, Na and Al (from left to right).}
\label{fig:comp2}
\end{figure*}

In Fig. \ref{fig:alpha_tracks} we compare the [$\alpha$/Fe] vs. [Fe/H] plane for all our models. The upper left panel shows the fiducial yield set and to the right of it we compare the three models varying physical parameters of the SSP while the bottom row shows the four alternative yield sets tested. In order to guide the eye we indicate solar values with gray lines and overlay observational results from the GALAH survey. Note, for this work we apply only a simple spatial selection to the simulated data and refrain from modelling in detail the GALAH selection function (e.g. accounting for that fact that GALAH mainly observes cool dwarfs). Additionally, we accompany this plot by figures \ref{fig:comp} and \ref{fig:comp2} which compare single element (columns) abundance tracks among all models (rows). For this first exploratory work we have decided to focus on a differential comparison between yield sets. Therefore we refrain from exactly modelling the GALAH selection function and decided to simply select stars within an annulus of $6.5<R<9.5$ kpc from the simulations.

In all panels, two sequences can be seen: an extended primary sequence with the highest stellar mass densities reaching supersolar [Fe/H] and showing a bimodality at the highest metallicities, and a secondary sequence at low [Fe/H] as well as low [$\alpha$/Fe] (Fig \ref{fig:alpha_tracks}). The primary sequence are in-situ stars of the main galaxy, whereas the secondary sequence originates from accreted satellite galaxies in these simulations \citep[see][for more details]{Buck2020}. Recent literature \citep{Hayes2018,Das2020,Buder2020} have found similarly pronounced differences between accreted and in-situ stars also for the MW. Their lower abundances in $\alpha$-elements as well as Na and Al agree qualitatively with our study. However, our predictions for [C/Fe] are above the primary sequence, whereas they have found to be below the primary sequence in observations \citep[e.g.][]{Hayes2020} found [C/Fe] to be below the primary sequence at the same [Fe/H] for the MW.

Subsequently, we first focus on the primary sequence, before addressing the secondary sequence.

%%%%%%%%%%%%%%%%%% FIGURE 12 %%%%%%%%%%%%%%%%%%%%%%%%%%%
%%%%%%%%%%%%%%%%%%%%%%%%%%%%%%%%%%%%%%%%%%%%%%%%%%%%

Figure \ref{fig:alpha_tracks} confirms our previous notion of distinctive differences among different yield sets. We find a range of shapes for the $\alpha$-bimodality ranging from a single sequence evolving from high to low [$\alpha$/Fe] (\textit{alt}) over a less separated bimodality ({\textit{fiducial model}}) up to well separated sequences (all other models). Especially the amount of $\alpha$-enhancement differs between different yield sets as would be expected from their different predictions for the element release (e.g. Fig. \ref{fig:yield_comparison}).   
Notably the \textit{steepIMF} model which uses the same yield set as the fiducial model heavily underproduces the $\alpha$-abundance at all metallicities. This is simply caused by a reduction of the number of high mass stars which contribute most to the $\alpha$-element synthesis. Contrary, the \textit{alt3} model overproduces $\alpha$-elements and never reaches solar values despite a clear bimodality. The strongest separation of the two sequences can be seen in the \textit{longDelay} model which shifts the onset of SNIa to $\sim160$ Myrs which is a factor of 4 longer delay time compared to the fiducial model. The \textit{highNorm} run fits the overall [$\alpha$/Fe] values of the MW better than the fiducial run but the larger number of SN~Ia leads to slightly too large iron abundance for the low-$\alpha$ stars.

Finally, the double knee seen in model \textit{alt2} is due to a slightly modified SFH in this model which leads to lower values of the peak SFR between redshift $z=2-1$ ($\sim 3-5$ Gyr after the big bang) as can be seen from Fig. \ref{fig:sfh} in the appendix. 

Let us turn to compare the intrinsic scatter seen in the simulation data to the observed broad sequences. In each panel the gray cross in the lower right corner highlights the median observational uncertainty of the GALAH abundances. While accurately forward modelling the simulation data to account for the exact survey selection function and the observational uncertainties is outside the scope of this work we still recognise that the simulated sequences exhibit a narrower intrinsic scatter. Comparing those narrow tracks with the median observational uncertainty (the gray cross) it seems that even when we would account for the broad observational uncertainty in future work the simulated tracks are still too narrow. This poses a challenge for our theoretical models to explain chemical enrichment in MW mass galaxies. 

Figure \ref{fig:comp} and \ref{fig:comp2} add a more diverse view to this as different elements X obey different slopes in the [X/Fe] vs. [Fe/H] plane both in the models and in the GALAH data. Especially the very steep slope observed for O is not reproduced by any of our models\footnote{It is still debated, if this steep trend is the true astrophysical one, or introduced by the observational analysis. E.g. APOGEE measures a shallower trend in the IR compared to optical studies \citep[see e.g. Fig. 22 in][]{Buder2018}. The cause for this discrepancy is still not clear and explanations range from atomic data or 3D NLTE effects in the IR to other not explained effect in the optical.}. C is relatively similar for all models with less variation compared to O or Mg for example. Similar to C, Ca tracks are relatively similar in all our models except for the model \textit{alt} which shows way stronger Ca enhancement at low metallicities than all other models. For Si, Na and Al we find similar strong differences among yield sets. Only for Ti those differences are less pronounced. In fact, Si is almost always overproduced while Ti is under-produced in all models which is a known problem of the theoretical models \citep[see e.g. discussion of Fig. 23 in][]{Kobayashi2020} which might be resolved in 2 or 3D models of nucleosynthesis. 

Comparing the scatter of the model galaxies with the observations shows that also for individual elements the simulations reveal quite narrow tracks. For example, the Mg track of the fiducial model agrees quite well with the centroid of the GALAH data but when taking into account the observational uncertainty of $\sim0.25$ dex the simulated data would still be narrower than the observations. One reason for this miss-match might be that observational uncertainties are underestimated. Future re-analysis of stellar spectra might decrease the observational scatter and allow to better constrain the intrinsic abundance scatter in the MW \citep[see also][for a determination of the intrinsic abundance scatter in the MW]{Bedell2018,Sharma2020,Yuxi2021}. Another reason might certainly be an under-estimation of the intrinsic scatter in the simulations. Several model simplifications might artificially decrease the intrinsic scatter: A simplistic feedback prescription might fail to reproduce the intrinsic scatter observed for the MW. Assuming a fully sampled IMF might in fact drive the chemical enrichment to the mean of an SSP neglecting the effects of stochasticity on the enrichment process. Similarly, over-estimating the effect of metal mixing in the ISM and underestimating the effect of radial migration decreases the intrinsic abundance scatter at fixed radius. The primary focus of this work is to study different yield sets and evaluate their viability in reproducing the observational constraints. Therefore we did not model observational uncertainties in the simulation as this would have hindered our comparison. However, future work will address this point and model observational selection effects and uncertainties. Exactly pin-pointing if there is indeed a discrepancy between simulation results and MW observations as well as researching the exact cause for it, may it be observational or theoretical, is certainly needed and will help us understand galaxy formation and the MW in more depth.

%%%%%%%%%%%%%%%%%%%%%%%%%%%%%%%%%%%%%%%%%%%%%%%%%%%
\section{Conclusion}
\label{sec:conclusion}
%%%%%%%%%%%%%%%%%%%%%%%%%%%%%%%%%%%%%%%%%%%%%%%%%%%

By coupling a flexible chemical evolution/stellar population model with a well tested galaxy formation model we are not only able to study in more depth physical processes of galaxy formation, but more importantly, we are able to constrain stellar evolution and elemental synthesis models. Our chemical evolution model allows us to flexibly explore stellar evolution parameters in a full cosmological galaxy formation model. This helps to put constraints on valid parameter combinations by comparing simulation results to chemical abundance structures as observed in the MW and local galaxies. In particular, we show that at fixed SSP parameters (i.e. stellar lifetimes, mass ranges for CC-SN and SN Ia, etc.) different yield sets from the literature show significant difference in their ability to enrich the galaxies with different chemical species. Especially the $\alpha$-element abundances of different yield sets show significant differences in dwarf and MW mass galaxies. This in combination with exquisite observational data for the MW and its dwarf galaxies could potentially be used to discriminate between different models. Similarly there exist difference in the chemical abundances for slight variations of the SN~Ia delay time or the high mass slope of the IMF.    

Our results are summarised as follows:
\begin{itemize} 
\item We have presented a new, mass and metallicity dependent, time resolved chemical enrichment model for cosmological simulations that is able to follow any element evolution self-consistently. This model enables a flexible parameter combination with the potential of parameter inference from observational data via the \C~package prior to the simulation run, i.e. optimisation with a computational cheap model.
 
\item We implement our new model in the \G~code within the NIHAO feedback model. Figure \ref{fig:mass_metal} and \ref{fig:oxygen} show that this new model reproduces stellar mass metallicity and gas phase oxygen abundance relations generally well. We find a slight tension of the models around $10^7-10^8$ $\Msun$. Investigating the reason for this requires a larger sample of galaxies at these stellar masses but it might point towards a failure of the current feedback model at those stellar masses as previously shown by \citet{Agertz2020a}. Future work will investigate this point in more depth.  

\item We find that total metallicity and iron abundances (Fig. \ref{fig:mass_metal} and \ref{fig:oxygen}) as well as their distribution functions (Fig. \ref{fig:metal_hist} and \ref{fig:z_hist}) do not differ much between different yield sets when used with the NIHAO feedback model. We find that for all yield sets tested no recalibration of the feedback model is necessary in order to match observational constraints on the mass-metallicity relations.

\item The $\alpha$-element distribution function (Fig. \ref{fig:alpha_hist}) as well as the [$\alpha$/Fe] vs. [Fe/H] plane (Fig. \ref{fig:alpha_tracks}) show significant diversity between the models and can help putting constraints on valid yield set combinations or SSP parameter choices such as SN~Ia delay time or IMF shape/slopes, when compared with observational data, as done in Figs. \ref{fig:alpha_tracks}, \ref{fig:comp} and \ref{fig:comp2}. 

\item We find that none of our models is able to simultaneously match observational constraints for dwarf galaxies and the MW when looking at the $\alpha$ element distribution function (Fig. \ref{fig:alpha_hist}). While in the dwarf galaxy regime the models \textit{alt, alt3} and \textit{longDelay} seem to best match the \citep{Kirby2019} data for the MW the best matching models are the \textit{fiducial} and \textit{highNorm} models. While most of our simulated $\alpha$ element distribution functions show an increasing median with stellar mass, the observations are consistent with the opposite trend, thus signalling an unique {\em diversity of abundance patterns}, which we find difficult to reconcile in a single chemical enrichment model.

\item For the MW model we find a narrow intrinsic scatter around the tracks in the [$\alpha$/Fe] vs. [Fe/H] plane (Fig. \ref{fig:alpha_tracks}) and similarly for individual elements (Fig. \ref{fig:comp} and \ref{fig:comp2}) which compared to the observational scatter is in slight tension with MW observations, the {\em problem of matching the scatter in chemical abundance space}. Possible solutions to this might be an underestimation of stochasticity from the assumption of a fully sampled IMF and a possible under-estimation of observational uncertainties. However, for a quantitative conclusion a more in depth analysis of the simulations including the exact modelling of the observational selection function is needed and will be done in future work.

\item Within the framework of the NIHAO feedback model the best parameter combination we have found are tables by \citet{Karakas2016} for AGB stars with tables from \citet{Chieffi2004} for CC-SN and yields from \citet{Seitenzahl2013} for SN\,Ia with a power law delay time distribution function following \citet{Maoz2012} with a delay time of $\sim40$ Myr and normalisation of $-2.9$. Those are combined with stellar lifetimes from \citet{Argast2000} and a fiducial \citet{Chabrier2003} IMF where the maximum mass of stars exploding as CC-SN is set to 40 \Msun above which we assume direct black hole collapse. The minimum stellar mass exploding as CC-SN is set to $8 \Msun$, see table \ref{tab:params} for more details. A nearly identical parameter combination has been inferred from MW data previously by \citet{Philcox2018}. Note, for the dwarf galaxy we have found that either a steeper IMF or a higher SN~Ia normalisation brings its iron abundance closer to the observed relation of \citet{Kirby2013}. This might be an indication that a variable IMF model might actually be preferable. 
\end{itemize}

The combination of a generic parameter inference tool with fully cosmological galaxy simulations can help us improve our understanding of galaxy formation in general and the chemo-kinematics of our own MW in particular. 
We are confident that our approach of flexibly varying essential stellar evolution and chemical enrichment parameters will positively contribute to our understanding of galaxies and might further help to improve nucleosynthetic yields and feedback prescriptions. 

\section{Outlook}

The chemical enrichment prescription presented in this paper offers several new pathways of studying galaxy formation across cosmic time. Fig. \ref{fig:yieldhist} shows the relative contribution of different sources (AGB stars, SN Ia and CC-SN) to the abundance of different chemical elements. Specifically studying those elements which result predominantly from a single source might help us gain deeper insight into the (astro)physical mechanisms of galaxy formation and evolution. Especially, in this way observations of emission and absorption lines of the gaseous components of galaxies and their surroundings at different cosmic times might offer new insights into the mechanisms of galactic winds enriching the CGM and ISM.

Our new model further offers the possibility to study the relative contribution of different enrichment channels in L$^\star$ galaxies (SNII vs. AGB vs. SNIa) which is not possible from observational data alone. Similarly the MW is the only galaxy to date for which we have quantitative resolved spectroscopic observations of single stars to study its chemo-kinematics. Assessing to which extent the features we see in the MW are a generic outcome of galaxy formation or the result of a particular formation history is challenging to answer. Studying the chemo-kinematic properties of a larger ensemble of simulated L$^\star$ galaxies where the survey specific selection functions are taken into account will certainly boost our knowledge.
Therefore a more in depth and quantitative comparison to MW data will be performed in a follow-up study.

At the dwarf galaxy scale we find opposite trends with stellar mass for the median [$\alpha$/Fe] values in simulations and observations. Certainly the star formation efficiency in dwarf galaxies plays a large role in setting their $\alpha$-enhancement. Whether the $\alpha$-element distributions of simulated dwarf galaxies in fact in disagreement with the observed diversity of abundance tracks in Local Group dwarf galaxies remains to be seen. Therefore future work will thoroughly compare of model dwarf galaxy abundance tracks with observed element distributions and help us understand the physical processes of galaxy formation in more depth.

%%%%%%%%%%%%%%%%%%%%%%%%%%%%%%%%%%%%%%%%%%%%%%%%%%%
\section*{acknowledgments}
%%%%%%%%%%%%%%%%%%%%%%%%%%%%%%%%%%%%%%%%%%%%%%%%%%%
TB thanks Martin Rey and Cristina Chiappini for helpful discussions on an earlier version of this draft. 
TB and CP acknowledge support from the European Research Council under ERC-CoG grant CRAGSMAN-646955. 
JR was funded by the DLR (German space agency) via grant 50\,QG\,1403.
AO is funded by the Deutsche Forschungsgemeinschaft (DFG, German Research Foundation) –- 443044596. This research made use of {\sc{pynbody}} \citet{pynbody}, {\sc{SciPy}} \citep{scipy}, {\sc{NumPy}} \citep{numpy}, {\sc{IPython}} \citep{ipython} and {\sc{Jupyter}} \citep{jupyter} and used {\sc{matplotlib}} \citep{matplotlib} to display all figures in this work. We gratefully acknowledge the Gauss Centre for Supercomputing e.V. (www.gauss-centre.eu) for funding this project by providing computing time on the GCS Supercomputer SuperMUC at Leibniz Supercomputing Centre (www.lrz.de).
This research was carried out on the High Performance Computing resources at New York University Abu Dhabi. We greatly appreciate the contributions of all these computing allocations.
All simulations performed for this study including test runs and failed simulations used about 2.5M core-hours which under the assumption of 30W per core-hour and a CO2 intensity of
electricity of $\sim 600$ kgCO2/MWh amounts to a carbon footprint of $\sim45$t of CO2.

\section*{Data availability}

The data underlying this article will be shared on reasonable request to the corresponding author. Python scripts and data to reproduce all plots of this paper can be found at: \url{https://github.com/TobiBu/chemical_enrichment.git}.

\bibliography{astro-ph.bib}

\appendix

\section{Star formation history}

%%%%%%%%%%%%%%%%%% FIGURE A1 %%%%%%%%%%%%%%%%%%%%%%%%%%%
\begin{figure}
\begin{center}
\includegraphics[width=.9\columnwidth]{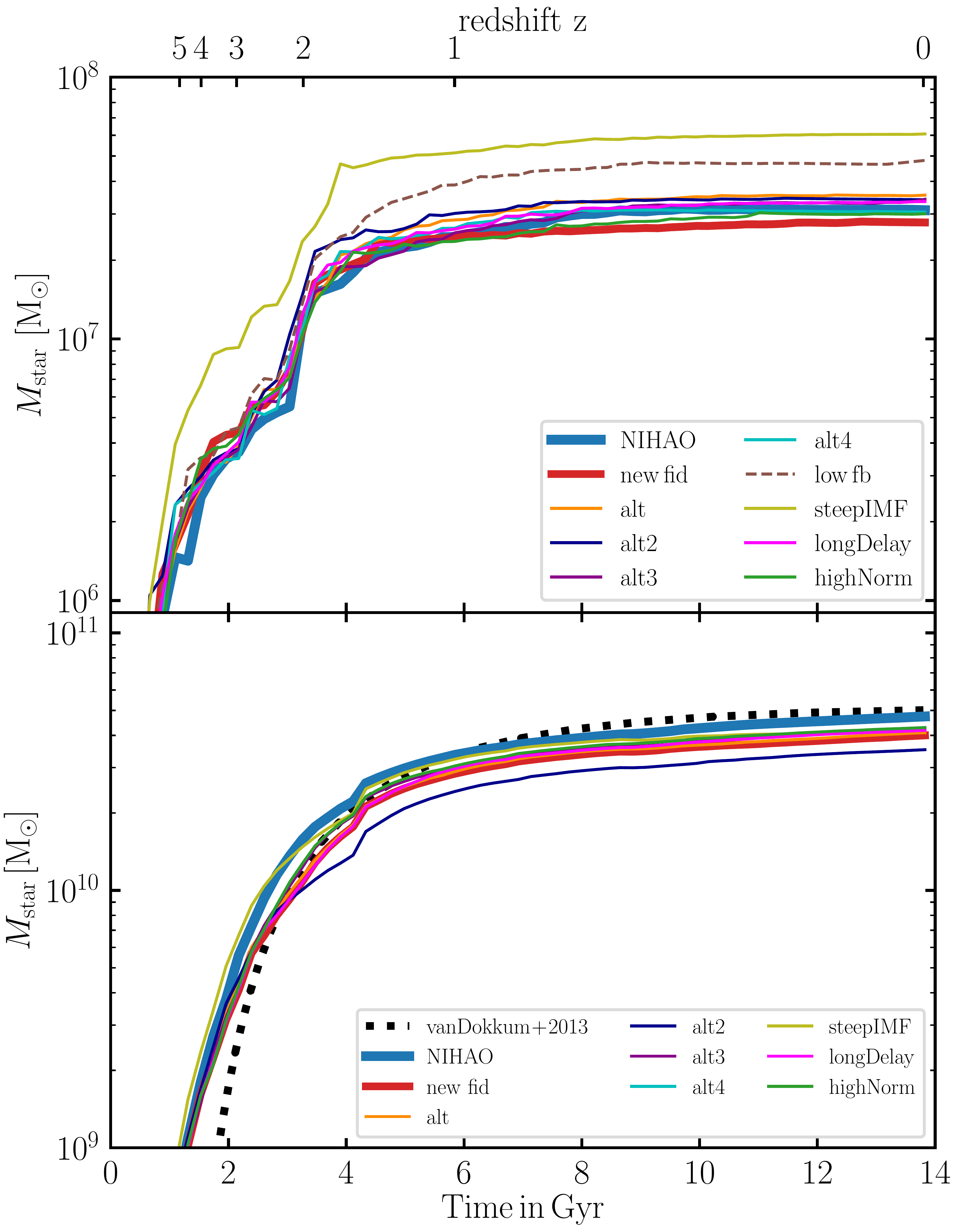}
\end{center}
\vspace{-.35cm}
\caption{Stellar mass growth for g2.83e10 (upper panel) and g8.26e11 (lower panel) for all model runs performed for this work. The thick blue line shows the NIHAO simulation. In the lower panel the black dashed line shows the observationally inferred stellar mass growth for MW progenitor galaxies from \citet{vanDokkum2013}. In order to enhance the slight difference at lower redshift we cropped the plot at a lower mass limit of $10^9 \Msun$.}
\label{fig:sfh}
\end{figure}
%%%%%%%%%%%%%%%%%%%%%%%%%%%%%%%%%%%%%%%%%%%%%%%%%%%%

Figure \ref{fig:sfh} shows the cumulative mass growth history of the dwarf galaxy \texttt{g2.83e10} (upper panel) and the MW mass galaxy \texttt{g8.26e11} (lower panel) for all yield sets studied in this work. For the dwarf galaxy the \textit{low fb} and the \textit{steepIMF} runs lead to stronger mass growth in the early universe because of less feedback. This effect is not so much seen in the MW mass model. 

\section{Metallicity distribution function}

%%%%%%%%%%%%%%%%%% FIGURE A3 %%%%%%%%%%%%%%%%%%%%%%%%%%%
\begin{figure}
\begin{center}
\includegraphics[width=0.42\columnwidth]{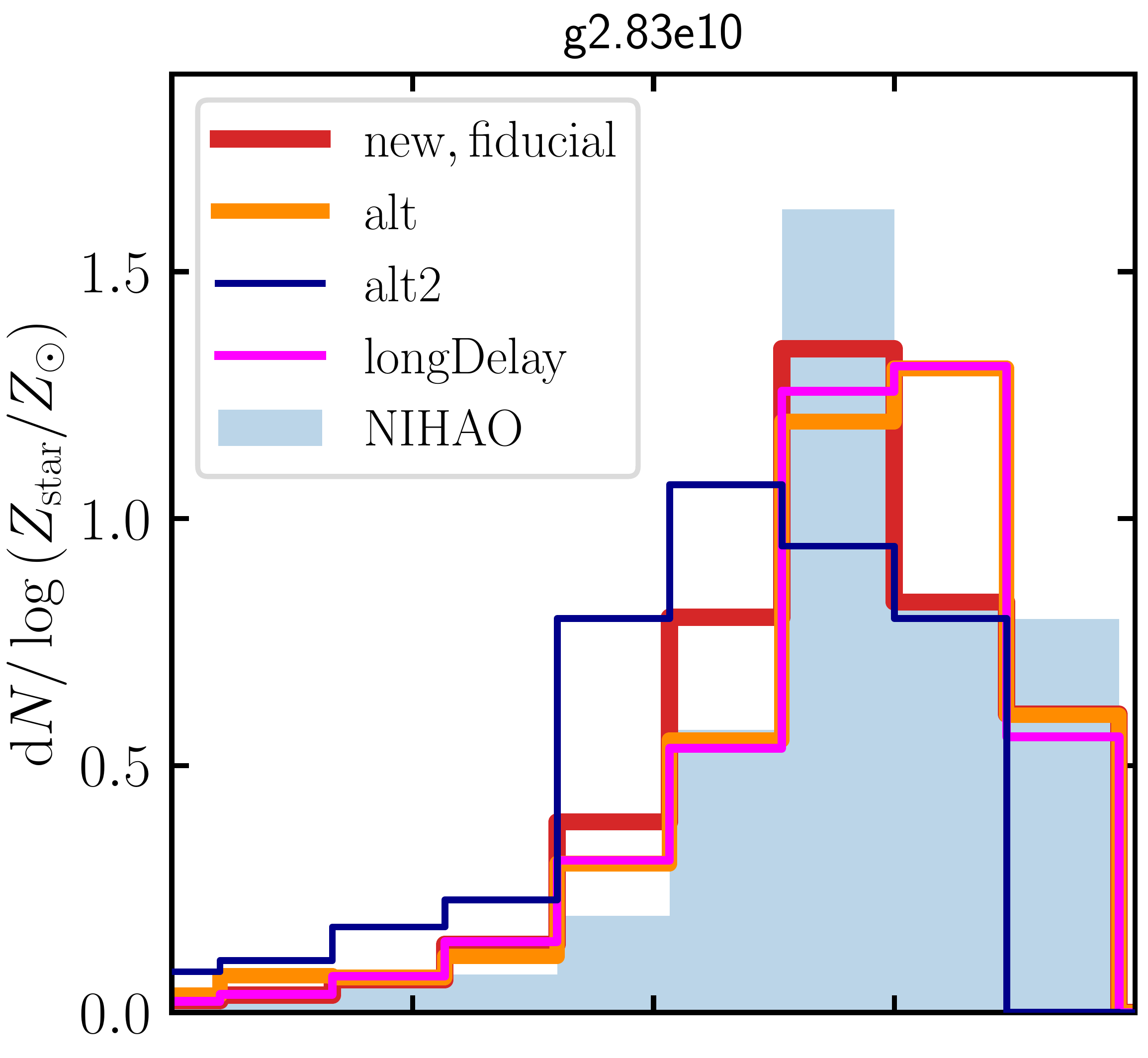}
\includegraphics[width=0.42\columnwidth]{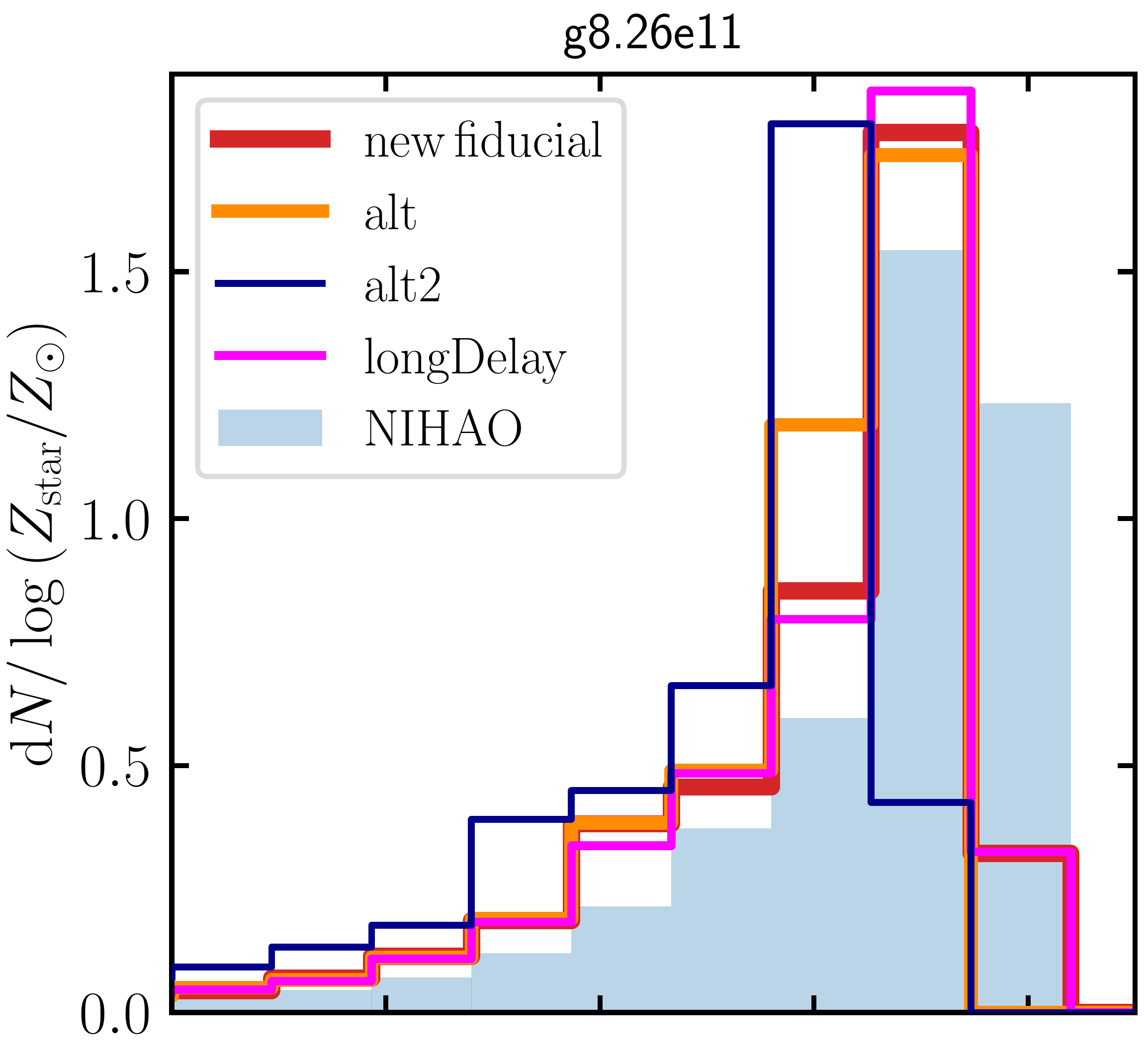}
\includegraphics[width=0.42\columnwidth]{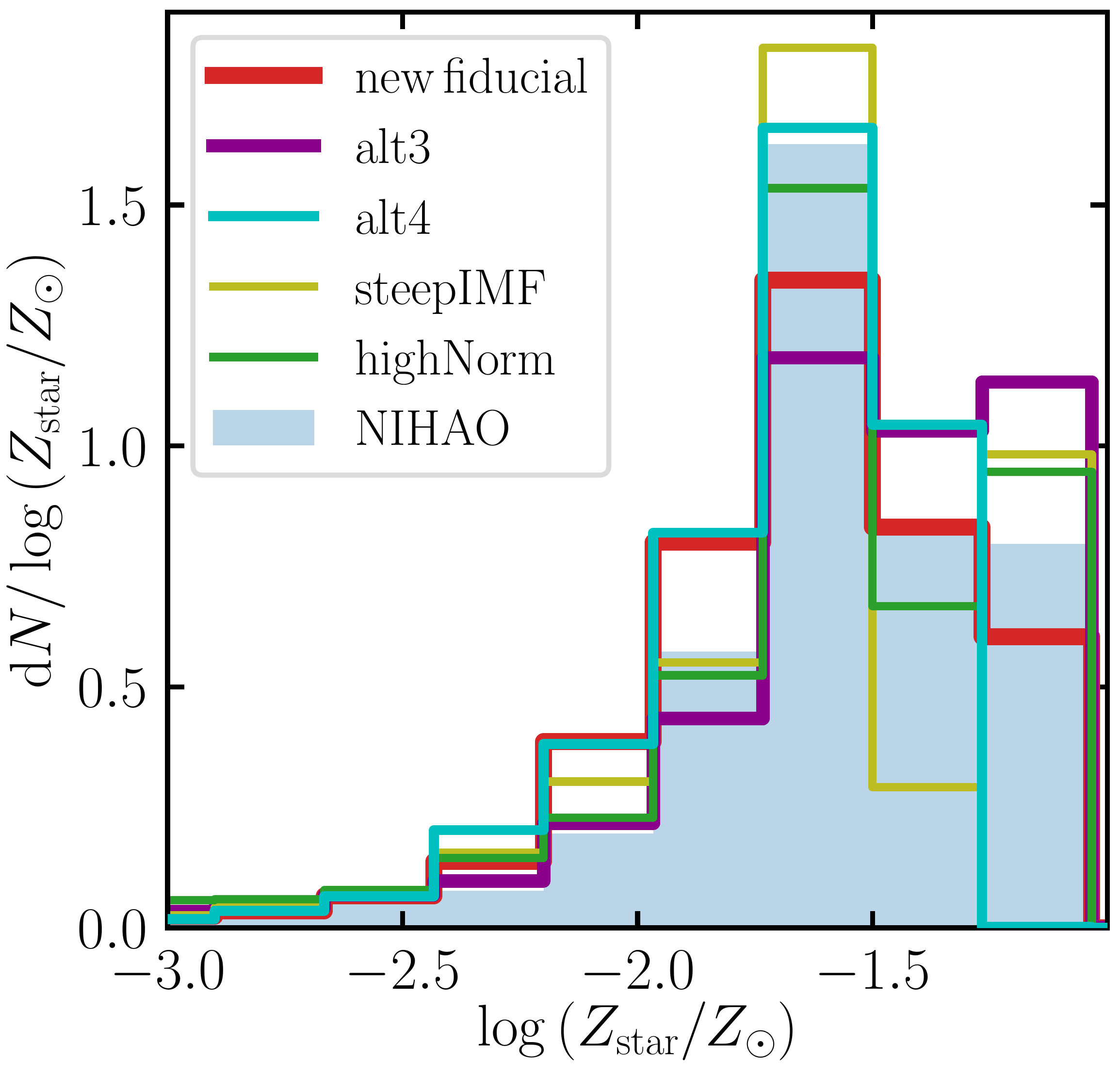}
\includegraphics[width=0.42\columnwidth]{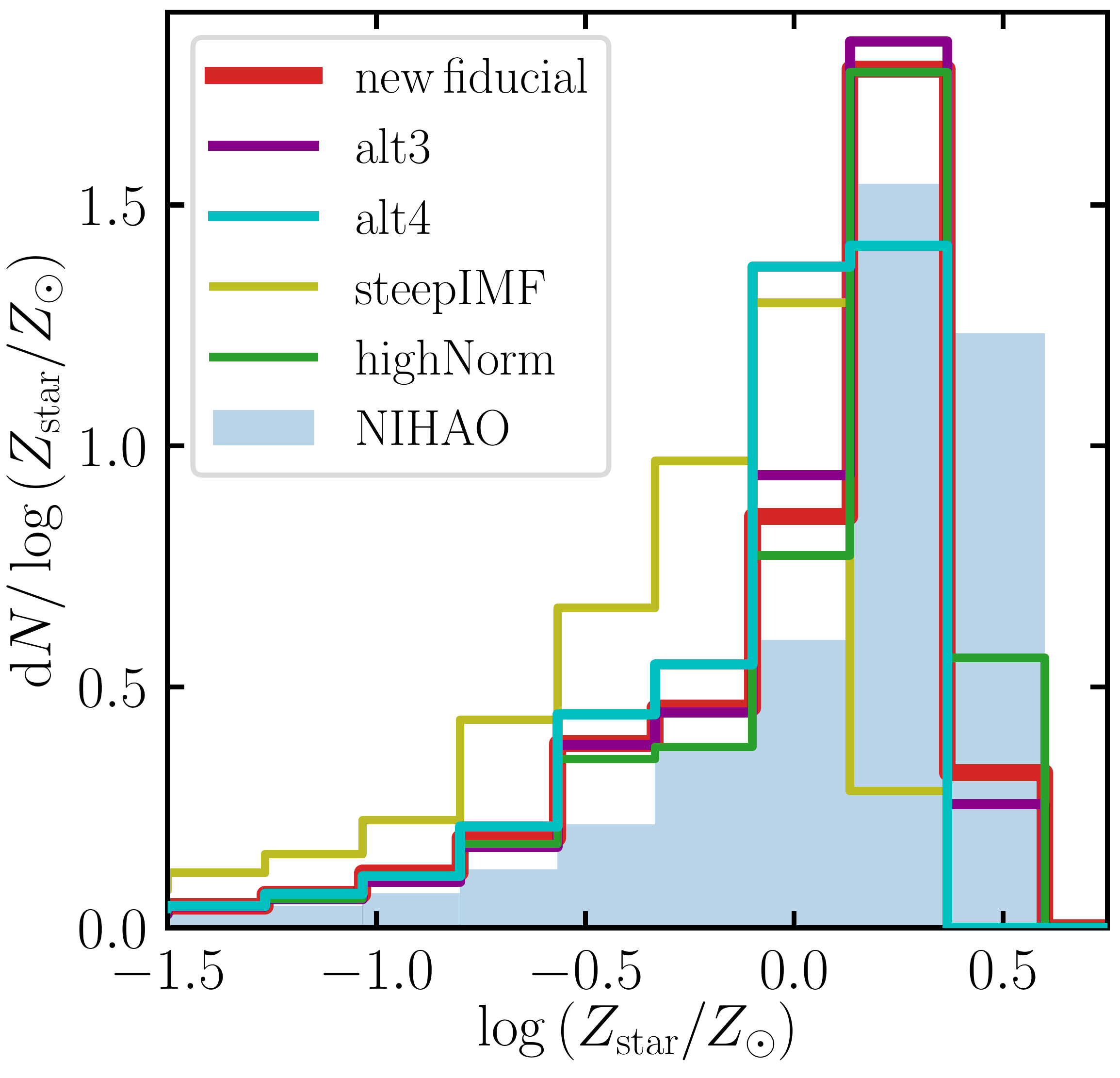}
\end{center}
\vspace{-.35cm}
\caption{Distribution functions of total stellar metallicity for the dwarf galaxy g2.83e10 (left panels) and the MW mass galaxy g8.26e11 (right panels) for various different yield sets. Blue filled histograms show the results for the NIHAO model.}
\label{fig:z_hist}
\end{figure}
%%%%%%%%%%%%%%%%%%%%%%%%%%%%%%%%%%%%%%%%%%%%%%%%%%%%

%%%%%%%%%%%%%%%%%% FIGURE A4 %%%%%%%%%%%%%%%%%%%%%%%%%%%
\begin{figure}
\begin{center}
\includegraphics[width=0.4225\columnwidth]{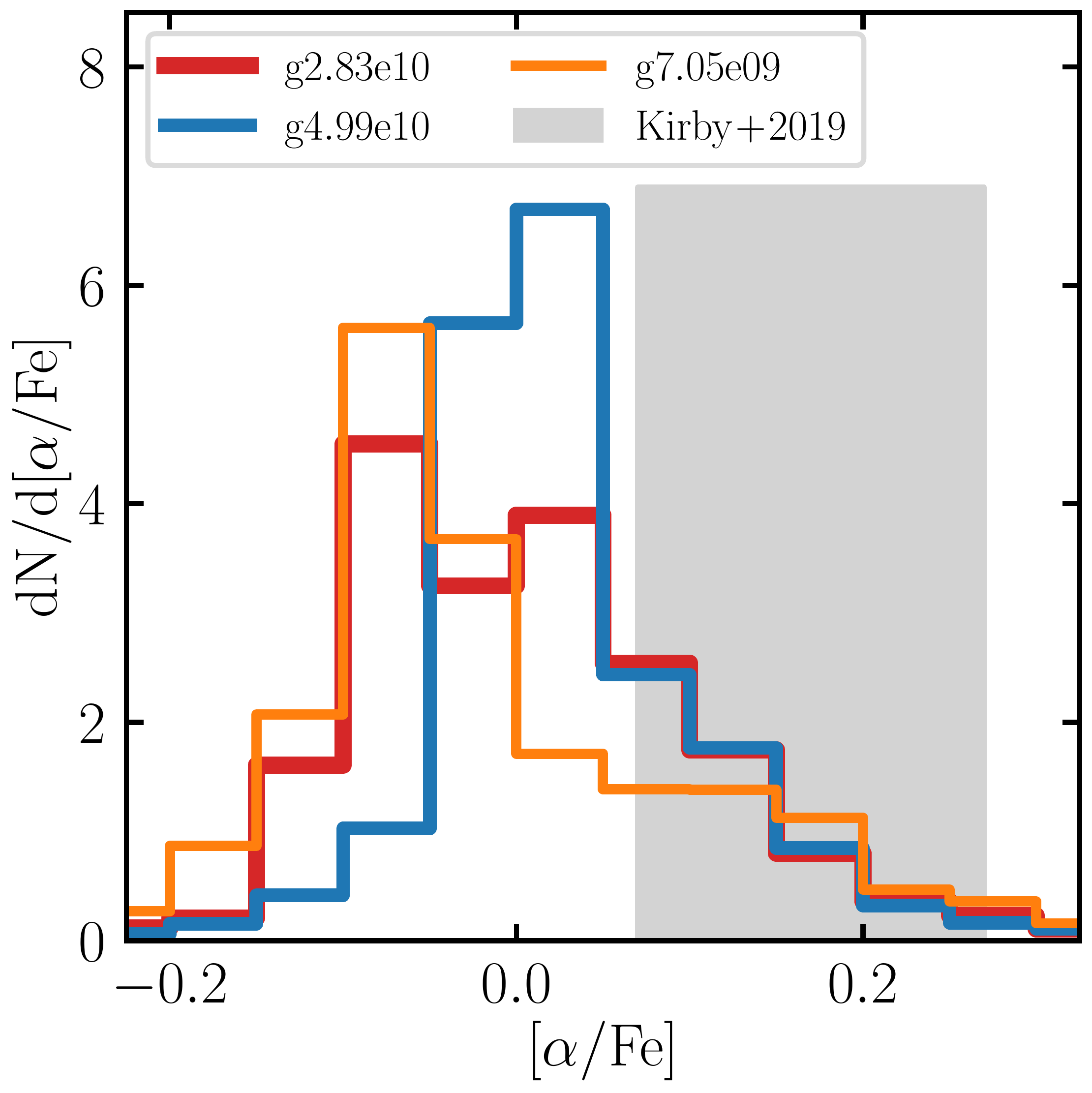}
\includegraphics[width=0.45\columnwidth]{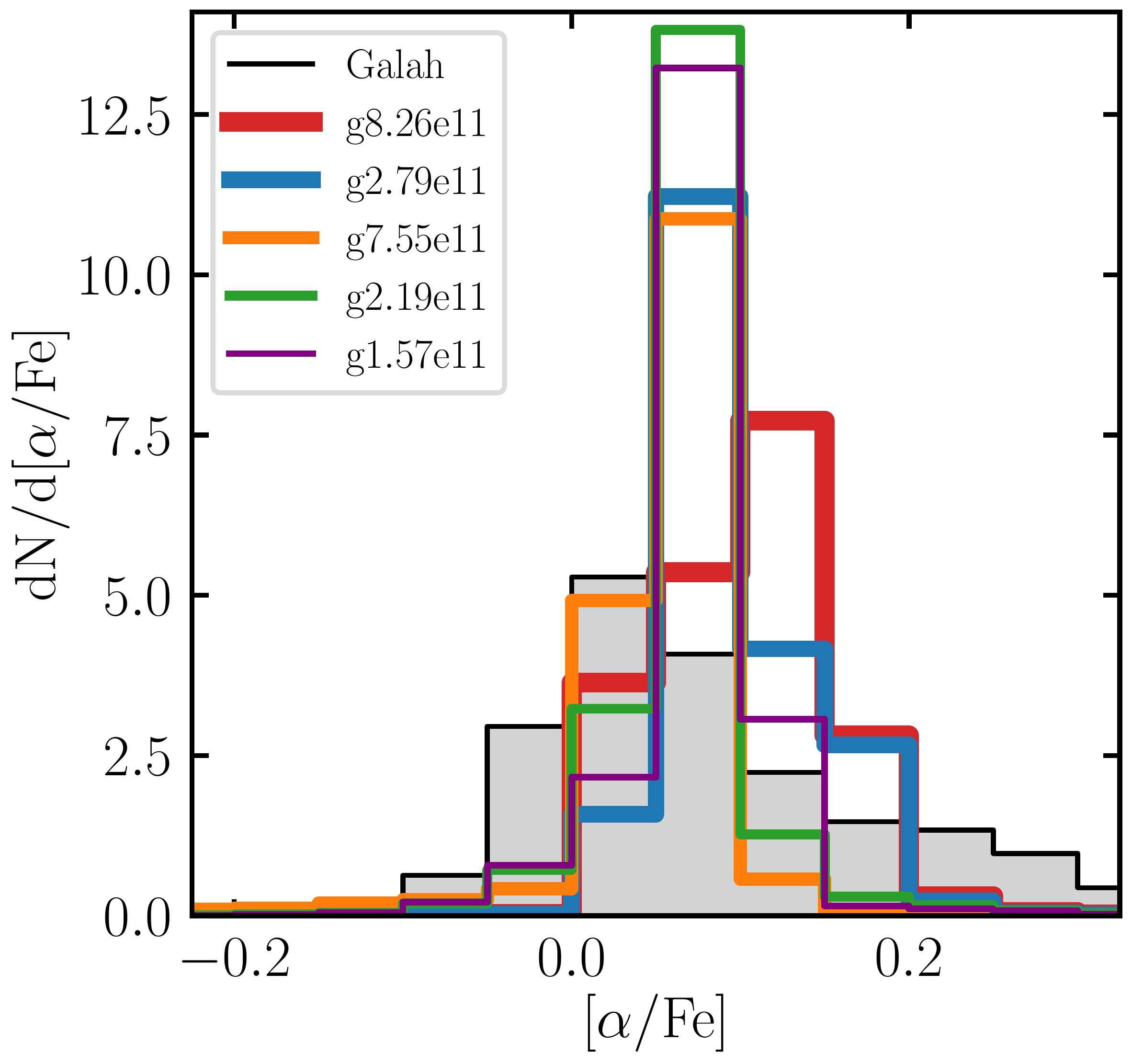}
\end{center}
\vspace{-.35cm}
\caption{Distribution functions of $\alpha$-element abundance for the dwarf galaxies (left panel) and more massive galaxies (right panel) for the fiducial yield set. The gray bar in the left panel shows observational data from \citet{Kirby2019} and the filled histogram in the right panel show the GALAH data.}
\label{fig:mass_comp_hist}
\end{figure}
%%%%%%%%%%%%%%%%%%%%%%%%%%%%%%%%%%%%%%%%%%%%%%%%%%%%

\begin{figure*}
\begin{center}
\includegraphics[width=0.42\columnwidth]{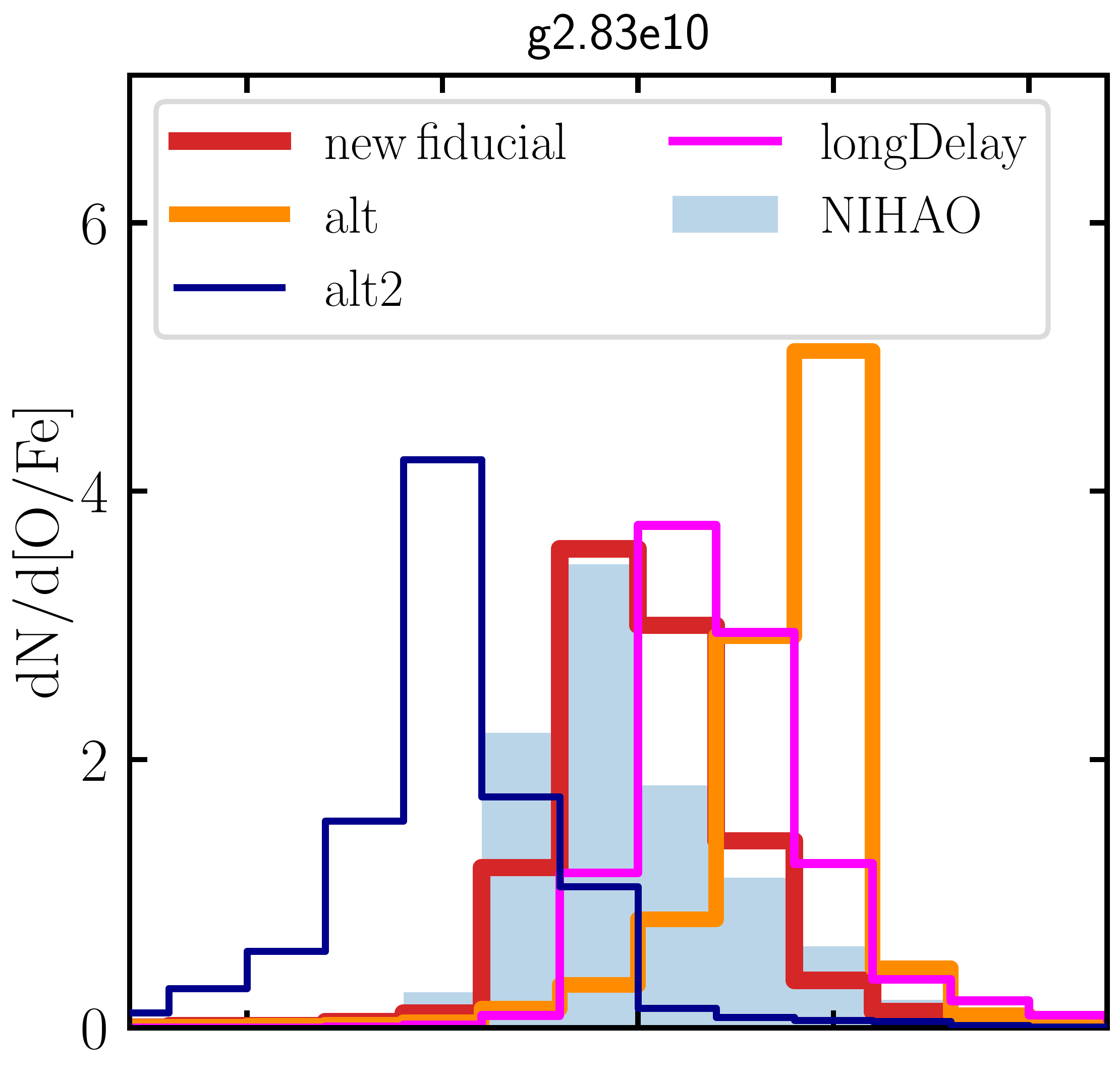}
\includegraphics[width=0.42\columnwidth]{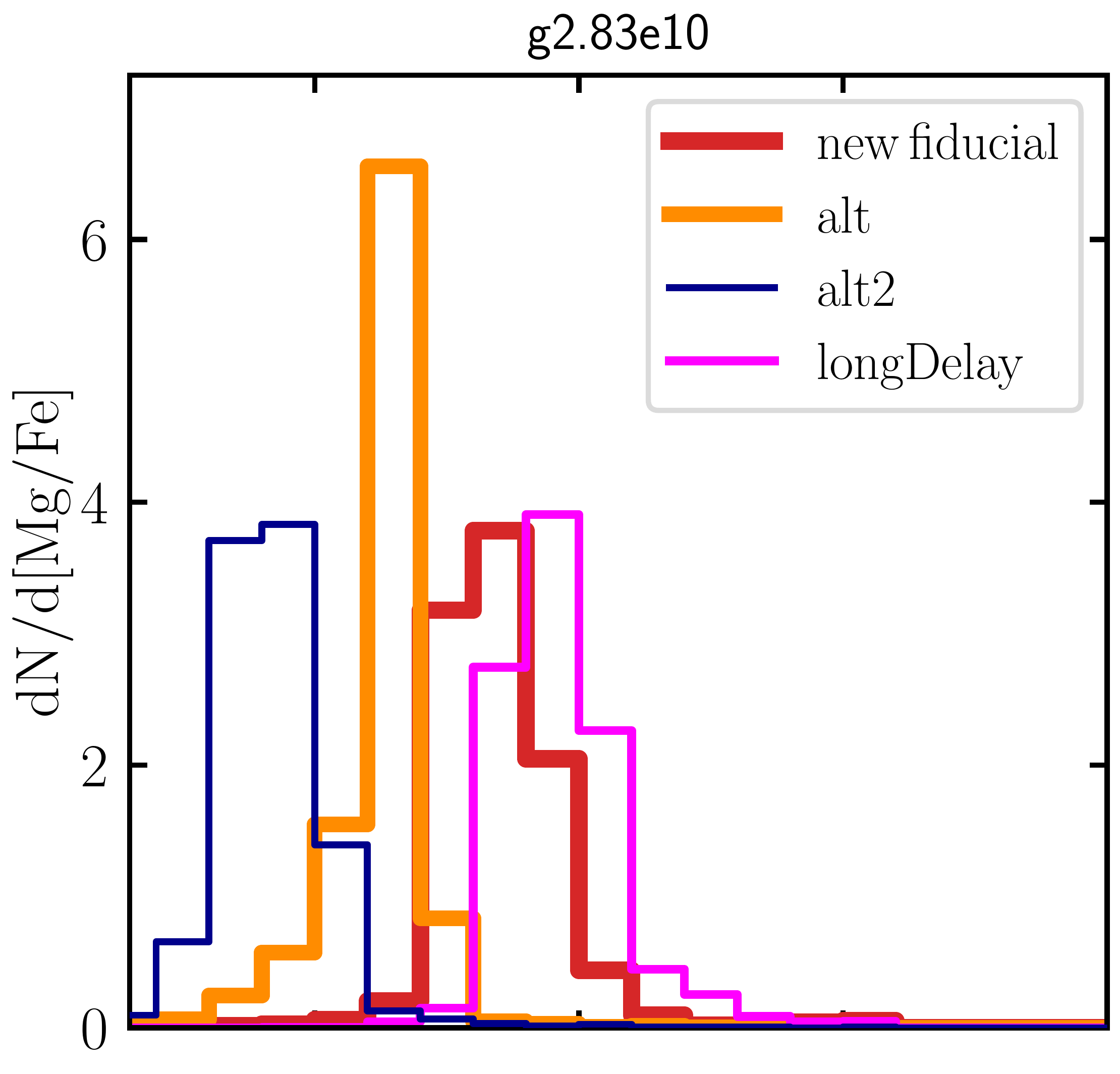}
\includegraphics[width=0.43\columnwidth]{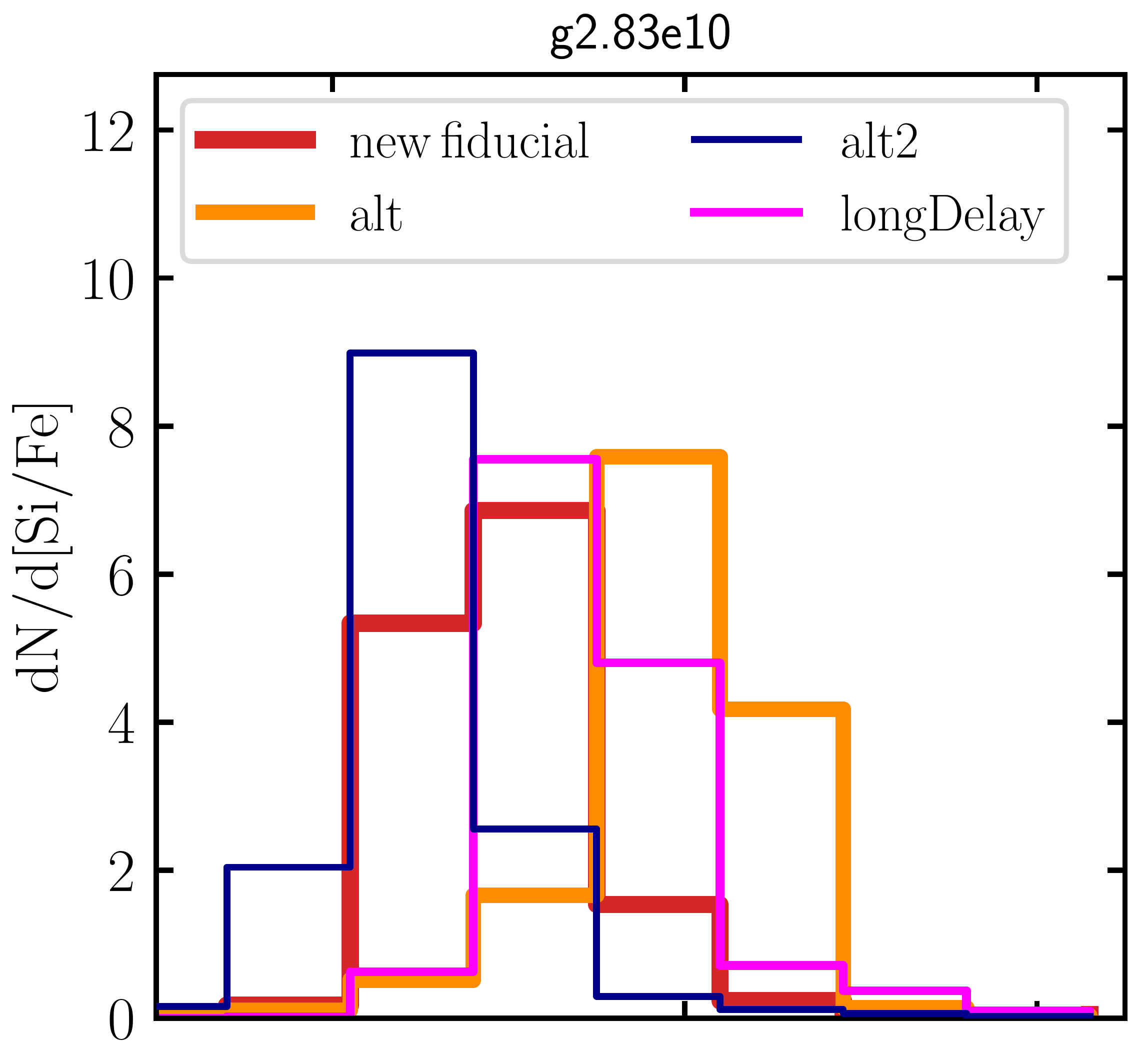}
\includegraphics[width=.42\columnwidth]{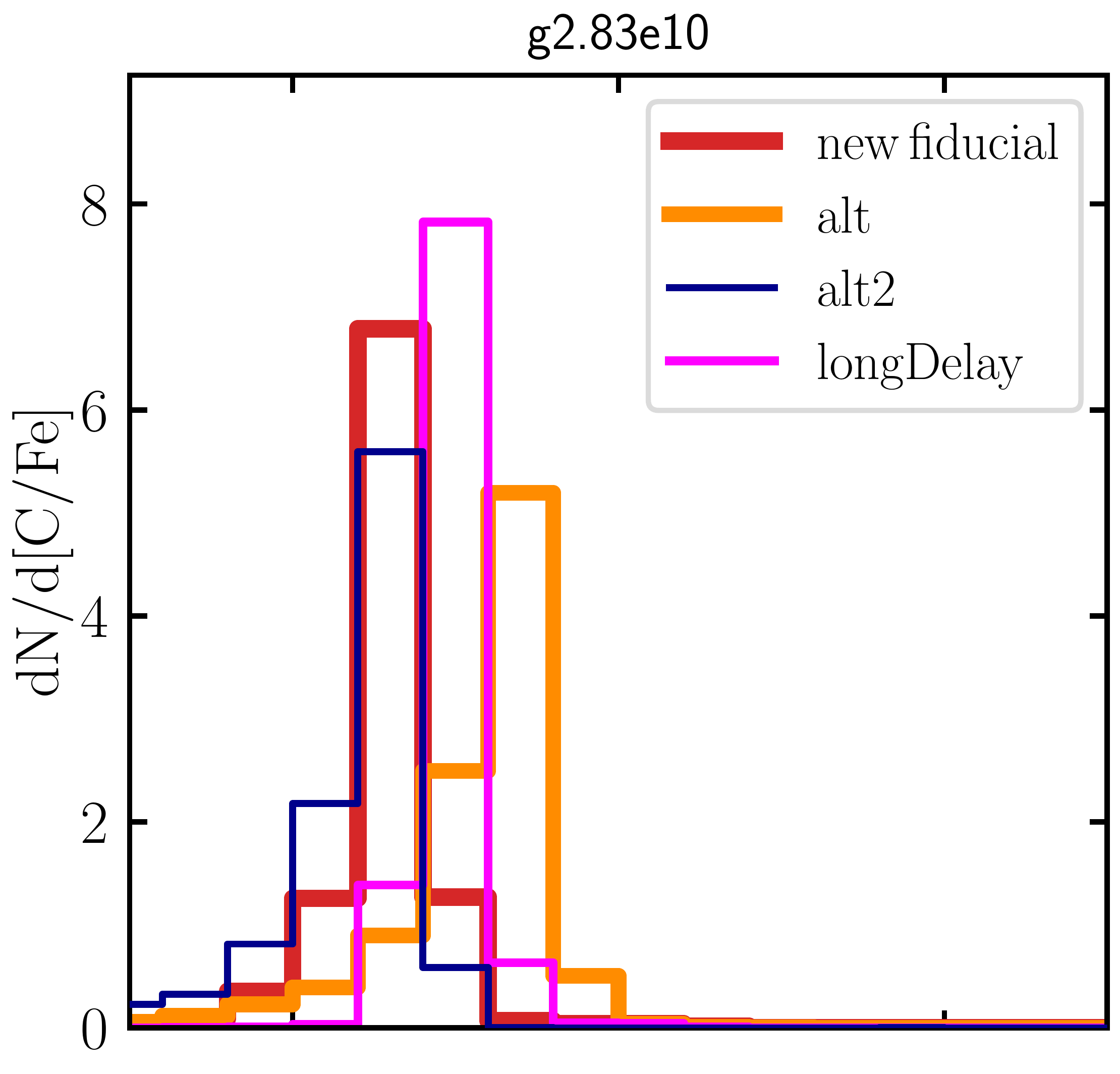}

\includegraphics[width=0.42\columnwidth]{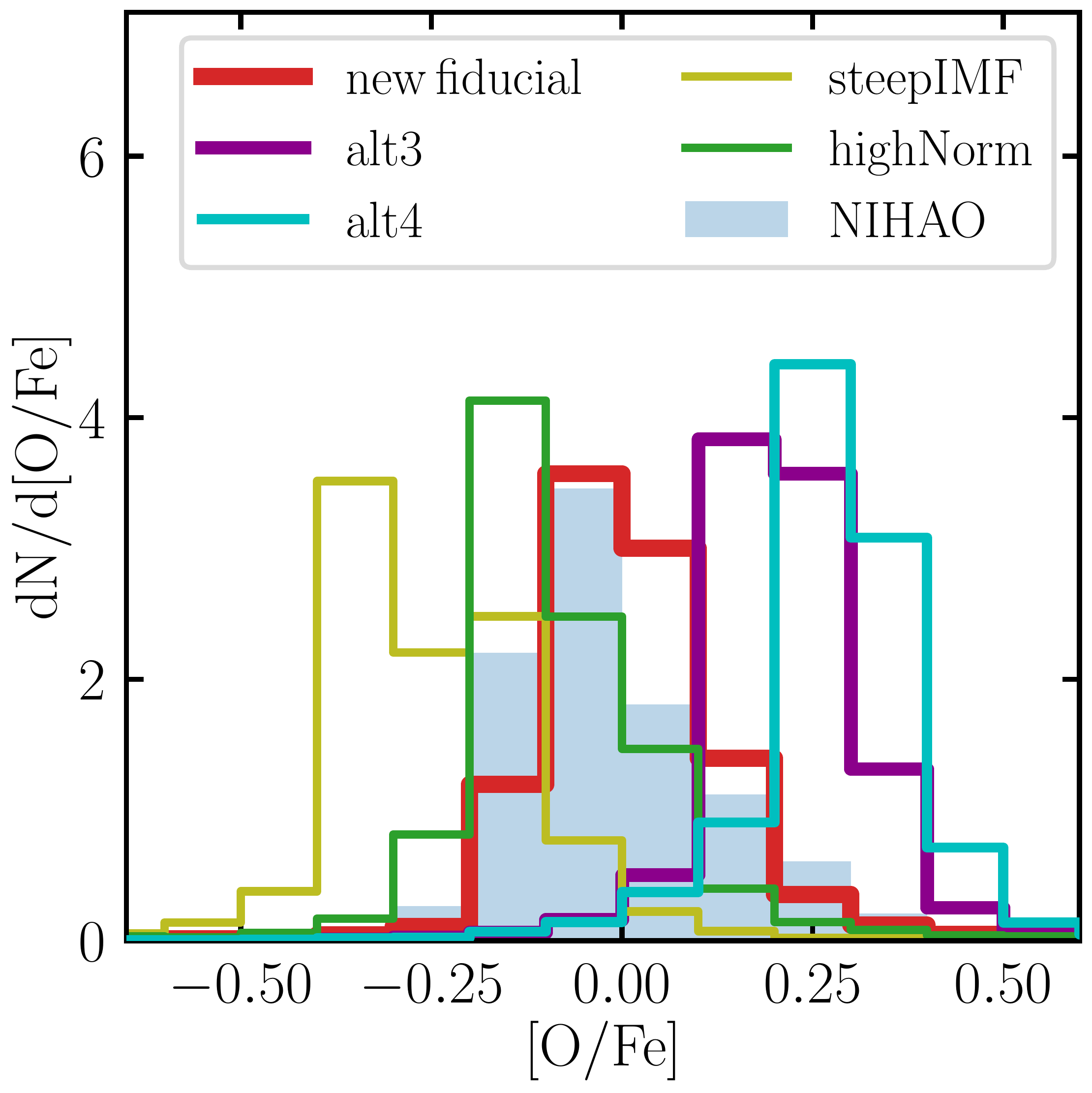}
\includegraphics[width=0.43\columnwidth]{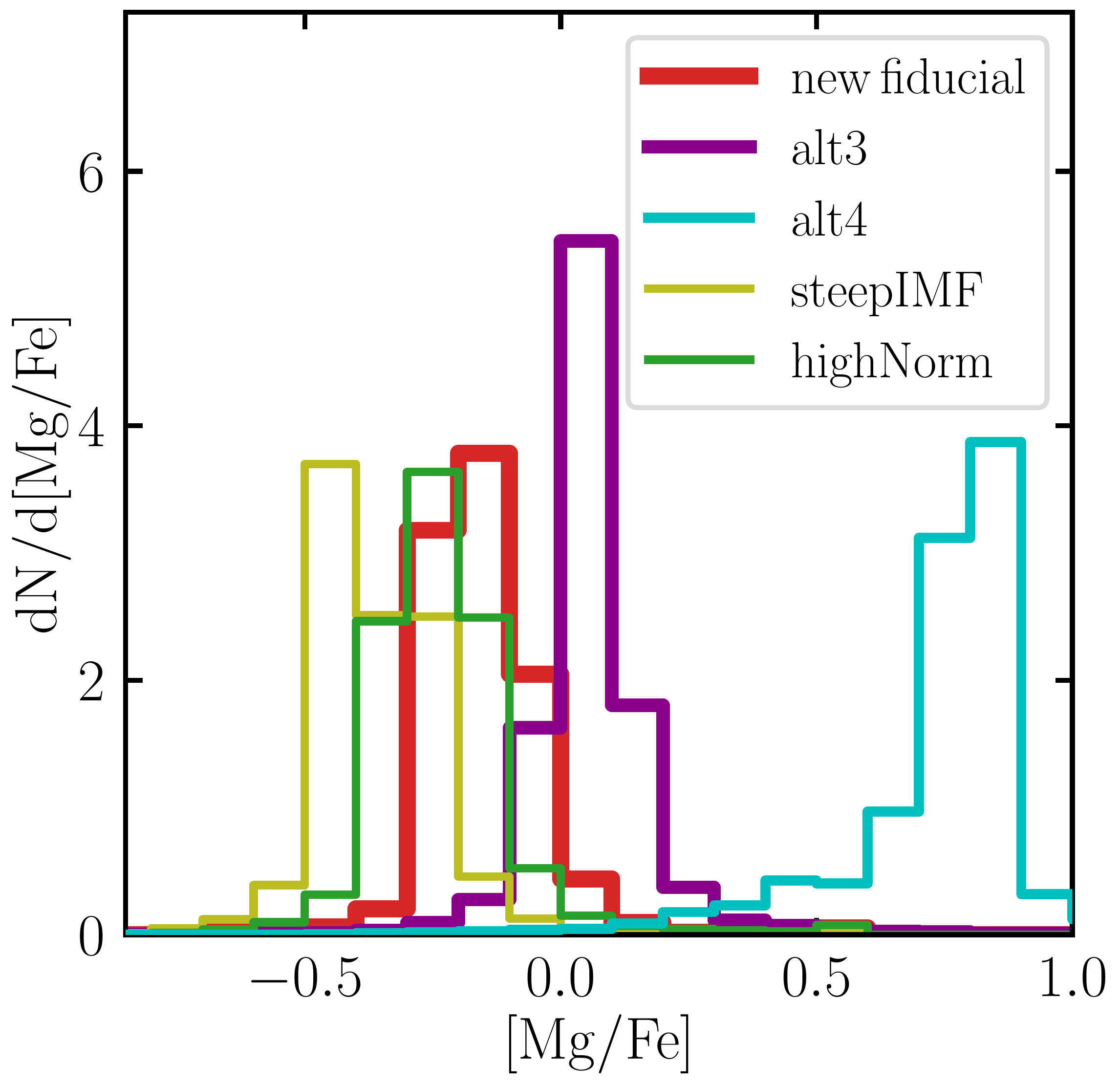}
\includegraphics[width=0.43\columnwidth]{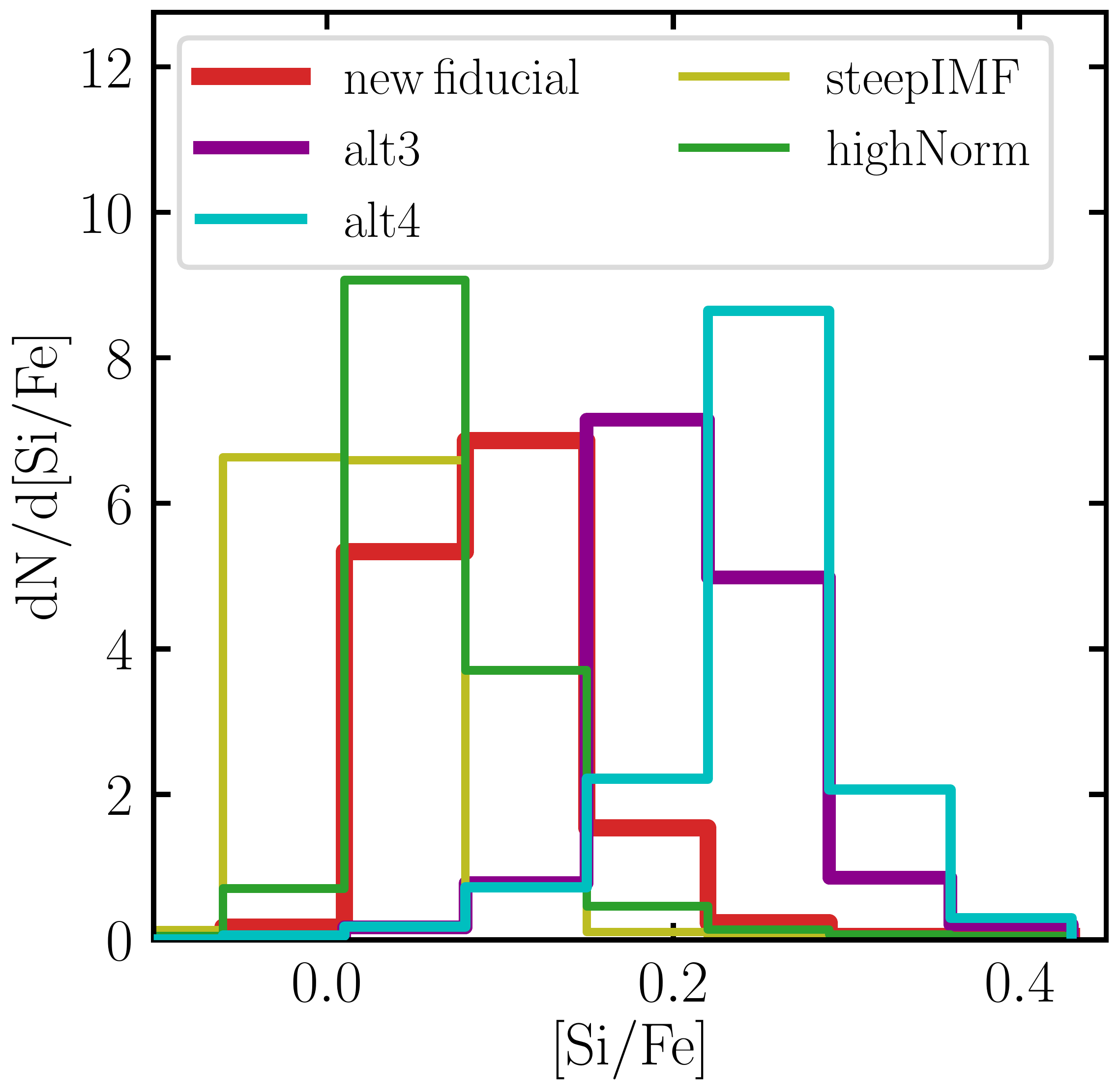}
\includegraphics[width=.42\columnwidth]{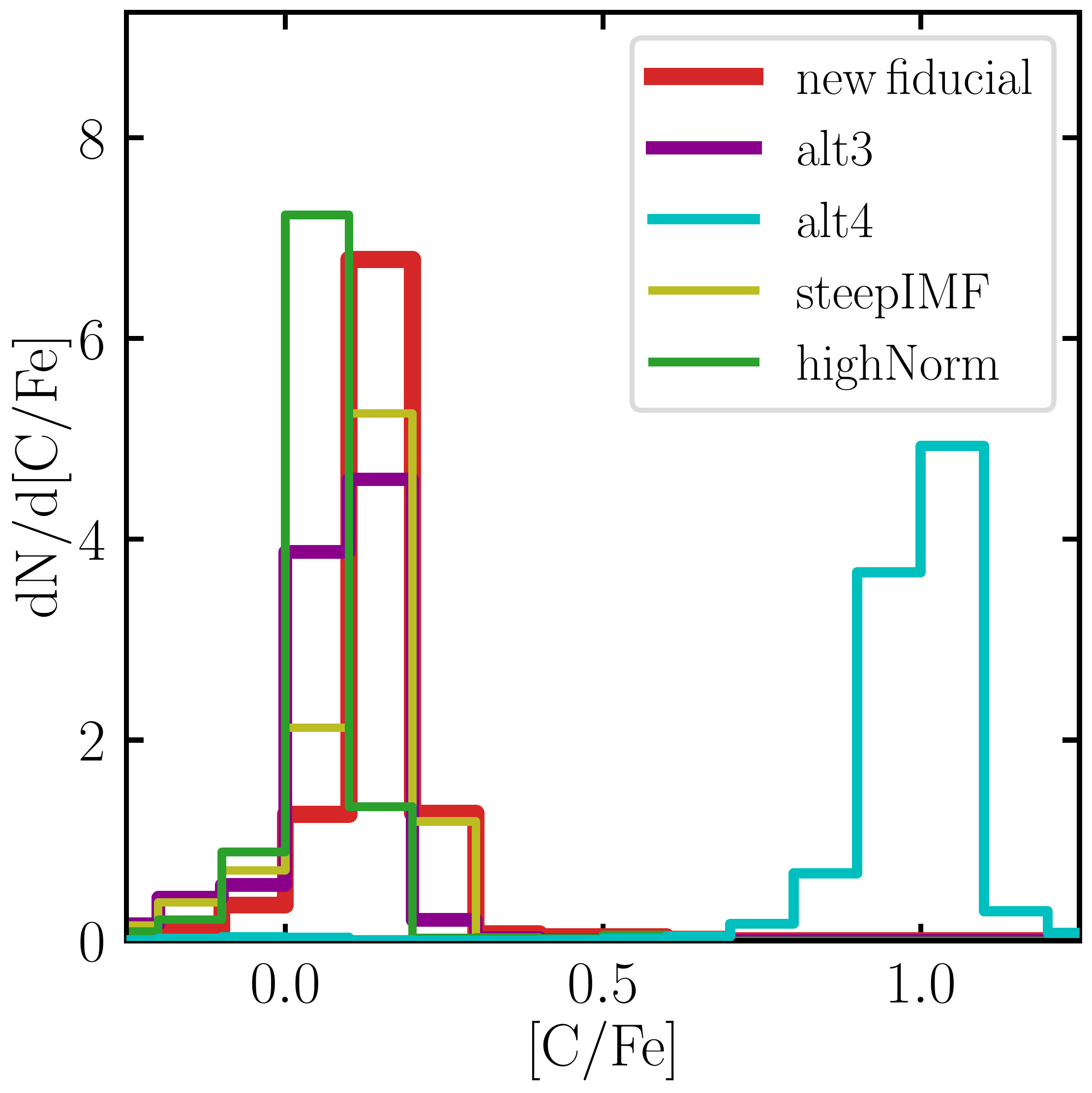}
\end{center}
\vspace{-.35cm}
\caption{Distribution functions of O (left), Mg (second column), Si (third column) and C (right) for the dwarf galaxy g2.83e10 for various different yield sets tested in this work. The gray filled histogram in the left panels show the results for the fiducial NIHAO implementation.}
\label{fig:omgsis_hist}
\end{figure*}

\begin{figure*}
\begin{center}
\includegraphics[width=0.42\columnwidth]{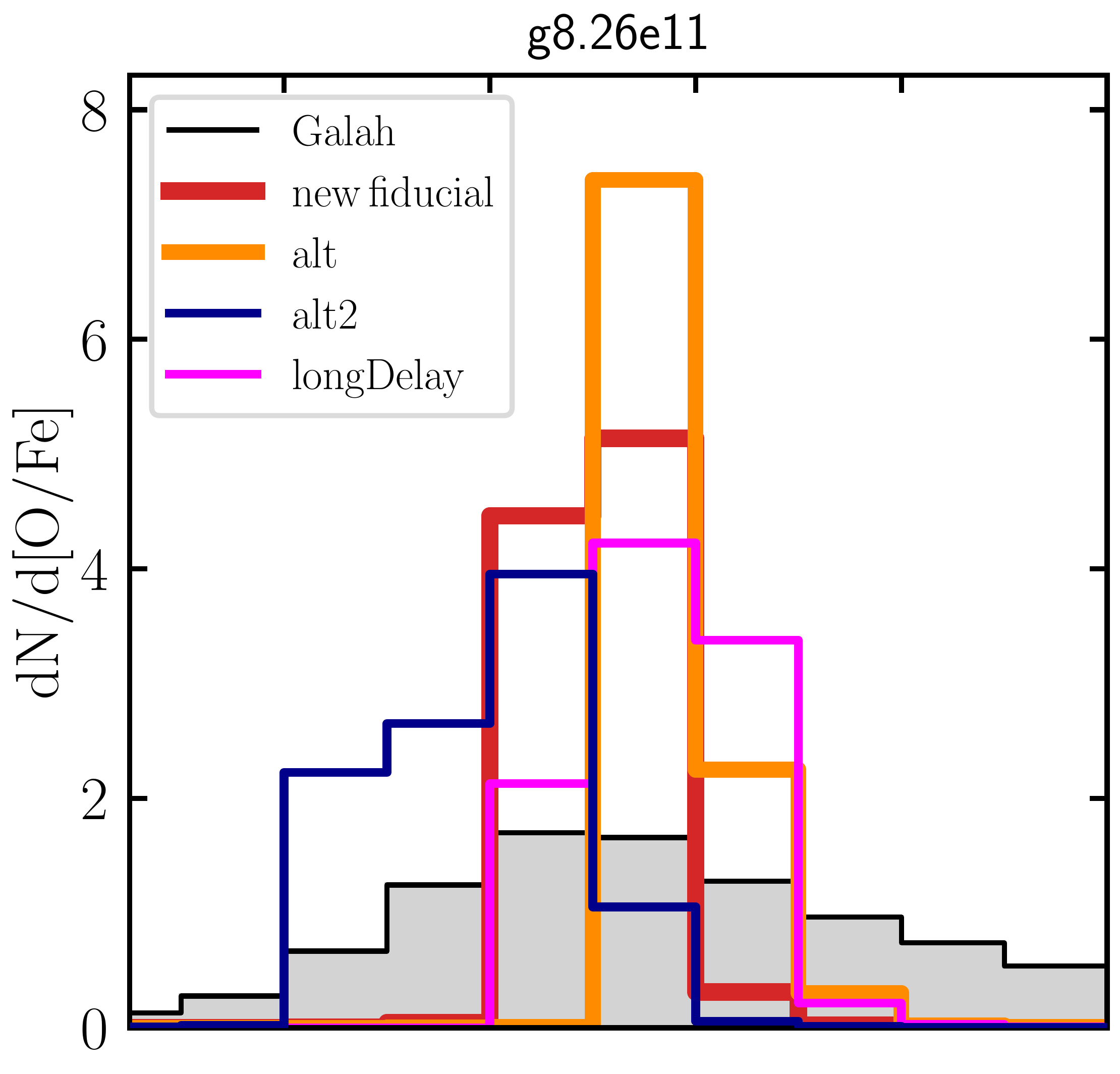}
\includegraphics[width=0.42\columnwidth]{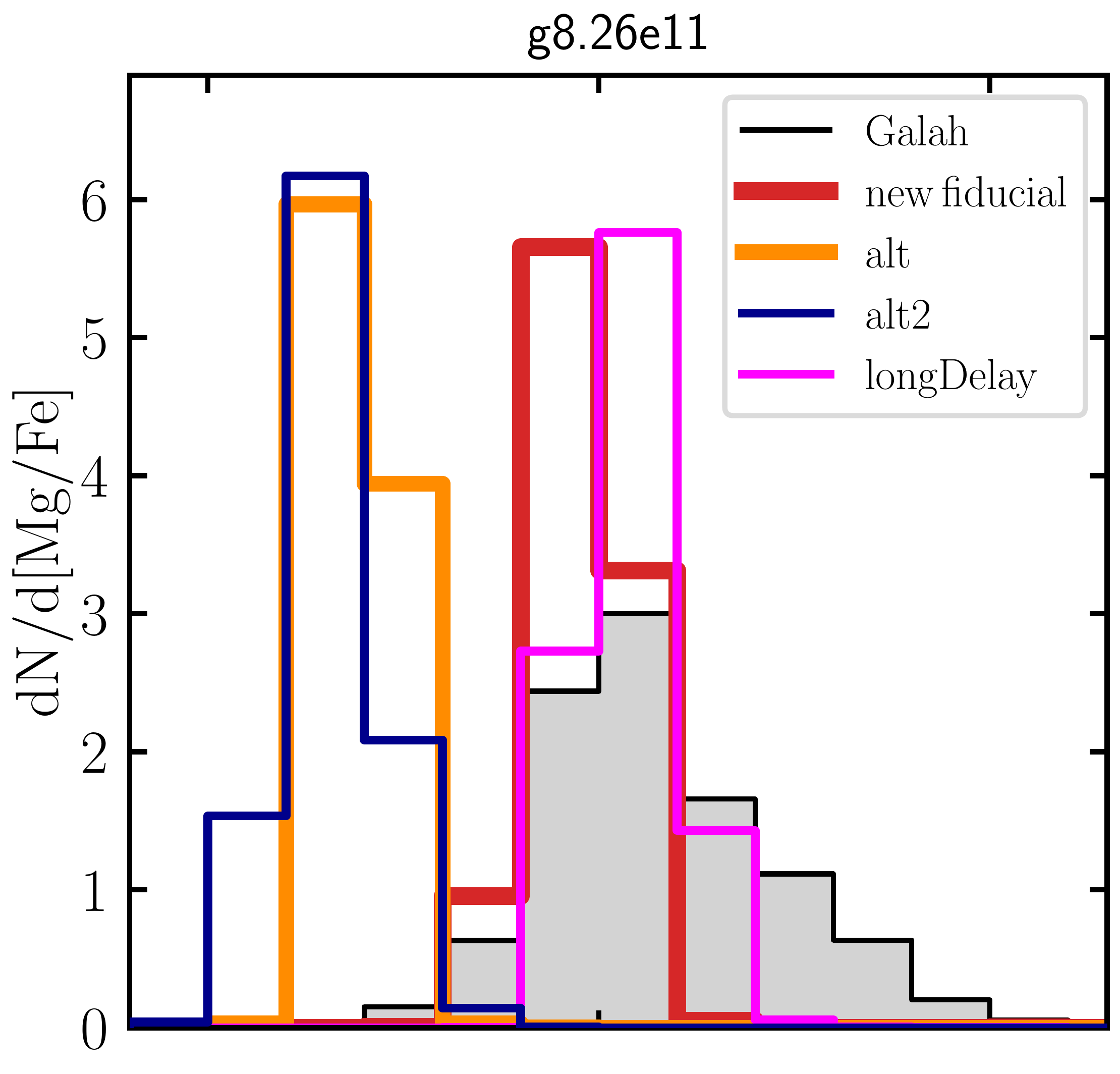}
\includegraphics[width=0.445\columnwidth]{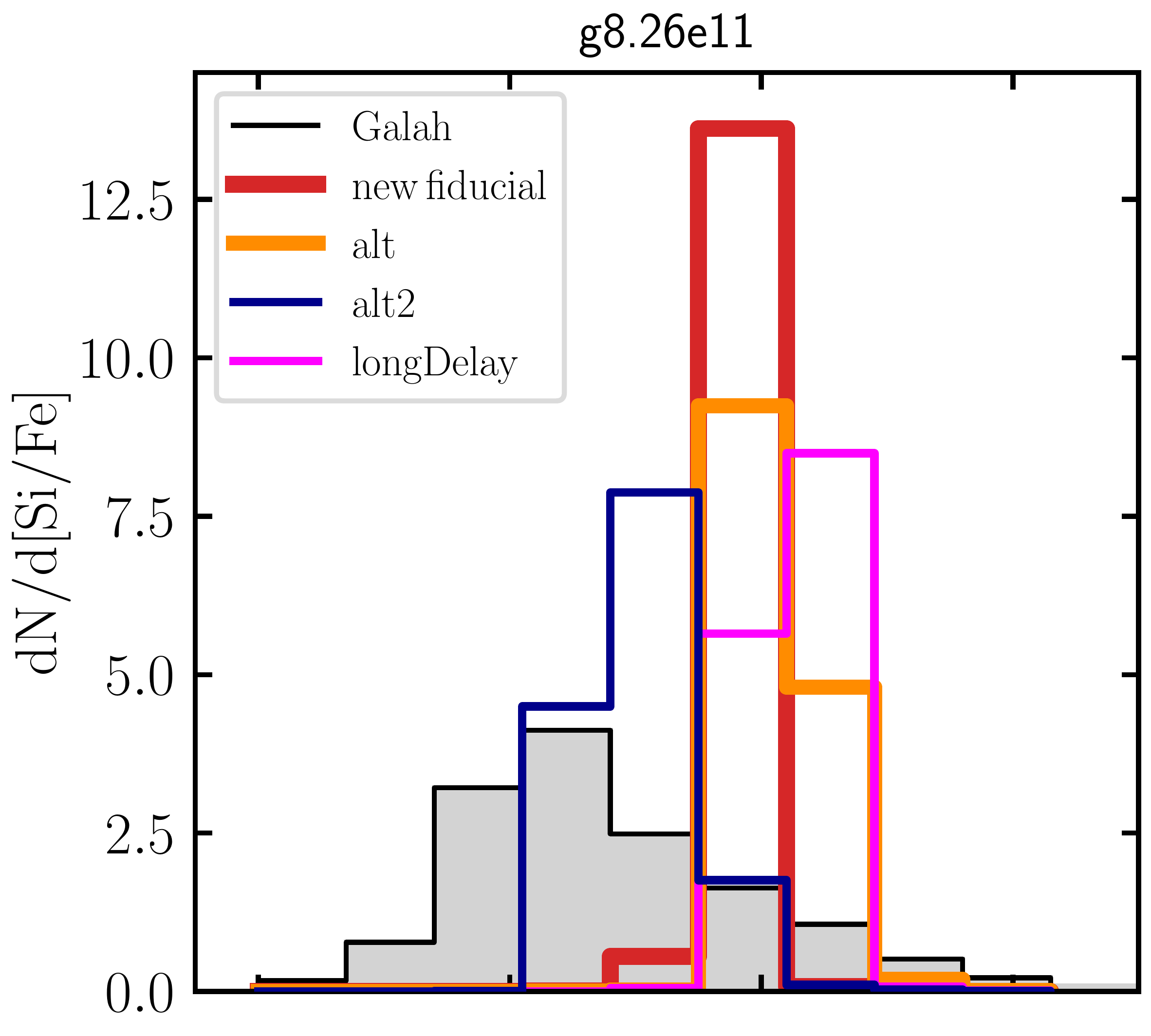}
\includegraphics[width=0.42\columnwidth]{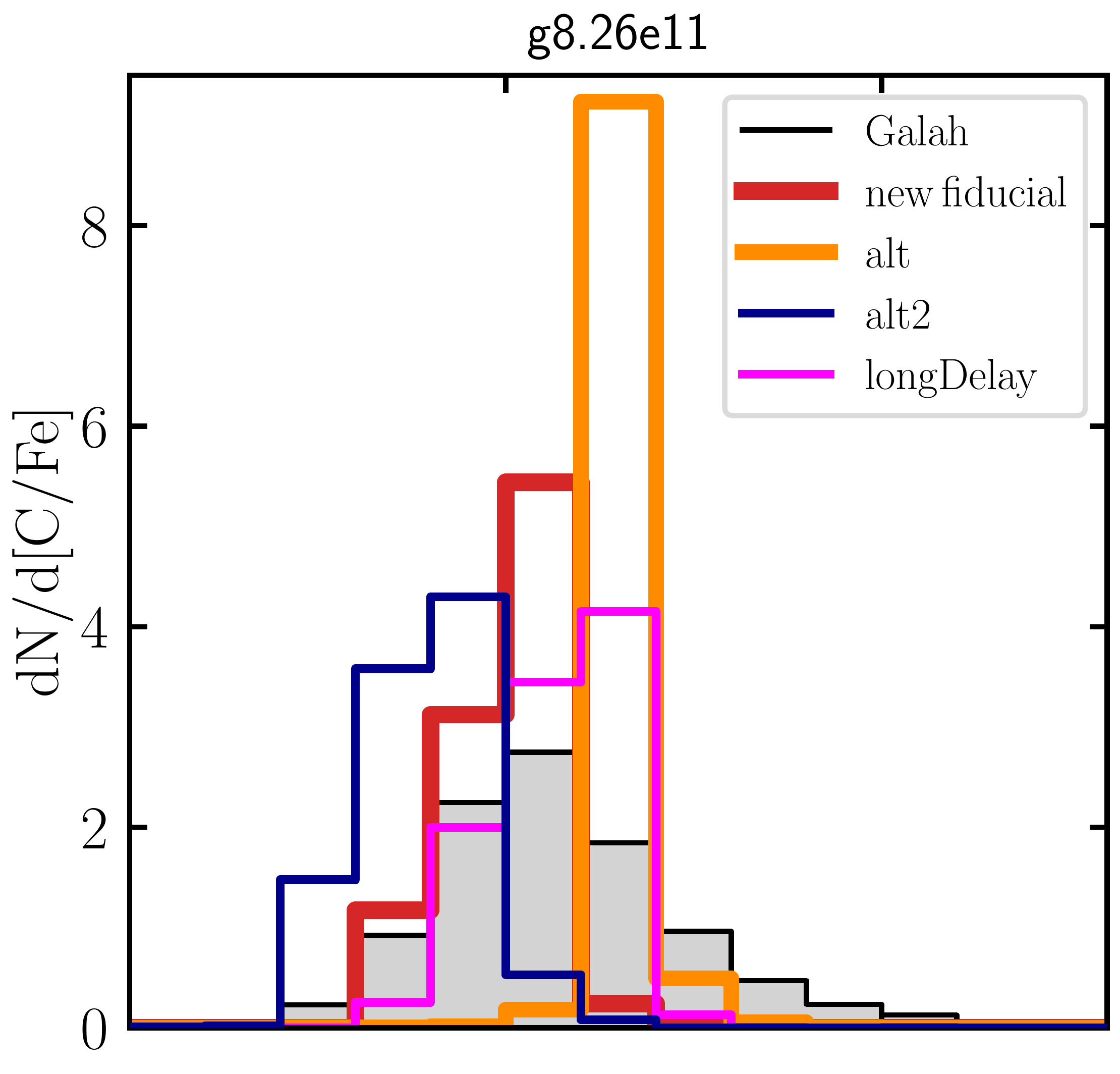}

\includegraphics[width=0.435\columnwidth]{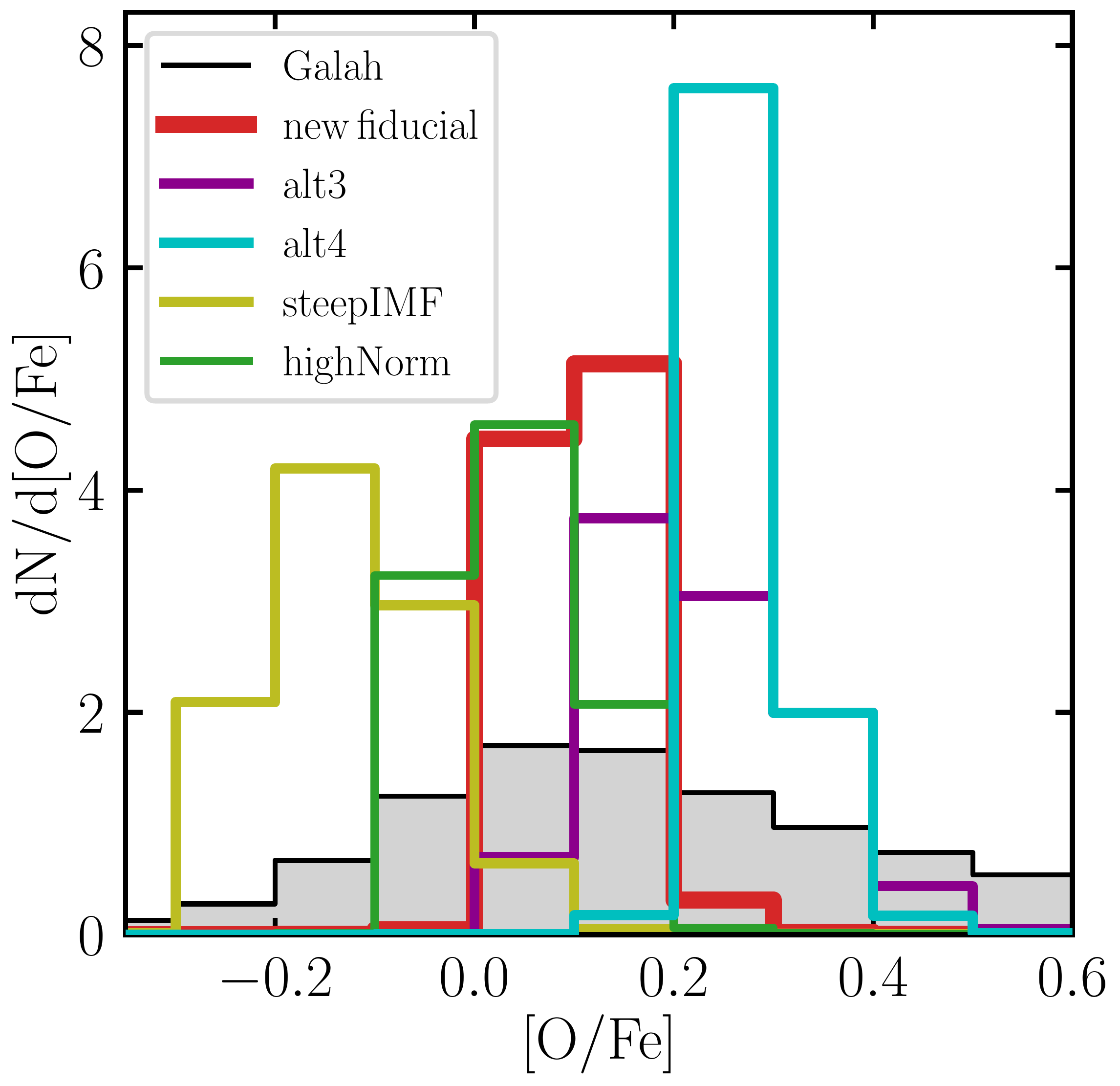}
\includegraphics[width=0.42\columnwidth]{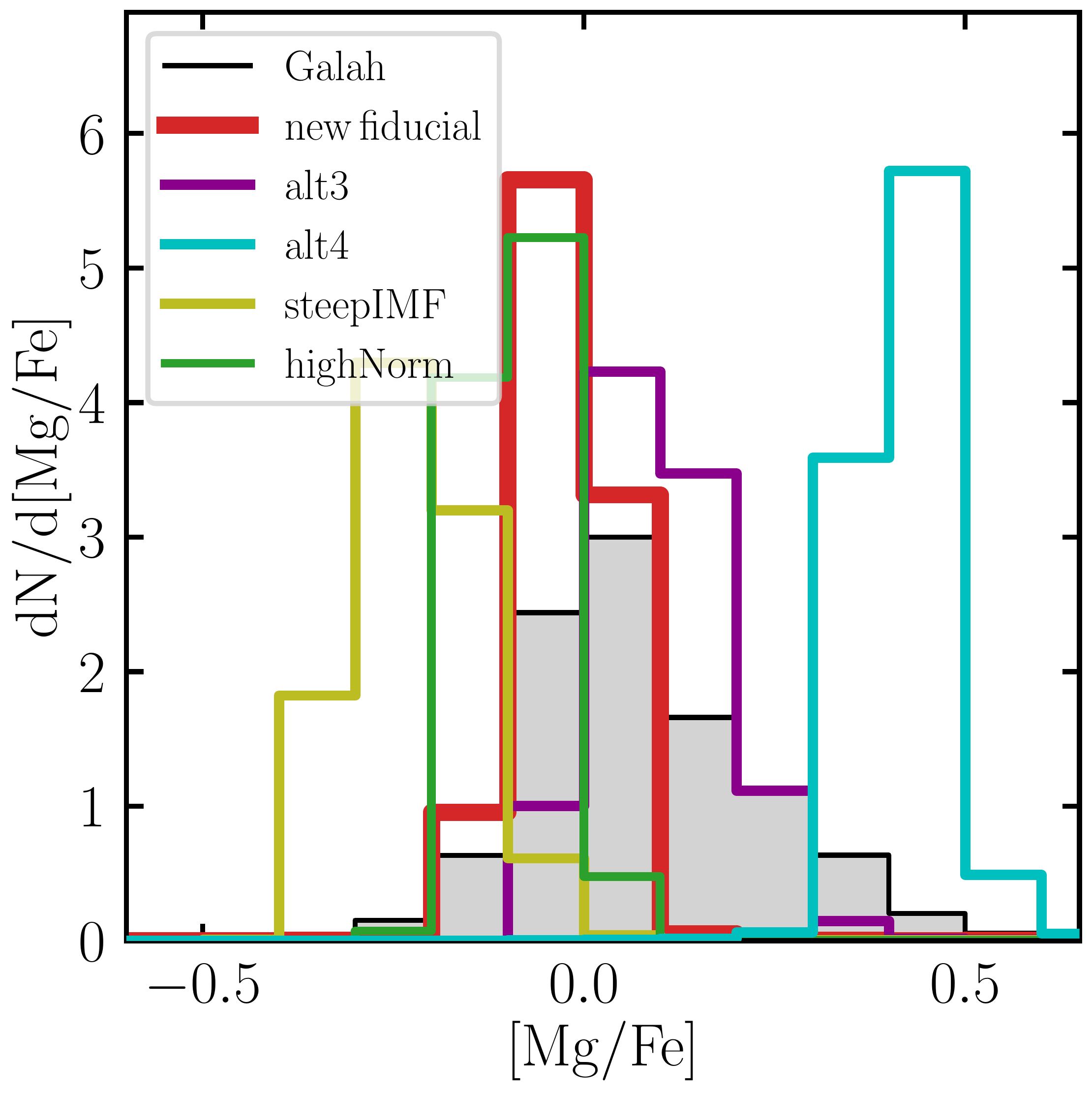}
\includegraphics[width=0.445\columnwidth]{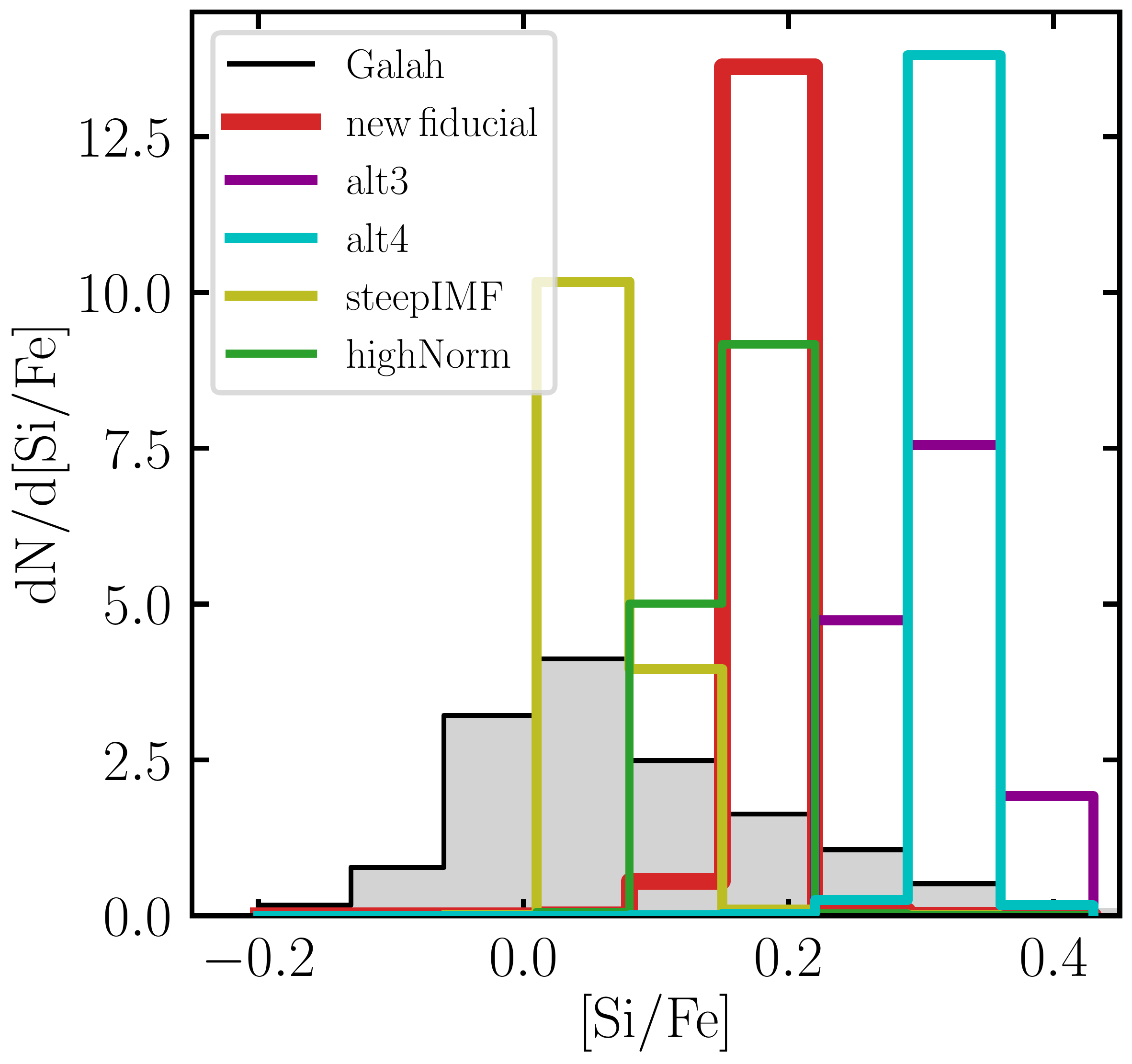}
\includegraphics[width=0.42\columnwidth]{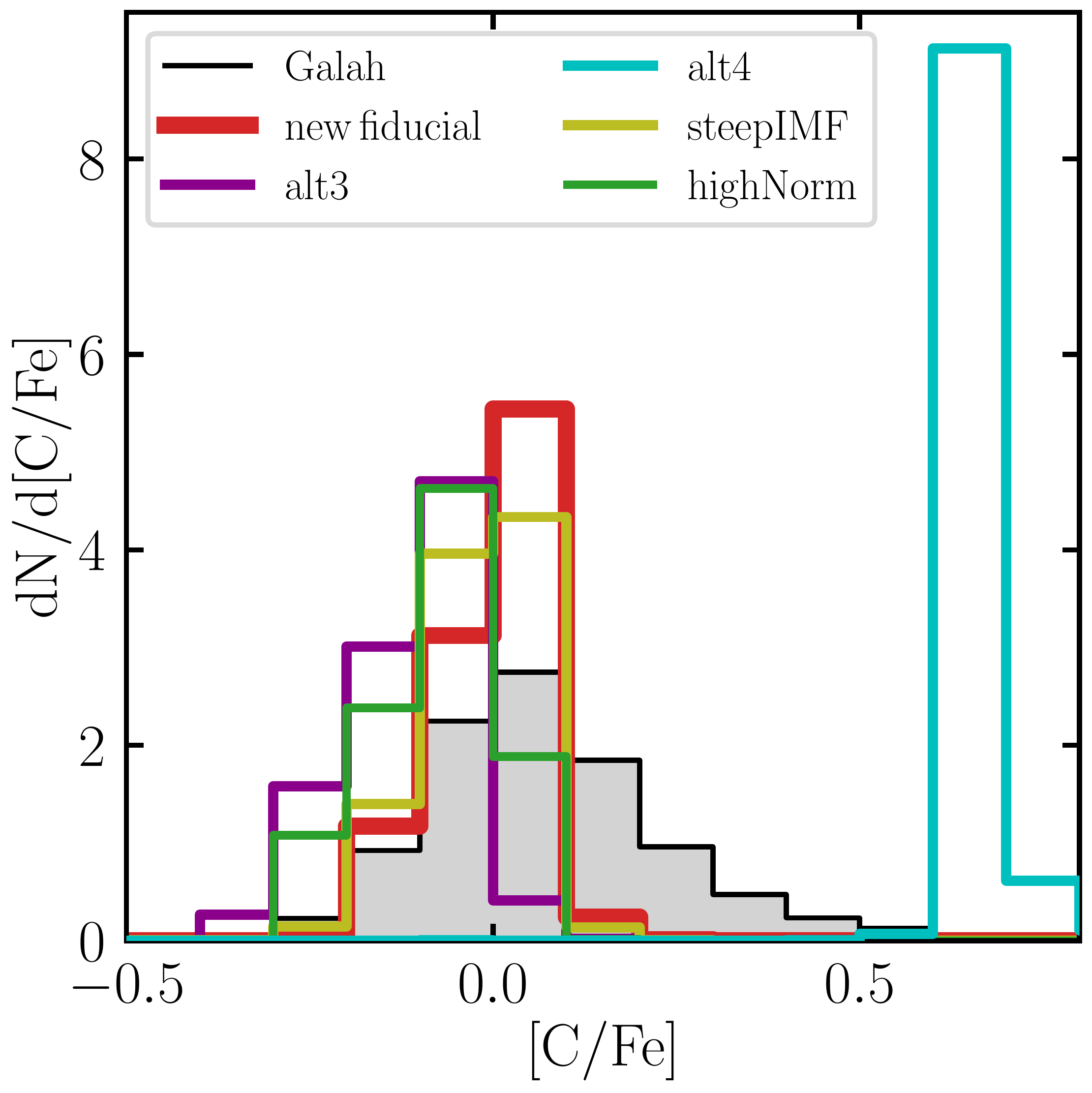}
\end{center}
\vspace{-.35cm}
\caption{Same as Fig. \ref{fig:omgsis_hist} but for the MW mass galaxy g8.26e11 for various different yield sets tested in this work. The gray filled histogram shows observational data for the MW from the GALAH survey.}
\label{fig:omgsis_hist2}
\end{figure*}

Figure \ref{fig:z_hist} shows the distribution functions of total stellar metallicity for the dwarf galaxy \texttt{g2.83e10} (left panels) and the MW mass galaxy \texttt{g8.26e11} (right panels) for various different yield sets tested in this work. Figure \ref{fig:omgsis_hist} and \ref{fig:omgsis_hist2} show additional distribution functions for the elements O, Mg, Si and C.

Figure \ref{fig:mass_comp_hist} shows the $\alpha$-element distribution function for our fiducial model comparing all model galaxies. The left panel shows the three dwarf galaxies while the right panel shows the more massive galaxies including the three L$^\star$ galaxies (\texttt{g2.79e12, g8.26e11, g7.55e11}). The left panel shows the expected dependence of average $\alpha$-enhancement on stellar mass for the dwarf galaxies where more massive galaxies (\texttt{g4.99e10}) show larger $\alpha$-enhancement. The overall width of the distribution is rather independent of stellar mass. On the other hand, the more massive galaxies do not show such a dependence of average $\alpha$-enhancement on stellar mass but they do show indications that the width of the distribution seems to broader for the more massive galaxies (\texttt{g2.79e12, g8.26e11}). 

%%%%%%%%%%%%%%%%%% FIGURE A5 %%%%%%%%%%%%%%%%%%%%%%%%%%%
%%%%%%%%%%%%%%%%%%%%%%%%%%%%%%%%%%%%%%%%%%%%%%%%%%%%

%%%%%%%%%%%%%%%%%% FIGURE A6 %%%%%%%%%%%%%%%%%%%%%%%%%%%
%%%%%%%%%%%%%%%%%%%%%%%%%%%%%%%%%%%%%%%%%%%%%%%%%%%%

\section{Additional CC SN yields}
%%%%%%%%%%%%%%%%%% FIGURE A7 %%%%%%%%%%%%%%%%%%%%%%%%%%%
\begin{figure*}
\begin{center}
\includegraphics[width=\textwidth]{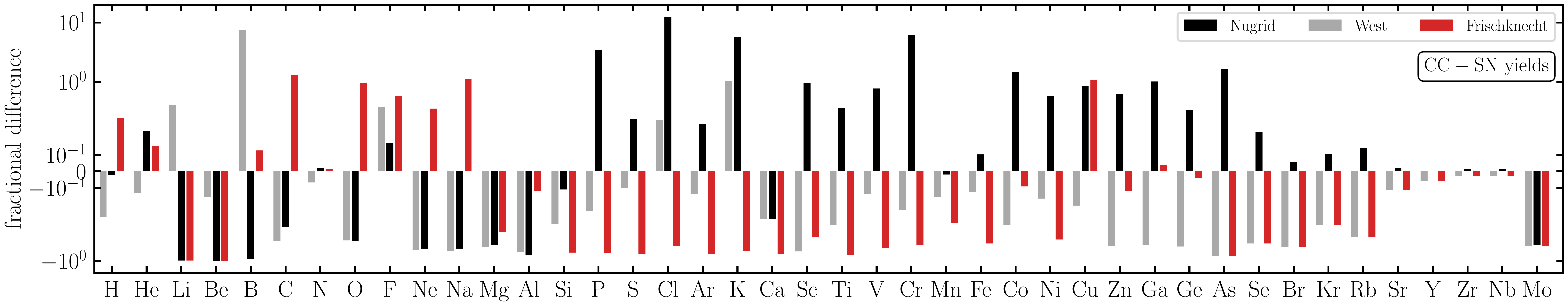}
\end{center}
\vspace{-.325cm}
\caption{Same as Fig. \ref{fig:yield_comparison} but this time comparing the CC SN yields from \citet{Frischknecht2016}, West \& Heger (in prep.) and from the Nugrid collaboration \citep{Ritter2018} to our fiducial yield-set.}
\label{fig:yield_comparison1}
\end{figure*}
%%%%%%%%%%%%%%%%%%%%%%%%%%%%%%%%%%%%%%%%%%%%%%%%%%%%

Figure \ref{fig:yield_comparison1} shows the same as Fig. \ref{fig:yield_comparison} in the main text but for three additional CC-SN yield sets from the literature. 

\label{lastpage}
\end{document}